\documentclass[twocolumn]{article}
\usepackage[latin1]{inputenc}
\usepackage{makeidx}
\usepackage{amsmath}
\usepackage{amstext}
\usepackage{amsthm}
\usepackage{amssymb}
\usepackage{rotating}
\usepackage{graphicx}
\usepackage{diagxy}
\usepackage{stmaryrd} 

\newtheorem*{definition}{Definition}

\newcommand{\TMP}[3]{T^{\phantom{i}#1}_{{#2}{#3}}M}
\newcommand{\STM}[2]{\Gamma T^{#1}_{#2}M}
\newcommand{\ST}{\Gamma T}
\newcommand{\STMV}[2]{\Gamma T^{\Vert #1}_{\phantom{\Vert} #2}M}
\newcommand{\STSigma}[2]{\Gamma T^{#1}_{#2}\Sigma}
\newcommand{\End}{\mathrm{End}}
\newcommand{\Trace}{\mathrm{Tr}}
\newcommand{\rank}[2]{(#1{,}#2)}
\newcommand{\EMT}{\mathbf{T}}
\newcommand{\Riem}{\mathbf{Riem}}
\newcommand{\RiemMet}{\mathrm{Riem}}
\newcommand{\Ricci}{\mathbf{Ric}}

\newcommand{\Ein}{\mathbf{Ein}}
\newcommand{\Scalar}{\mathbf{Scal}}
\newcommand{\Weyl}{\mathbf{Weyl}}
\newcommand{\Sec}{\mathbf{Sec}}
\newcommand{\Wein}{\mathbf{Wein}}
\newcommand{\kuno}{\owedge}
\newcommand{\Span}{\mathrm{span}}
\newcommand{\reals}{\mathbb{R}}
\newcommand{\integers}{\mathbb{Z}}
\newcommand{\Emb}{\mathcal{E}}
\newcommand{\Embeddings}{\mathrm{Emb}}
\newcommand{\ED}{\mathcal{E}}  
\newcommand{\ECD}{\mathcal{S}} 
\newcommand{\MD}{\mathcal{M}}  
\newcommand{\signature}{\varepsilon}
\newcommand{\Constraint}{\mathcal{C}}
\newcommand{\FL}{\mathrm{FL}}
\newcommand{\Fl}{\mathrm{Fl}}
\newcommand{\GL}{\mathrm{GL}}

\newcommand{\LDensity}{\mathcal{L}}
\newcommand{\HDensity}{\mathcal{H}}
\newcommand{\Lie}{\mathrm{Lie}}
\newcommand{\Diff}{\mathrm{Diff}}
\newcommand{\RP}{\mathbb{R}\mathrm{P}}
\newcommand{\D}{\mathcal{D}}
\newcommand{\Id}{\mathrm{Id}}
\newcommand{\Ad}{\mathrm{Ad}}
\newcommand{\ad}{\mathrm{ad}}
\newcommand{\Momentum}{\mathfrak{M}}
\newcommand{\Gau}{\mathrm{Gau}}
\newcommand{\Bij}{\mathrm{Bij}}
\newcommand{\Stab}{\mathrm{Stab}}
\newcommand{\Orb}{\mathrm{Orb}}
\makeindex

\begin{document}
\title{Dynamical and Hamiltonian formulation of General Relativity}
\author{Domenico Giulini                       \\
        Institute for Theoretical Physics      \\
        Riemann Center for Geometry and Physics\\
        Leibniz University Hannover, Appelstrasse 2, D-30167      
        Hannover, Germany\\
        and\\ 
        ZARM Bremen, Am Fallturm, D-28359 Bremen, Germany}

\date{}

\twocolumn[
\begin{@twocolumnfalse}
\maketitle
\begin{abstract}
\noindent
This is a substantially expanded version of a chapter-contribution to  
\emph{The Springer Handbook of Spacetime}, edited by Abhay Ashtekar 
and Vesselin Petkov, published by Springer Verlag in 2014. It  
introduces the reader to the reformulation of Einstein's 
field equations of General Relativity as a constrained evolutionary 
system of Hamiltonian type and discusses some of its uses, 
together with some technical and conceptual 
aspects. Attempts were made to keep the presentation self 
contained and accessible to first-year graduate students. 
This implies a certain degree of explicitness and occasional 
reviews of background material. 
\vspace{1.0cm}
\end{abstract}
\end{@twocolumnfalse}
]

\begin{footnotesize}
\setcounter{tocdepth}{1}
\tableofcontents
\end{footnotesize}

\section{Introduction}
\label{sec:Introduction}
The purpose of this contribution is to explain how the field 
equations of General Relativity---often simply referred to as 
\index{Einstein!equations}%
\emph{Einstein's equations}---can be understood as dynamical 
system; more precisely, as a 
\index{constrained Hamiltonian system}
\emph{constrained Hamiltonian system}.

In General Relativity, it is often said, spacetime becomes 
dynamical.
\index{dynamical spacetime}
\index{spacetime (dynamical)}
This is meant to say that the geometric structure of spacetime
is encoded in a field that, in turn, is subject to local laws 
of propagation and coupling,  just as, e.g., the electromagnetic
field. It is \emph{not} meant to say that spacetime as a whole 
evolves. Spacetime does not evolve, spacetime just is.  
\index{evolution of space}%
\index{history of space}%
But a given spacetime (four dimensional) can be viewed as the 
evolution, or history, of space (three dimensional). There 
is a huge redundancy in this representation,
\index{redundancy!in representing spacetime}
in the sense that 
apparently very different evolutions of space represent the same 
spacetime. However, if the resulting spacetime is to satisfy 
Einstein's equations, the evolution of space must also obey 
certain well defined restrictions. Hence the task is to give 
precise mathematical expression to the redundancies
\index{redundancy!in representing spacetime}
in representation as well as the restrictions of evolution 
for this picture of spacetime as space's history. This will 
be our main task. 

This dynamical picture will be important for posing and solving
time-dependent problems in General Relativity, like the scattering 
of black holes with its subsequent generation and radiation of 
gravitational waves. Quite generally, it is a key technology to 
\begin{itemize}
\item
formulate and solve initial value problems; 
\item
integrate Einstein's equations by numerical codes;
\item
characterize dynamical degrees of freedom;
\item
characterize isolated systems and the association of
asymptotic symmetry groups, which will give rise to 
globally conserved `charges', like energy and linear 
as well as angular momentum (Poincar\'e charges).
\index{Poincar\'e!charges}
\end{itemize}
Moreover, it is also the starting point for the 
\emph{canonical quantization program}, 
\index{canonical!quantization}
\index{quantization, canonical}
which constitutes one main approach to the yet unsolved problem 
of \emph{Quantum Gravity}.
\index{Quantum Gravity}
In this approach one tries to make essential use of the 
Hamiltonian structure of the classical theory in formulating 
the corresponding quantum theory. This strategy has been applied 
successfully in the transition from classical to quantum mechanics
and also in the transition from classical to quantum electrodynamics. 
Hence the canonical approach to Quantum Gravity may be regarded 
as conservative, insofar as it tries to apply otherwise established 
rules to a classical theory that is experimentally and observationally 
extremely well tested. The underlying hypothesis here is that we may 
quantize interaction-wise. This distinguishes this approach from 
string theory, the underlying credo of which is that Quantum Gravity 
only makes sense on the basis of a unified descriptions of all 
interactions.  

Historically the first paper to address the problem of how to 
put Einstein's equations into the form of a Hamiltonian dynamical 
system was Dirac's \cite{Dirac:1958-b} from 1958. He also 
noticed its constrained nature and started to develop the 
corresponding generalization of \emph{constrained Hamiltonian systems}
\index{constrained Hamiltonian system} 
in \cite{Dirac:1958-a} and their quantization \cite{Dirac:LQM}.
On the classical side, this developed into the more geometric 
\emph{Dirac-Bergmann theory} of constraints \cite{Gotay.Nester.Hinds:1978}
and on the quantum side into an elaborate theory of quantization 
of systems with gauge redundancies; 
\index{redundancy!gauge}
see \cite{HenneauxTeitelboim:QGS} for a comprehensive account. 

Dirac's attempts were soon complemented by 
an extensive joint work of Richard Arnowitt, Stanley Deser, and Charles 
Misner - usually and henceforth abbreviated by ADM.
\index{ADM} 
Their work started in 1959 by a paper \cite{Arnowitt.Deser:1959-a}
of the first two of these authors and continued in the series 
\cite{Arnowitt.Deser.Misner:1959-b}
\cite{Arnowitt.Deser.Misner:1960-a}
\cite{Arnowitt.Deser.Misner:1960-b}
\cite{Arnowitt.Deser.Misner:1960-c}
\cite{Arnowitt.Deser.Misner:1960-d}
\cite{Arnowitt.Deser.Misner:1960-e}
\cite{Arnowitt.Deser.Misner:1960-f}
\cite{Arnowitt.Deser.Misner:1960-g}
\cite{Arnowitt.Deser.Misner:1960-h}
\cite{Arnowitt.Deser.Misner:1961-a}
\cite{Arnowitt.Deser.Misner:1961-b}
\cite{Arnowitt.Deser.Misner:1961-c}
of 12 more papers by all three. A comprehensive summary of their work 
was given in 1962 in \cite{Arnowitt.Deser.Misner:1962}, which was 
republished in 2008 in \cite{Arnowitt.Deser.Misner-GoldenOldie:2008}; 
see also the editorial note \cite{Pullin:2008} with short biographies 
of ADM. 

A geometric discussion of Einstein's evolution equations in terms of 
infinite-dimensional symplectic geometry has been worked out by 
Fischer and Marsden in \cite{Fischer.Marsden:1972}; see also their 
beautiful summaries and extended discussions in 
\cite{Fischer.Marsden:1979-1} and \cite{Fischer.Marsden:1979-2}.
More on the mathematical aspects of the initial-value problem,
including the global behavior of gravitational fields in General 
Relativity, can be found in \cite{Choquet-Bruhat:GR}, 
\cite{ChruscielFriedrich:EinsteinEquations}, and 
\cite{Rendall:PDEinGR}.
Modern text-books on the 3+1 formalism and its application to 
physical problems and their numerical solution-techniques are 
\cite{Baumgarte.Shapiro:NumericalRelativity,Gourgoulhon:ThreePlusOneGR}.
The Hamiltonian structure and its use in the canonical quantization 
program for gravity is discussed in
\cite{Bojowald:CanonicalGravity,Kiefer:QuantumGravity,Rovelli:QuantumGravity,Thiemann:MCQGR}.

\section{Notation and conventions}
\label{sec:NotationConventions}
From now on ``General Relativity'' will be abbreviated by ``GR''.
Spacetime is a differentiable manifold $M$ of dimension $n$,
endowed with a metric $g$ of signature $(\signature,+,\cdots,+)$.
In GR $n=4$ and $\signature=-1$ and it is implicitly understood
that these are the ``right'' values. However, either for the 
sake of generality and/or particular interest, we will 
sometimes state formulae for general $n$ and $\signature$, 
where usually $n\geq 2$ (sometimes $n\geq 3$) and either 
$\signature=-1$ (Lorentzian metric) or  $\signature=+1$
 \index{Lorentz!metric}\index{metric!Lorentzian}
(Riemannian metric; also called Euclidean metric).
\index{Riemannian metric}\index{metric!Riemannian}
\index{Euclidean metric}\index{metric!Euclidean} 
The case $\signature=1$ has been extensively considered in 
path-integral approaches to Quantum Gravity, then referred 
to as \emph{Euclidean Quantum Gravity}. 
\index{path integral}
\index{Euclidean Quantum Gravity}

The tangent space of $M$ at point $p\in M$ will be denoted by 
$T_pM$, the cotangent space by $T^*_pM$, and the tensor product 
of $u$ factors of $T_pM$ with $d$ factors of $T^*_pM$ by $\TMP{u}{p}{d}$. 
(Mnemonic in components: $u=$ number of indices ``upstairs'', 
$d=$ number of indices ``downstairs''.) 
An element $T$ in $\TMP{u}{p}{d}$ is called a tensor of 
contravariant rank $u$ and covariant rank $d$ at point 
$p$, or simply a tensors of rank $\rank{u}{d}$ at $p$.
\index{tensor}
$T$ is called \emph{contravariant} 
\index{contravariant tensor}
\index{tensor!contravariant}
if $d=0$ and $u>0$, and \emph{covariant} if $u=0$ and $d>0$.
\index{covariant!tensor}
\index{tensor!covariant}
A tensor with $u>0$ and $d>0$ is then referred to as of
\emph{mixed type}. 
\index{mixed tensor}
\index{tensor!mixed}
Note that $T_pM=\TMP{1}{p}{0}$ and  $T^*_pM=\TMP{0}{p}{1}$.
The set of tensor \emph{fields}, 
\index{tensor!fields}
i.e. smooth assignments of an element in $\TMP{u}{p}{d}$
for each $p\in M$, are denoted by $\STM{u}{d}$. Unless 
stated otherwise, \emph{smooth} means $C^\infty$, i.e. 
continuously differentiable to any order. For $t\in\STM{u}{d}$ 
we denote by $t_p\in\TMP{u}{p}{d}$ the evaluation of $t$ at 
$p\in M$. $C^\infty(M)$ denotes the set of all $C^\infty$ real-valued 
functions on $M$, which we often simply call smooth functions. 

If $f: M\rightarrow N$ is a diffeomorphism between manifolds
$M$ and $N$, then $f_{*p}:T_pM\rightarrow T_{f(p)}N$ denotes the 
differential at $p$. The transposed (or dual) of the latter map 
is, as usual, denoted by $f^*_p:T^*_{f(p)}N\rightarrow T^*_pM$.  
If $X$ is a vector field on $M$ then $f_*X$ is a vector field on $N$, 
called the \emph{push forward} 
\index{push forward}
of $X$ by $f$. It is defined by $(f_*X)_q:=f_{*f^{-1}(q)}X_{f^{-1}(q)}$, 
for all $q\in N$. If $\alpha$ is a co-vector field on $N$ then 
$f^*\alpha$ is a co-vector field on $M$, called the \emph{pull back} 
\index{pull back}
of $\alpha$ by $f$. It is defined by 
$(f^*\alpha)_p:=\alpha_{f(p)}\circ f_{*p}$, for all $p\in M$.
For these definitions to make sense we see that $f$ need 
generally not be a diffeomorphism; $M$ and $N$ need not even be 
of the same dimension. More precisely, if $\alpha$ is a smooth 
field of co-vectors that is defined at least on the image of 
$f$ in $N$, then $f^*\alpha$, as defined above, is always a 
smooth field of co-vectors on $M$. However, for the push 
forward $f_*$ of a general vector field on $M$ to result in a 
well defined vector field on the image of $f$ in $N$ we 
certainly need injectivity of $f$. If $f$ is a diffeomorphism 
we can not only push-forward vectors and pull back co-vectors, 
but also vice versa. Indeed, if $Y$ is a vector field on $N$ 
one can write $f^*Y:=(f^{-1})_*Y$ and call it the pull back of 
$Y$ by the diffeomorphism $f$. Likewise, if $\beta$ is a co-vector 
field on $M$, one can write $f_*\beta:=(f^{-1})^*\beta$ and call 
it the push forward of $\beta$. In this fashion we can define 
both, push-forwards and pull-backs, of general tensor fields
$T\in\STM{u}{d}$ by linearity and applying $f_*$ or $f^*$ 
tensor-factor wise.

Note that the general definition of \emph{metric} 
\index{metric!on manifold}
is as follows: $g\in\STM{0}{2}$, such that $g_p$ is a symmetric
non-degenerate bilinear form on $T_pM$. Such a metric provides 
isomorphisms (sometimes called the \emph{musical isomorphisms})
\index{musical isomorphisms}
\begin{subequations}
\label{eq:MusicalIsomorphisms}
\begin{alignat}{1}
\label{eq:MusicalIsomorphisms-a}
\flat: T_pM&\rightarrow T_p^*M\nonumber\\
          X&\mapsto X^\flat:=g(X,\,\cdot\,)\,,\\
\label{eq:MusicalIsomorphisms-b}
\sharp: T^*_pM&\rightarrow T_pM\nonumber\\
        \omega&\mapsto \omega^\sharp:=\flat^{-1}(\omega)\,.
\end{alignat}
\end{subequations}
Using $\sharp$ we obtain a metric $g_p^{-1}$ on $T^*_pM$ from 
the metric $g_p$ on $T_pM$ as follows:
\begin{equation}
\label{eq:DefInverseMetric}  
g_p^{-1}(\omega_1,\omega_2):=g_p(\omega^\sharp_1,\omega^\sharp_2)
=\omega_1(\omega_2^\sharp)\,.
\end{equation}
We also recall that the tensor space $\TMP{1}{p}{1}$ is naturally 
isomorphic to the linear space $\End(T_pM)$ of all endomorphisms 
(linear self maps) of $T_pM$. Hence it carries a natural structure 
as associative algebra,
\index{algebra!associative}
the product being composition of 
maps denoted by $\circ$. As usual, the \emph{trace}, denoted 
$\Trace$, and the \emph{determinant}, denoted $\det$, are 
the naturally defined real-valued functions on the space of 
endomorphisms. For purely co- or contravariant tensors the 
trace can be defined by first applying one of the isomorphisms
\eqref{eq:MusicalIsomorphisms}. In this case we write $\Trace_g$
to indicate the dependence on the metric $g$. 

Geometric representatives of curvature are often denoted by 
bold-faced abbreviations of their names, like $\Riem$ and $\Weyl$ 
for the (covariant, i.e. all indices down) Riemann and Weyl 
tensors, $\Sec$ for the sectional curvature, $\Ricci$ 
and $\Ein$ for the Ricci and Einstein tensors, $\Scalar$ for 
the scalar curvature, and $\Wein$ for the Weingarten map 
(which is essentially equivalent to the extrinsic curvature). 
This is done in order to highlight the geometric meaning behind 
some basic formulae, at least the simpler ones. 
Later, as algebraic expressions become more involved, we 
will also employ the standard component notation for 
computational ease.

\section{Einstein's equations}
\label{sec:EinsteinsEquations}
In $n$-dimensional spacetime Einstein's equations form a set of 
$\tfrac{1}{2}n(n+1)$ quasi-linear partial differential equations 
of second order for $\tfrac{1}{2}n(n+1)$ functions (the components 
of the metric tensor) depending on $n$ independent variables (the 
coordinates in spacetime). At each point of spacetime (event) they 
equate a purely geometric quantity to the distribution of energy 
and momentum carried by the matter. More precisely, this distribution 
comprises the local densities (quantity per unit volume) and 
current densities (quantity per unit area and unit time) of energy and 
momentum. The geometric quantity in Einstein's equations is the 
\index{Einstein!tensor}\index{tensor!Einstein}%
Einstein tensor $\Ein$, the matter quantity is the 
\index{energy-momentum tensor}\index{tensor!energy-momentum}%
energy-momentum tensor $\EMT$. Both tensors are 
of second rank, symmetric, and here taken to be covariant
(in components: ``all indices down''). Their number of 
independent components in $n$ spacetime dimensions is $\tfrac{1}{2}n(n+1)$ 

Einstein's equations (actually a single tensor equation, but throughout 
we use the plural to emphasize that it comprises several component equations)
\index{Einstein!equations}
state the simple proportionality of $\Ein$ with~$\EMT$
\begin{equation}
\label{eq:EinsteinTensorEq}
\Ein=\kappa\,\EMT\,,
\end{equation}
where $\kappa$ denotes the dimensionful constant of proportionality.  
Note that no explicit reference to the dimension $n$ of spacetime 
enters \eqref{eq:EinsteinTensorEq}, so that even if $n\ne 4$ it is 
usually referred to as Einstein's equations. We could have explicitly 
added a \emph{cosmological constant} term
\index{cosmological constant}
$g\Lambda$ on the left-hand side, where $\Lambda$ is a constant 
the physical dimension of which is the square of an inverse length.
However, as long as we write down our formulae for general $\EMT$
we may absorb this term into $\EMT$ where it accounts for a 
contribution $\EMT_\Lambda=-g\Lambda/\kappa$. This has to be kept in 
mind when explicit models for $\EMT$ are used and when we speak of 
``vacuum'', which now means:
\begin{equation}
\label{eq:T-Vacuum} 
\EMT_{\rm vacuum}=\EMT_\Lambda:=-\kappa^{-1}g\Lambda\,.
\end{equation}  
The signs are chosen such that a positive $\Lambda$ accounts for 
a positive energy density and a negative pressure if 
the spacetime is Lorentzian ($\signature=-1$). 

There is another form of Einstein's equations which is sometimes 
advantageous to use and in which $n$ explicitly enters:  
\begin{equation}
\label{eq:EinsteinTensorEq-Alt}
\Ricci=\kappa\,\Bigl(\EMT-\tfrac{1}{n-2}g\,\Trace_g({\EMT})\Bigr)\,.
\end{equation}
These two forms are easily seen to be mathematically equivalent 
by the identities 
\begin{subequations}
\label{eq:EinsteinRicciRelation}
\begin{alignat}{2}
\label{eq:EinsteinRicciRelation-a}
& \Ein&&\,=\,\Ricci-\tfrac{1}{2}g\,\Trace_g(\Ricci)\,,\\
\label{eq:EinsteinRicciRelation-b}
& \Ricci&&\,=\,\Ein-\tfrac{1}{n-2}g\,\Trace_g(\Ein)\,.
\end{alignat}
\end{subequations}
With respect to a local field of basis vectors 
$\{e_0,e_1,\cdots, e_{n-1}\}$ we write 
$\Ein(e_\mu,e_\nu)=:G_{\mu\nu}$, $\EMT(e_\mu,e_\nu)=:T_{\mu\nu}$,
and $\Ricci(e_\mu,e_\nu)=:R_{\mu\nu}$. Then \eqref{eq:EinsteinTensorEq}
and \eqref{eq:EinsteinTensorEq-Alt} take on the component forms  
\begin{equation}
\label{eq:EinsteinComponentEq}
G_{\mu\nu}=\kappa\,T_{\mu\nu}
\end{equation}
and 
\begin{equation}
\label{eq:EinsteinComponentEq-Alt}
R_{\mu\nu}=\kappa\,
\left(T_{\mu\nu}-\tfrac{1}{n-2}g_{\mu\nu}T^\lambda_\lambda\right)
\end{equation}
respectively. Next we explain the meanings of the symbols 
in Einstein's equations from left to right. 

\subsection{What aspects of geometry?}
\label{sec:EinstEqGeometry}
The left-hand side of Einstein's equations comprises certain 
measures of curvature. As will be explained in detail in  
Section\,\ref{sec:CurvatureTensors}, all curvature information 
in dimensions higher than two can be reduced to that of \emph{sectional curvature}.
\index{sectional curvature}
The sectional curvature at a point $p\in M$ tangent to 
$\Span\{X,Y\}\subset T_pM$ is the Gaussian curvature 
\index{Gaussian curvature}
\index{curvature!Gaussian}
at $p$ of the submanifold spanned by the geodesics in $M$ 
emanating from $p$ tangent to $\Span\{X,Y\}$. The Gaussian 
curvature is defined 
as the product of two principal curvatures, each being measured 
in units of an inverse length (the inverse of a principal radius). 
Hence the Gaussian curvature is measured in units of an inverse 
length-squared. 

At each point $p$ in spacetime the Einstein and Ricci tensors are 
symmetric bilinear forms on $T_pM$. Hence $\Ein_p$ and $\Ricci_p$ 
are determined by the values $\Ein_p(W,W)$ and $\Ricci_p(W,W)$ for 
all $W\in T_pM$. By continuity in $W$ this remains true if we 
restrict $W$ to the open and dense set of vectors which are 
not null, i.e. for which $g(W,W)\ne 0$. As we will see later on, we 
then have 
\begin{alignat}{3}
\label{eq:EinDefSecCurv}
&\Ein(W,W)&&  \,=\,-&&g(W,W)\sum_{\perp W}^{N_1}\Sec\,,\\
\label{eq:RicciDefSecCurv}
&\Ricci(W,W)&&\,=\,+&&g(W,W)\sum_{\Vert W}^{N_2}\Sec\,.
\end{alignat}
For the Einstein tensor the sum on the right-hand side is over 
any complete set of $N_1=\frac{1}{2}(n-1)(n-2)$ sectional curvatures 
of pairwise orthogonal planes in the orthogonal complement of $W$ 
in $T_pM$. For the Ricci tensor it is over any complete set of 
$N_2=n-1$ sectional curvatures of pairwise orthogonal planes 
containing $W$. If $W$ is a timelike unit vector representing 
an observer, $\Ein(W,W)$ is simply $(-\signature)$ times an 
equally weighted sum of spacelike sectional curvatures, whereas 
$\Ricci(W,W)$ is $\signature$ times an equally weighted sum 
of timelike sectional curvatures. In that sense we may say that,
e.g., $\Ein(W,W)$ at $p\in M$ measures the mean Gaussian curvature of 
the (local) hypersurface in $M$ that is spanned by geodesics 
emanating from $W$ orthogonal to $W$. It, too, is measured in 
units of the square of an inverse length.  

\subsection{What aspects of matter?}
\label{sec:EinstEqMatter}
Now we turn to the right-hand side of Einstein's equations.
We restrict to four spacetime dimensions, though much 
of what we say will apply verbatim to other dimensions. 
The tensor $\EMT$ on the right-hand side of 
\eqref{eq:EinsteinTensorEq} is the \emph{energy-momentum tensor}
\index{energy-momentum tensor}
of matter. With respect to an orthonormal basis 
$\{e_0,e_1,\cdots,e_{n-1}\}$ with timelike $e_0$ the 
components  $T_{\mu\nu}:=\EMT(e_\mu,e_\nu)$ form 
a symmetric $4\times 4$ matrix, which we represent as 
follows by splitting off terms involving a time component: 
\begin{equation}
\label{eq:EM-TensorSplit}
T_{\mu\nu}=
\begin{pmatrix}
\ED&-c\vec\MD\\
-\frac{1}{c}\vec\ECD & T_{mn}  
\end{pmatrix}\,.
\end{equation}
Here all matrix elements refer to the matter's energy 
momentum distribution relative to the rest frame of the 
observer who momentarily moves along $e_0$ (i.e. with 
four-velocity $u=ce_0$) and uses the basis $\{e_1,e_2,e_3\}$
in his/her rest frame.  Then $\ED=T_{00}$ is the energy density,
\index{energy!density}
$\vec\ECD=(s_1,s_2,s_3)$ the (components of the) energy 
current-density, 
\index{energy!current-density}
i.e. energy per unit surface area and 
unit time interval, $\vec\MD$ the momentum density, 
\index{momentum!density}
and finally $T_{mn}$ the (component of the) momentum 
current-density,
\index{momentum!current-density} 
i.e. momentum per unit of area and unit time interval. Note that 
symmetry $T_{\mu\nu}=T_{\nu\mu}$ implies a simple relation 
between the energy current-density and the momentum density
\begin{equation}
\label{eq:SymmetryT}
\vec\ECD=c^2\,\vec\MD\,.
\end{equation}   
The remaining relations $T_{mn}=T_{nm}$ express equality 
of the $m$-th component of the current density for 
$n$-momentum with the $n$-th component of the current 
density for $m$-momentum. 
\index{symmetry, of energy-momentum tensor}
Note that the two minus signs in front of the mixed 
components of \eqref{eq:EM-TensorSplit} would have 
disappeared had we written down the contravariant
components $T^{\mu\nu}$.  In flat spacetime, the four 
equations $\partial^\mu T_{\mu\nu}$ express the local  
conservation of energy and momentum. In curved 
spacetime (with vanishing torsion) we have the 
identity (to be proven later; 
compare \eqref{eq:BianchiIdentityComp-second})
\begin{equation}
\label{eq:Bianchi2Contracted}
\nabla^{\mu}G_{\mu\nu}\equiv 0
\end{equation}
implies via \eqref{eq:EinsteinComponentEq}
\begin{equation}
\label{eq:DivergencelessT}
\nabla^{\mu}T_{\mu\nu}=0\,,
\end{equation}
which may be interpreted as expressing a local conservation 
of energy and momentum for the matter \emph{plus} the 
gravitational field, though there is no such thing as a 
separate energy-momentum tensor on spacetime for the 
gravitational field.  

Several positivity conditions can be imposed on the 
energy momentum tensor $\EMT$. The simplest is 
known as \emph{weak energy-condition} 
\index{energy condition!weak}
and reads $\EMT(W,W)\geq 0$ for all timelike $W$. It is equivalent 
to the requirement that the energy density measured by any local 
observer is non negative. For a perfect fluid of rest-mass density $\rho$ and
pressure $p$ the weak energy-condition is equivalent to both conditions 
$\rho\geq 0$ and $p\geq -c^2\rho$. The strong energy-condition 
\index{energy condition!strong}
says that $\bigl(\EMT-\tfrac{1}{2}g\Trace_g(\EMT)\bigr)(W,W)\geq 0$
again for all timelike $W$. This neither follows nor implies the weak 
energy-condition. For a perfect fluid it is equivalent to both conditions 
$p\geq -c^2\rho$ and $p\geq -c^2\rho/3$, i.e. to the latter alone if 
$\rho$ is positive and to the former alone if $\rho$ is negative (which 
is not excluded here). Its significance lies in the fact that it ensures 
attractivity of gravity as described by Einstein's equations. It must, for 
example, be violated if matter is to drive inflation. Note that upon imposing 
Einstein's equations the weak and the strong energy-conditions read 
$\Ein(W,W)\geq 0$ and $\Ricci(W,W)\geq 0$ respectively. From 
\eqref{eq:EinDefSecCurv} and \eqref{eq:RicciDefSecCurv} we can see 
that for fixed $W$ these imply conditions on complementary sets of 
sectional curvatures. 
For completeness we mention the \emph{condition of energy dominance},
\index{energy condition!energy dominance}
which states that $\EMT(W,W)\geq\vert\EMT(X,X)\vert$ for any pair 
of orthonormal vectors $W,X$ where $W$ is timelike (and hence 
$X$ is spacelike). It is equivalent to the weak energy-condition
supplemented by the requirement that $(i_W\EMT)^\sharp$ be non 
spacelike for all timelike $W$. The second requirement ensures 
locally measured densities of energy currents and momenta of 
matter to be non spacelike. 

\subsection{How do geometry and matter\\ relate quantitatively?}
\label{sec:EinstEqConstant}
We return to Einstein's equations and finally discuss the 
constant of proportionality $\kappa$ on the 
right-hand side of \eqref{eq:EinsteinTensorEq}. Its physical
dimension is that of curvature ($m^{-2}$ in SI units) divided 
by that of energy density ($J\cdot m^{-3}$ in SI units, where 
$J=$ Joule). It is given by  
\begin{equation}
\label{eq:KappaConst}
\kappa:=\frac{8\pi G}{c^4}\approx 2.1\times 10^{-43}\,
\frac{m^{-2}}{J\cdot m^{-3}}\,,
\end{equation}
where 
$G\approx 6.67384(80)\times 10^{-11}\mathrm{m^3\cdot kg^{-1}\cdot s^{-2}}$
is Newton's constant.
\index{Newton's constant}
\index{gravitational constant}
It is currently (March 2013) known with a relative 
standard uncertainty of $1.2\times 10^{-4}$ and is 
thus by far the least well known of the fundamental 
physical constants. $c=299\,792\,458\,\mathrm{m\cdot s^{-1}}$ 
is the vacuum speed of light 
\index{speed of light}
\index{velocity of light}
whose value is exact, due to the SI-definition of 
meter (``the meter is the length of the path traveled 
by light in vacuum during a time interval of $1/299\,792\,458$
of a second''). 

The physical dimension of $\kappa$ is 
$\mathrm{time^2/(mass\cdot length)}$, 
that is in SI-units $\mathrm{s^2\cdot kg^{-1}\cdot m^{-1}}$ or 
$\mathrm{m^{-2}/(J\cdot m^{-3})}$, where $\mathrm{J}$ = Joule = 
$\mathrm{kg\cdot m^2\cdot s^{-2}}$. It converts the common 
physical dimension of all components $T_{\mu\nu}$, which 
is that of an energy density (Joule per cubic meter in 
SI-units) into that of the components of $\Ein$, which is 
that of curvature (in dimension $\geq 2$), i.e., the square of
an inverse length (inverse square-meter in SI-units).
   
If we express energy density as mass density times $c^2$,
the conversion factor is $\kappa c^2=8\pi G/c^2$. 
It can be expressed in various units that give a feel for the
local ``curving power'' of mass-densities. 
\index{curvature!of spacetime}
\index{curvature!caused by matter}
For that of water, $\rho_W\approx 10^3\,\mathrm{kg\cdot m^{-3}}$, 
and nuclear matter in the core of a neutron star (which 
is more than twice that of atomic nuclei), 
$\rho_N\approx 5\times 10^{17}\,\mathrm{kg\cdot m^{-3}}$, we get,
respectively: 
\begin{equation}
\label{eq:KappiInUnits}
\kappa c^2
\approx\left(\frac{1}{1.5\,{\rm AU}}\right)^2\cdot\rho_W^{-1}
\approx\left(\frac{1}{10\,{\rm km}}\right)^2\cdot\rho_N^{-1}\,,
\end{equation}
where ${\rm AU}=1.5\times 10^{11}\,\mathrm{m}$ is the astronomical unit 
(mean Earth-Sun distance). Hence, roughly speaking, 
matter densities of water cause curvature radii of the 
order of the astronomical unit, whereas the highest 
known densities of nuclear matter cause curvature radii 
of tens of kilometers. The curvature caused by mere 
mass density is that expressed in $\Ein(W,W)$ when $W$ is 
taken to be the unit timelike vector characterizing the 
local rest frame of the matter: It is a mean of spatial 
sectional curvatures in the matter's local rest frame. 
Analogous interpretations can be given for the curvatures
caused by momentum densities (energy current-densities) 
and momentum current-densities (stresses).

\subsection{Conserved energy-momentum tensors
and globally conserved  quantities}
\label{sec:DigrSymm}
In this subsection we briefly wish to point out that energy-momentum 
tensors $\EMT$ whose divergence vanishes \eqref{eq:DivergencelessT} 
give rise to conserved quantities 
\index{conserved quantity}
in case the spacetime $(M,g)$ admits non-trivial isometries. 
We will stress the global nature of these quantities and clarify 
their mathematical habitat. 

Conservation laws for the matter alone result in the 
presence of symmetries, more precisely, if \emph{Killing fields}
\index{Killing fields}
for $(M,g)$ exist. Recall that a vector field $V$ is called a Killing 
field  iff $L_V g=0$, where $L_V$ is the Lie derivative with respect to 
$V$. Recall that the Lie derivative can be expressed in terms of the 
Levi-Civita covariant derivative with respect to $g$, in which case we 
get the component expression: 
\begin{equation}
\label{eq:KillingEquation}
(L_V g)_{\mu\nu}=\nabla_\mu V_\nu+\nabla_\nu V_\mu=0\,.
\end{equation}

We consider the one-form $J_V$ that results from 
contracting $\EMT$ with $V$:  
\begin{equation}
\label{eq:DefEnMomFourCurrent}
J_V:=i_V \EMT=V^\mu T_{\mu\nu}dx^\nu\,.
\end{equation}
As a result of Killing's equation \eqref{eq:KillingEquation}
it is divergence free, 
\begin{equation}
\label{eq:EnMomFourCurrent}
\nabla_\mu J_V^\mu=0\,.
\end{equation}
This may be equivalently expressed by saying that the 3-form 
$\star J_V$, which is the Hodge dual of the 1-form $J_V$, 
is closed:
\begin{equation}
\label{eq:EnMomThreeForm}
d\star J_V=0\,.
\end{equation}
Integrating $\star J_V$ over some 3-dimensional submanifold 
$\Sigma$ results in a quantity 
\begin{equation}
\label{eq:Q-Map1}
Q[V,\Sigma]:=\int_\Sigma \star J_V
\end{equation}
which, because of \eqref{eq:EnMomThreeForm}, is largely 
independent of $\Sigma$. More precisely, if  $\Omega\subset M$ 
is an oriented domain with boundary $\partial\Omega=\Sigma_1-\Sigma_2$, 
then Stokes' theorem gives $Q[V,\Sigma_1]=Q[V,\Sigma_2]$. 

Suppose now that $V$ arises from a finite-dimensional Lie group
\index{Lie!group} 
$G$ that acts on $(M,g)$ by isometries. We will discuss general 
Lie-group actions on manifolds in the Appendix at the end of this 
contribution, containing detailed proofs of some relevent formulae.
But in order not to interrupt the argument too much, let us  
recall at this point that an \emph{action} 
\index{group action} 
of $G$ on $M$ is a map  
\begin{subequations}
\label{eq:DefGroupActionOnM}
\begin{equation}
\label{eq:DefGroupActionOnM-a}
\begin{split}
\Phi: 
G\times M&\rightarrow M\,,\\
(g,m)&\mapsto\Phi(g,m)=\Phi_g(m)\,,
\end{split}
\end{equation}
which satisfies 
\begin{alignat}{1}
\label{eq:DefGroupActionOnM-b}
\Phi_{e}&=\Id_M\,,\\
\label{eq:DefGroupActionOnM-c}
\Phi_g\circ\Phi_h&=\Phi_{gh}\,.
\end{alignat}
\end{subequations}
Here $e\in G$ denotes the neutral element, $\Id_M$ the 
identity map on $M$, and equation
\eqref{eq:DefGroupActionOnM-c} is valid for any two 
elements $g,h$ of G. In fact, equation \eqref{eq:DefGroupActionOnM-c}
characterizes a \emph{left action}.
\index{group action!left and right}
In contrast, for a \emph{right action}
we would have $\Phi_{hg}$ instead of $\Phi_{gh}$ on the 
right-hand side of \eqref{eq:DefGroupActionOnM-c}. 
Moreover, as the group acts by isometries for the metric $g$, 
we also have $\Phi_h^*g=g$ for all $h\in G$.
\index{group action!by isometries} 

Now, this action defines a map, $V$, from $\Lie(G)$, 
the Lie algebra 
\index{Lie!algebra}\index{algebra!Lie}%
of $G$, into the vector fields on $M$. The vector 
field corresponding to $X\in\Lie(G)$ is denoted $V^X$. 
Its value at a point $m\in M$ is defined by  
\begin{equation}
\label{eq:G-InducedVectorFields}
V^X(m):=\frac{d}{dt}\Big\vert_{t=0}\Phi\bigl(\exp(tX),m\bigr)\,.
\end{equation}
From this it is obvious that $V:\Lie(G)\rightarrow\STM{1}{0}$ is linear.
Moreover, one may also show (compare \eqref{eq:GroupAction_GenCommRightLeftAction-b} in Appendix)  that this 
map is a Lie anti-homomorphism, i.e. that
\index{Lie!anti-homomorphism} 
\begin{equation}
\label{eq:LieAntiHomo}
V^{[X,Y]}=-[V^X,V^Y]\,.
\end{equation}
(As shown in the Appendix, a right action would have resulted 
in a proper Lie homomorphism -- see
\index{Lie!homomorphism}
 \eqref{eq:GroupAction_GenCommRightLeftAction-a} --, i.e. 
without the minus sign on the right-hand side, which however 
is not harmful.)  The left action of $G$ on $M$ extends to a left 
action on all tensor fields by push forward. In particular, the 
push forward of $V^X$ by $\Phi_g$ has a simple expression
(see \eqref{eq:GroupActions_AdEquivLeftRightaction-a} in Appendix) :  
\begin{equation}
\label{eq:Ad-EquivarianceVF-1}
\Phi_{g*}V^X=V^{\Ad_g(X)}\,,
\end{equation}
where $\Ad$ denotes the adjoint representation of $G$ on $\Lie(G)$. 
\index{group representation!adjoint}
In fact, relation \eqref{eq:Ad-EquivarianceVF-1} 
can be directly deduced from definition \eqref{eq:G-InducedVectorFields}. 
Indeed, writing $\Phi(g,p)=g\cdot p$ for notational simplicity, we have 
(see Appendix for more explanation) 
\begin{equation}
\label{eq:Ad-EquivarianceVF-2}
\begin{split}
(\Phi_{g*}V^X)(g\cdot p)
&=\frac{d}{dt}\Big\vert_{t=0}\bigl(g\,\exp(tX)\cdot p\bigr)\\
&=\frac{d}{dt}\Big\vert_{t=0}\bigl(g\,\exp(tX)\,g^{-1}g\cdot p\bigr)\\
&=\frac{d}{dt}\Big\vert_{t=0}\Bigl(\exp\bigl(t\Ad_g(X)\bigr)\cdot (g\cdot p)\Bigr)\\
&=V^{\Ad_g(X)}(g\cdot p)\,.
\end{split}
\end{equation}
This leads to \eqref{eq:Ad-EquivarianceVF-1} which we shall use 
shortly.

Returning to the expression~\eqref{eq:Q-Map1} we see that, for 
fixed $\Sigma$, it becomes a linear map from $\Lie(G)$ to 
$\reals$:
\begin{equation}
\label{eq:Q-Map2}
\Momentum:\Lie(G)\rightarrow\reals\,,\quad
\Momentum (X):=Q[V^X,\Sigma]\,.
\end{equation}
Hence each hypersurface $\Sigma$ defines an element 
$\Momentum\in\Lie^*(G)$ in the vector space that is 
dual to the Lie algebra,
\index{Lie!algebra, dual of}%
given that the integral over $\Sigma$ converges. This is the 
case for spacelike $\Sigma$ and energy momentum tensors with 
spatially compact support 
(or at least sufficiently rapid fall off). The same argument as above 
using Stokes' theorem and \eqref{eq:EnMomThreeForm} then shows that 
$\Momentum$ is independent of the choice of spacelike $\Sigma$. 
In other words, we obtain a conserved quantity 
\index{conserved quantity}
$\Momentum\in\Lie^*(G)$ for $G$-symmetric spacetimes $(M,g)$ and 
covariant divergence free tensors $\EMT$. 

So far we considered a fixed spacetime $(M,g)$ and a fixed
energy-momentum tensor $\EMT$, both linked by Einstein's 
equations. In this case the vanishing divergence 
\eqref{eq:DivergencelessT} is an integrability 
condition for Einstein's equation and hence automatic. 
However, it is also of interest to consider the more 
general case where $(M,g)$ is merely a background for 
some matter represented by energy-momentum tensors $\EMT[\mu]$,
all of which are divergence free \eqref{eq:DivergencelessT}
with respect to the background metric $g$. Note that we do 
not assume $(M,g)$ to satisfy Einstein's equations with any of 
the $\EMT[\mu]$ on the right-hand side. The $\mu$ stands for 
some matter variables which may be fundamental fields and/or 
of phenomenological nature. In any case, we assume the 
isometric action \eqref{eq:DefGroupActionOnM} to extend to 
an action of $G$ on the set of matter variables $\mu$, 
which we denote by $\mu\mapsto\Phi_{g*}\mu$, like the 
push-forward on tensor fields. This is also meant to 
indicate that we assume this to be a left action, 
i.e. $\Phi_{g*}\circ\Phi_{h*}=\Phi_{gh*}$. 

We regard the energy-momentum tensor $\EMT$ as a map from the 
space of matter variables to the space of symmetric second-rank 
covariant tensor fields on $M$. We require this map to satisfy 
the following covariance property:
\begin{equation}
\label{eq:EMT-covariance}
\EMT[\Phi_{g*}\mu]=\Phi_{g*}\EMT[\mu]:=\Phi_{g^{-1}}^*\EMT\,,
\end{equation}
where $\Phi_{g*}$ is the ordinary push-forward of the tensor 
$\EMT$. Since we take $\EMT$ to be covariant, its push forward 
is the pull back by the inverse diffeomorphism, as indicated by 
the second equality in \eqref{eq:EMT-covariance}. 

For each specification $\mu$ of matter variables we can compute 
the quantitty $Q[V^X,\Sigma,\mu]$ as in \eqref{eq:Q-Map1}.
Note that we now indicate the dependence on $\mu$ explicitly. 
We are interested in computing how $Q$ changes as $\mu$ is acted 
on by $g\in G$. This is done as follws:  
\begin{equation}
\label{eq:Q-covariance}
\begin{split}
Q\bigl[V^X,\Sigma,\Phi_{g*}\mu\bigr]
&=\int_\Sigma\star\,i_{V^X}\,\EMT[\Phi_{g*}\mu]\\
&=\int_\Sigma\star\,i_{V^X}\,\Phi^*_{g^{-1}}\EMT[\mu]\\
&=\int_\Sigma\star\,\Phi^*_{g^{-1}}\bigl(i_{\Phi_{{g^{-1}}*}V^X}\EMT[\mu]\bigr)\\
&=\int_\Sigma\star\,\Phi^*_{g^{-1}}\bigl(i_{V^{\Ad_{g^{-1}}(X)}}\EMT[\mu]\bigr)\\
&=\int_\Sigma\Phi^*_{g^{-1}}\bigl(\star i_{V^{\Ad_{g^{-1}}(X)}}\EMT[\mu]\bigr)\\
&=\int_{\Phi_{g^{-1}}(\Sigma)} \star\,i_{V^{\Ad_{g^{-1}}(X)}}\EMT[\mu]\\
&=Q\bigl[V^{\Ad_{g^{-1}}(X)},\Phi_{g^{-1}}(\Sigma),\mu\bigr]\,.
\end{split}
\end{equation}
Here we used \eqref{eq:EMT-covariance} in the second equality, 
the general formula $i_Vf^*T=f^*(i_{f_*V}T)$ (valid for any diffeomorphism 
$f$, vector field $V$, and covariant tensor field $T$) in the third equality,
\eqref{eq:Ad-EquivarianceVF-1} in the fourth equality, the formula 
$\star\,f^*F=f^*\star F$ in the fifth equality (valid for any 
orientation preserving isometry $f$ and any form-field $F$; here 
we assume $M$ to be oriented), and finally the general formula
for the integral of the pull back of a form in the sixth equality. 

Our final assumption is that $Q$ does not depend on which 
hypersurface $\Phi_g(\Sigma)$ it is evaluated on. Since we assume 
\eqref{eq:DivergencelessT} this is guaranteed if all 
$\Phi_g(\Sigma)$ are in the same homology class or, more 
generally, if any two hypersurfaces $\Sigma$ and $\Phi_g(\Sigma)$
are homologous to hypersurfaces in the complement of the support 
of $\EMT$. A typical situation arising in physical applications 
is that of a source $\EMT[\mu]$ with spatially compact support;
then any two sufficiently extended spacelike slices through the 
timelike support-tube of $\EMT[\mu]$ is homologous to the timelike 
cylindrical hypersurface outside this support-tube. In this case 
we infer from \eqref{eq:Q-covariance} that 
\begin{equation}
\label{eq:Q-covariance-2}
 Q\bigl[V^X,\Sigma,\Phi_{g*}\mu\bigr]e
=Q\bigl[V^{\Ad_{g^{-1}}(X)},\Sigma,\mu\bigr]\,.
\end{equation}

Recall from \eqref{eq:Q-Map2} that for fixed $\Sigma$ and $\EMT$
we have $\Momentum\in\Lie^*(G)$. Given the independence on $\Sigma$ 
and the depencence of $\EMT$ on $\mu$, we now regard $\Momentum$ as 
a map from the matter variables $\mu$ to $\Lie^*(G)$. This map may be 
called the \emph{momentum map}. (Compare the notion of a momentum map 
in Hamiltonian mechanics; cf. Section~\ref{sec:ConstrainedHamiltonianSystems}.)
\index{momentum!map}
Equation \eqref{eq:Q-covariance-2} then just states the $\Ad^*$-equivariance 
of the momentum map:   
\begin{equation}
\label{eq:Ad-Equivariance}
\Momentum\circ \Phi_{g*}=\Ad^*_g\circ\Momentum\,.
\end{equation} 
Here $\Ad^*$ denote the co-adjoint representation of $G$ on 
$\Lie(G)$, which is defined by $\Ad_g^*(\alpha)=\alpha\circ\Ad_{g^{-1}}$.
\index{group representation!co-adjoint}
From all this we see that the conserved ``momentum'' that we 
obtain by evaluating $\Momentum$ on the matter configuration $\mu$
is a conserved quantity that is \emph{globally} associated to 
all of spacetime, not a particular region or point of it. It is 
an element of the vector space ${\Lie}^*(G)$ which carries 
the co-adjoint representation of the symmetry group $G$. 
\index{group representation!co-adjoint}

In particular this applies to Special Relativity, where $M$ is 
the four-dimensional real affine space with associated 
(four-dimensional real) vector space $V$ and $g$ a bilinear,
symmetric, non-degenerate form of signature $(-,+,+,+)$ [the
signature does not matter in what follows]. The 
linear isometries of $(V,\eta)$ form the Lorentz group
\index{Lorentz!group} 
$\mathit{Lor}\subset\mathit{GL(V)}$ and the isometries $G$ 
of $(M,g)$ can be (non-naturally) identified with the semi-direct product 
$V\rtimes\mathit{Lor}$, called the Poincar\'e group, $\mathit{Poin}$.
Using $g$ we can identify $\mathrm{Lie}^*(\mathit{Poin})$ with 
$V\oplus (V\wedge V)$. The co-adjoint action of 
$(a,A)\in\mathit{Poin}$ on $(f,F)\in\mathrm{Lie}^*(\mathit{Poin})$ 
is then given by 
\index{group representation!co-adjoint}
\begin{equation}
\label{eq:PoincareCoAdjointAction}
\Ad^*_{(a,A)}(f,F)=\Bigl(Af\,,\,(A\otimes A)F-a\wedge Af\Bigr)\,.
\end{equation}  
Note that, e.g., the last term on the right hand side includes 
the law of change of angular momentum under spatial translations. 
In contrast, the adjoint representation on 
$\mathrm{Lie}(\mathit{Poin})$, the latter also identified with
$V\oplus (V\wedge V)$, is given by
\index{group representation!adjoint}  
\begin{equation}
\label{eq:PoincareAdjointAction}
\Ad_{(a,A)}(f,F)=\Bigl(Af-\bigl((A\otimes A)F\bigr)a\,,\,(A\otimes A)F\Bigr)\,,
\end{equation}  
where the application of an element in $V\wedge V$ to an element in 
$V$ is given by $(u\wedge v)(w):=u\,g(v,w)-v\,g(u,w)$, and linear 
extension.  Note the characteristic difference between \eqref{eq:PoincareCoAdjointAction} and \eqref{eq:PoincareAdjointAction},
which lies in the different actions of the subgroup of
translations, whereas the subgroup of Lorentz transformations 
\index{Lorentz!transformations}
acts in the same fashion. Physical momenta transform as in \eqref{eq:PoincareCoAdjointAction}, as already exemplified by the 
non-trivial transformation behavior of angular momentum under spatial 
translations. For a detailed discussion of the proper group-theoretic
setting and the adjoint and co-adjoint actions, see the 
recent account~\cite{Giulini:EMTaMinSRT2015}.

\section{Spacetime decomposition}
\label{sec:DecompSpaceTime}
In this section we explain how to decompose a given 
spacetime $(M,g)$ into ``space'' and `time''. For this to be
possible we need to make the assumption that $M$ is 
diffeomorphic to the product of the real line $\reals$ 
and some 3-manifold $\Sigma$:
\begin{equation}
\label{eq:GlobHypTop}
M\cong\reals\times\Sigma\,.
\end{equation}
This will necessarily be the case for \emph{globally hyperbolic}
 \index{globally hyperbolic}
spacetimes, i.e. spacetimes admitting a \emph{Cauchy surface}
\cite{Geroch-DomainOfDep:1970}. 
We assume $\Sigma$ to be orientable, for, if it were not, we could 
take the orientable double cover of it instead. Orientable 
3-manifolds are always parallelizable \cite{Steenrod:ToFB}
\index{parallelizable}, i.e. admit three globally defined and 
pointwise linearly independent vector fields. This is equivalent 
to the triviality of the tangent bundle. In the closed case this 
is known as \emph{Stiefel's theorem} (compare 
\cite{Milnor.Stasheff:CharacteristicClasses}, problem 12-B)
and in the open case it follows, e.g., from the well known 
fact that every open 3-manifold can be immersed in 
$\reals^3$ \cite{Whitehead:1961}. Note that orientability 
is truly necessary; e.g., $\reals\mathrm{P}^2\times S^1$ is 
not parallelizable. Since Cartesian products of parallelizable 
manifolds are again parallelizable, it follows that a 4-dimensional 
product spacetime \eqref{eq:GlobHypTop} is also parallelizable. 
This does, of course,  not generalize to higher dimensions. 
Now, for non-compact four-dimensional spacetimes it is 
known from \cite{Geroch-SpinorStructure-1:1968} that parallelizability 
is equivalent to the existence of a \emph{spin structure}, 
\index{spin structure}
without which spinor fields could not be defined on spacetime.
So we see that the existence of spin structure is already implied 
by \eqref{eq:GlobHypTop} and hence does not pose any further 
topological restriction. Note that the only other potential 
topological restriction at this stage is that imposed from the 
requirement that a smooth Lorentz metric 
\index{Lorentz!metric}
is to exist everywhere on 
spacetime. This is equivalent to a vanishing Euler 
characteristic (see, e.g., \S\,40 in \cite{Steenrod:ToFB}) which
in turn is equivalently to the global existence of a continuous, 
nowhere vanishing vector field (possibly up to sign) on spacetime. 
But such a vector field clearly exists on any Cartesian product with 
one factor being $\reals$. We conclude that existence of a 
Lorentz metric and a spin structure on an orientable spacetime 
$M=\reals\times\Sigma$ pose no restrictions on the topology of 
an orientable $\Sigma$. As we will see later on, even Einstein's 
equation poses \emph{no} topological restriction on $\Sigma$, 
in the sense that \emph{some} (physically reasonable) solutions 
to Einstein's equations exist for any given $\Sigma$. Topological 
restrictions may occur, however, if we ask for solution with 
special properties (see below).

Now, given $\Sigma$, we consider a one-parameter family of 
embeddings
\begin{equation}
\label{eq:Embeddings}
\Emb_s:\Sigma\rightarrow M\,,\quad
\Sigma_s:=\Emb_s(\Sigma)\subset M\,.
\end{equation}

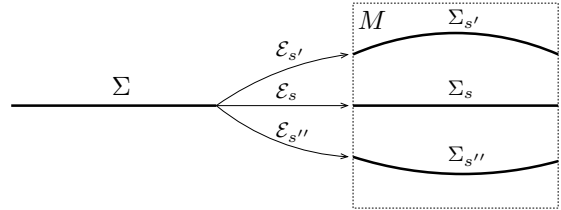
\begin{figure}[ht]
\centering
\includegraphics[width=0.95\linewidth]{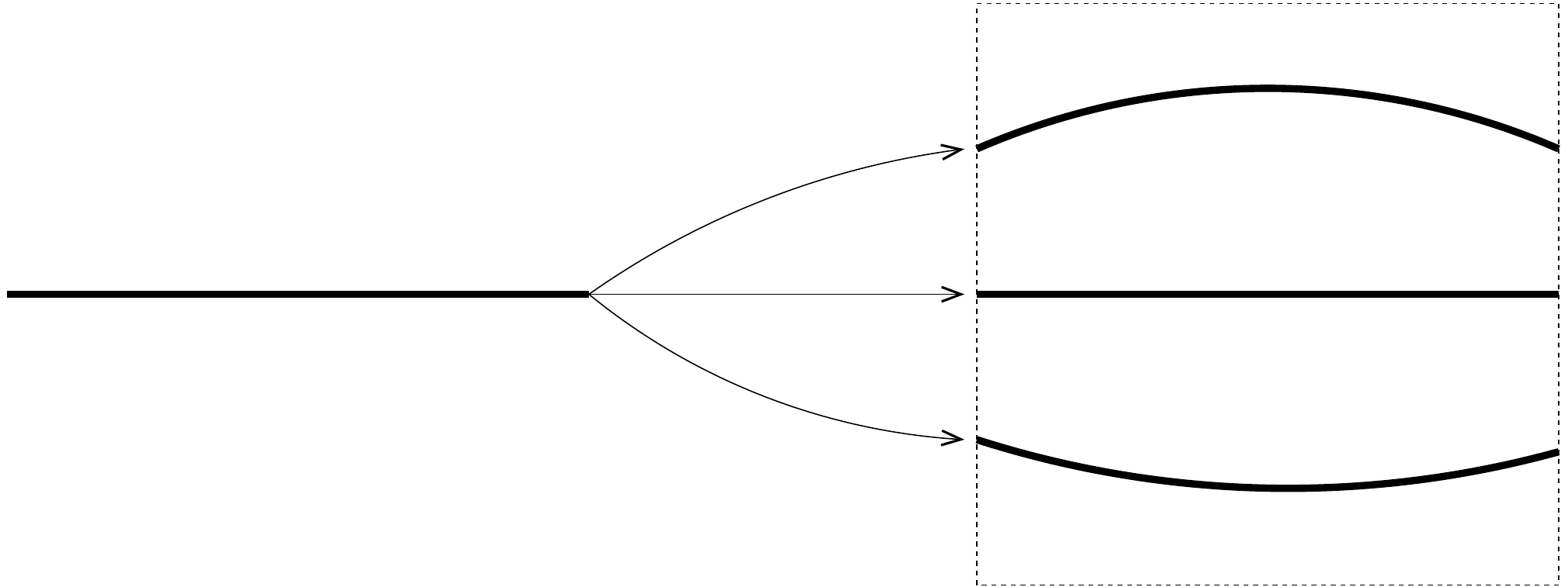}
\put(-77,68){$M$}
\put(-170,43){$\Sigma$}
\put(-108,58){\footnotesize $\Emb_{s'}$}
\put(-108,42){\footnotesize $\Emb_{s}$}
\put(-108,27){\footnotesize $\Emb_{s''}$}
\put(-43,70){\footnotesize $\Sigma_{s'}$}
\put(-43,43){\footnotesize $\Sigma_{s}$}
\put(-43,18){\footnotesize $\Sigma_{s''}$}
\caption{\label{fig:Embeddings}\footnotesize
Spacetime $M$ is foliated by a one-parameter family of 
spacelike embeddings of the 3-manifold $\Sigma$. Here the 
image $\Sigma_{s'}$ of $\Sigma$ under $\Emb_{s'}$ lies to the 
future (above) and $\Sigma_{s''}$ to the past 
(below) of $\Sigma_s$ if $s''<s<s'$. `Future' and `past' 
refer to the time function $t$ which has so far not been 
given any metric significance.}
\end{figure}

We distinguish between the abstract 3-manifold 
$\Sigma$ and its image $\Sigma_s$ in $M$. The 
latter is called the leaf corresponding to the 
value $s\in\reals$. Each point in $M$ is 
contained in precisely one leaf. Hence there is 
a real valued function $t:M\rightarrow\reals$ that 
assigns to each point in $M$ the parameter value 
of the leaf it lies on:
\begin{equation}
\label{eq:TimeFunction}
t(p)=s\Leftrightarrow p\in\Sigma_s\,.
\end{equation}

\index{foliation!of spacetime (by spacelike hypersurfaces)}
So far this is only a foliation of spacetime by 3-dimensional 
leaves. For them to be addressed as ``space'' the metric 
induced on them must be positive definite, that is, the 
leaves should be spacelike submanifolds. This means that 
the one-form $dt$ is timelike:
\begin{equation}
\label{eq:TimelikeOneForm}
g^{-1}(dt,dt)<0\,.
\end{equation}
The normalized field of one-forms is then 
\begin{equation}
\label{eq:NormalisedTimelikeOneForm}
n^\flat:=\frac{dt}{\sqrt{-g^{-1}(dt,dt)}}\,.
\end{equation}
As explained in section\,\ref{sec:NotationConventions}, we 
write $n^\flat$ since we think of this one form as the 
image under $g$ of the normalized vector field perpendicular 
to the leaves: 
\begin{equation}
\label{eq:NormalVF}
n^\flat=g(n,\,\cdot\,)\,.
\end{equation}

The linear subspace of vectors in $T_pM$ which are 
tangent to the leaf through $p$ is denoted by 
$T_p^\Vert M$; hence 
\begin{equation}
\label{eq:ParallelTangentSpace}
T_p^\Vert M:=\{X\in T_pM : dt(X)=0\}\,.
\end{equation}
The orthogonal complement is just the span of $n$ at $p$,
which we denote by $T_p^\perp M$. This gives, at each point
$p$ of $M$, the $g$-orthogonal direct sum 
\begin{equation}
\label{eq:DirectSum}
T_pM=T_p^\perp M\oplus T_p^\Vert M\,.
\end{equation}
and associated projections (we drop reference to the 
point $p$) 
\begin{subequations}
\label{eq:Projectors}
\begin{alignat}{3}
\label{eq:Projectors-a}
&P^\perp: &&TM&&\rightarrow T^\perp M\,,\nonumber\\
& &&X &&\mapsto \signature\,g(X,n)\,n\,,\\
\label{eq:Projectors-b}
&P^\Vert: &&TM&&\rightarrow T^\Vert M\,,\nonumber\\
& &&X &&\mapsto X-\signature g(X,n)\,n\,.
\end{alignat}
\end{subequations}
As already announced in Section\,\ref{sec:NotationConventions}, 
we introduced the symbol 
\begin{equation}
\label{eq:SignSymbol}
\signature=g(n,n)
\end{equation} 
in order to keep track of where the signature matters.
Note that the projection operators \eqref{eq:Projectors} 
are self-adjoint with respect to $g$, so that for all 
$X,Y\in TM$ we have
\begin{subequations}
\label{eq:SelfAdjointProjections}
\begin{alignat}{1}
\label{eq:SelfAdjointProjections-a}
g\bigl(P^\perp X,Y\bigr)&=g\bigl(X,P^\perp Y\bigr)\,,\\
\label{eq:SelfAdjointProjections-b}
g\bigl(P^\Vert X,Y\bigr)&=g\bigl(X,P^\Vert Y\bigr)\,.
\end{alignat}
\end{subequations}

A vector is called \emph{horizontal} iff it is in the 
kernel of $P^\perp$, which is equivalent to being 
invariant under $P^\Vert$. It is called \emph{vertical}
iff it is in the kernel of $P^\Vert$, which is equivalent 
to being invariant under $P^\perp$. 

All this can be extended to forms. We define \emph{vertical and 
horizontal forms} as those annihilating horizontal and vertical 
vectors, respectively: 
\begin{subequations}
\label{eq:VertHorForms}
\begin{alignat}{2}
\label{eq:VertHorForms-a}
&T_p^{*\perp} M&&:=
\{\omega\in T^*_pM : \omega(X)=0\,,\ \forall X\in T_p^\Vert M\}\,,\\
\label{eq:VertHorForms-b}
&T_p^{*\Vert} M&&:=
\{\omega\in T^*_pM : \omega(X)=0\,,\ \forall X\in T_p^\perp M\}\,.
\end{alignat}
\end{subequations}
Using the `musical' isomorphisms \eqref{eq:MusicalIsomorphisms},
the self-adjoint projection maps \eqref{eq:Projectors} on 
vectors define self-adjoint projection maps on co-vectors 
(again dropping the reference to the base-point $p$)
\begin{subequations}
\label{eq:CoProjectionMaps}
\begin{alignat}{4}
\label{eq:CoProjectionMaps-a}
&P_*^\perp:=&&\flat\circ P^\perp\circ\sharp\, &&:\, T^*M&&\rightarrow T^{*\perp}M\,,\\
\label{eq:CoProjectionMaps-b}
&P_*^\Vert:=&&\flat\circ P^\Vert\circ\sharp\, &&:\, T^*M&&\rightarrow T^{*\Vert}M\,.
\end{alignat}
\end{subequations}
For example, letting the horizontal projection of the form 
$\omega$ act on the vector $X$, we get 
\begin{equation}
\label{eq:HorFormOnVector}
\begin{split}  
P_*^\Vert\omega(X)
&=(P^\Vert\omega^\sharp)^\flat (X)\\
&=g\bigl(P^\Vert\omega^\sharp,X\bigr)\\
&=g\bigl(\omega^\sharp,P^\Vert X\bigr)\\
&=\omega\bigl(P^\Vert X\bigr)\,,
\end{split}
\end{equation}
where we merely used the definitions \eqref{eq:MusicalIsomorphisms}
of $\flat$ and $\sharp$ in the second and fourth equality, 
respectively, and the self-adjointness 
\eqref{eq:SelfAdjointProjections-b} of $P^\Vert$ in the third 
equality. The analogous relation holds for $P_*^\perp\omega(X)$.
It is also straightforward to check that $P_*^\Vert$ and 
$P_*^\perp$ are self-adjoint with respect to $g^{-1}$
(cf. \eqref{eq:DefInverseMetric}). 

Having the projections defined for vectors and co-vectors, we 
can also define it for the whole tensor algebra of the underlying 
vector space, just by taking the appropriate tensor products
of these maps. All tensor products between $P^\Vert$ and 
$P_*^\Vert$ will then, for simplicity, just be denoted by 
$P^\Vert$, the action on the tensor being obvious. Similarly 
for $P^\perp$. (For what follows we need not consider mixed 
projections.) The projections being pointwise operations, we 
can now define vertical and horizontal projections of arbitrary 
tensor fields. Hence a tensor field $T\in\STM{u}{d}$ is called 
\emph{horizontal} if and only if $P^\Vert T=T$. The space of 
horizontal tensor fields of rank $\rank{u}{d}$ is denoted by 
$\STMV{u}{d}$. 

As an example, the horizontal projection of the metric $g$ is
\begin{equation}
\label{eq:ProjMetric}
h:=P^\Vert g:=g\bigl(P^\Vert\,\cdot\,,P^\Vert\,\cdot\,\bigr)
=g-\signature n^\flat\otimes n^\flat\,.
\end{equation}
Hence $h\in\STMV{0}{2}$. Another example of a horizontal 
vector field is the ``acceleration'' of the normal field 
$n$:
\begin{equation}
\label{eq:DefNormalAcceleration}
a:=\nabla_nn\,.
\end{equation}
Here $\nabla$ denotes the Levi-Civita covariant derivative 
with respect to $g$. An observer who moves perpendicular to 
the horizontal leaves has four-velocity $u=cn$ and 
four-acceleration $c^2a$. If $L$ denotes the Lie derivative, 
it is easy to show that the acceleration 1-form satisfies 
\begin{equation}
\label{eq:CovAcceleration}
a^\flat=L_nn^\flat\,.
\end{equation}
Moreover, as $n$ is hypersurface orthogonal it is 
irrotational, hence its 1-form equivalent satisfies 
\begin{subequations}
\label{eq:NoCurl}
\begin{equation}
\label{eq:NoCurl-a}
dn^\flat\wedge n^\flat=0\,,
\end{equation}
which is equivalent to the condition of 
vanishing horizontal curl:
\begin{equation}
\label{eq:NoCurl-b}
P^{\Vert}dn^\flat=0\,.
\end{equation}
\end{subequations}
Equation \eqref{eq:NoCurl-a} can also be immediately 
inferred directly from \eqref{eq:NormalisedTimelikeOneForm}. 
Taking the operation $i_n\circ d$ (exterior derivative followed 
by contraction with $n$) as well as the Lie derivative 
with respect to $n$ of \eqref{eq:CovAcceleration} shows
\begin{subequations}
\label{eq:AccelerationCurl}
\begin{equation}
\label{eq:AccelerationCurl-a}
da^\flat\wedge n^\flat=0\,,
\end{equation}
an equivalent expression being again the
vanishing of the horizontal curl of $a$:
\begin{equation}
\label{eq:AccelerationCurl-b}
P^{\Vert}da^\flat=0\,.
\end{equation}
\end{subequations}
This will be useful later on.

Note that $a^\flat$ is a horizontal co-vector field, i.e. an 
element of $\STMV{u=0}{d=1}$. More generally, for a purely 
covariant horizontal tensor field we have the following 
results, which will also be useful later on:
Let $T\in\STMV{0}{d}$, then 
\begin{subequations}
\label{eq:HorCovFields}
\begin{alignat}{2}
\label{eq:HorCovFields-1}
&P^\Vert L_nT&&\,=\,L_nT\,,\\
\label{eq:HorCovFields-2}
&L_{fn}T&&\,=\, fL_nT\,,
\end{alignat}
\end{subequations}
for all $f\in C^\infty(M)$. Note that \eqref{eq:HorCovFields-1} 
states that the Lie derivative in normal direction of a 
horizontal covariant tensor field is again horizontal. 
That this is not entirely evident follows, e.g., from the 
fact that a corresponding result does not hold for  
$T\in\STMV{u}{d}$ where $u>0$. The proofs of \eqref{eq:HorCovFields}
just use standard manipulations. 

A fixed space-point $q\in\Sigma$ defines the worldline 
(history of that point) $\reals\ni s\mapsto \Emb_s(q)$.
The collection of all worldlines of all space-points
define a foliation of $M$ into one-dimensional 
timelike leafs. 
\index{foliation!of spacetime (by timelike curves)}
Each leaf is now labeled uniquely by a 
space point. We can think of ``space'', i.e., the 
abstract manifold $\Sigma$, as the quotient $M/\!\!\sim$,
where $p\sim p'$ iff both points lie on the same worldline.
As any $\Sigma_s$ intersects each worldline exactly once, 
each $\Sigma_s$ is a representative of space. Instead 
of using the foliation by 3-dimensional spatial leaves 
\index{foliation!of spacetime (by spacelike hypersurfaces)}
\eqref{eq:Embeddings} we could have started with a 
foliation by timelike lines, 
\index{foliation!of spacetime (by timelike curves)}
plus the condition that these lines are vorticity free. 
These two concepts are equivalent. Depending on the 
context, one might prefer to emphasize one or the other. 

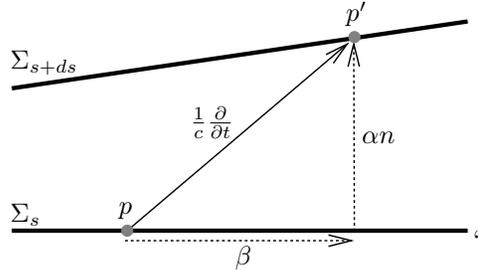
\begin{figure}[htb]
\centering\includegraphics%
[width=0.80\linewidth]{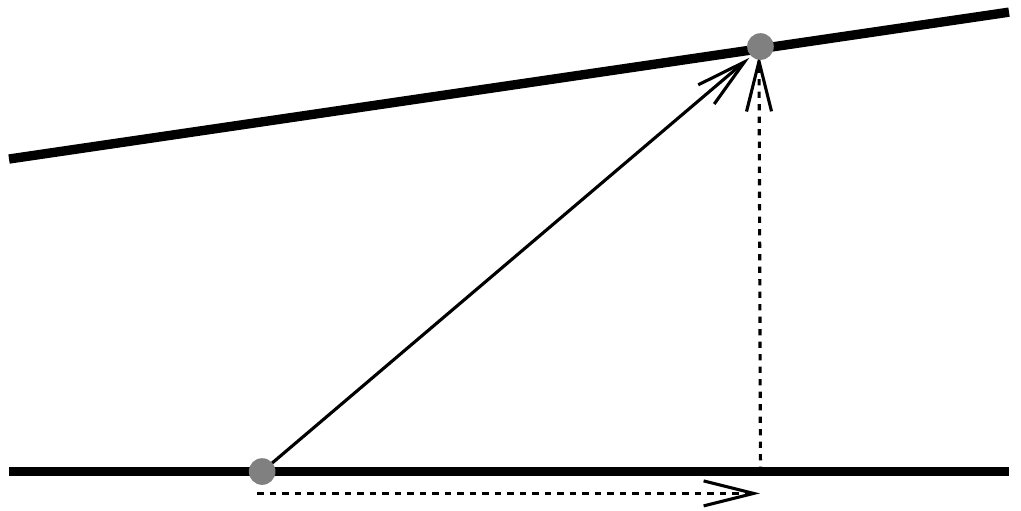}
\put(-175,10){$\Sigma_{s}$}
\put(-175,68){$\Sigma_{s+ds}$}
\put(-134,13){$p$}
\put(-48,86){$p'$}
\put(-90,-6){$\beta$}
\put(-42,40){$\alpha n$}
\put(-107,45){$\frac{1}{c}\frac{\partial}{\partial t}$}`
\caption{\label{fig:LapseShift}\footnotesize%
For fixed $q\in\Sigma$ its image points $p=\Emb_s(q)$ 
and $p'=\Emb_{s+ds}(q)$ for infinitesimal $ds$ are connected 
by the vector $\partial/\partial t\vert_p$, whose components 
normal to $\Sigma_s$ are $\alpha$ (one function, called lapse) 
and $\beta$ (three functions, called shift) respectively.}
\end{figure}

The vector parallel to the worldline at $p=\Emb_s(q)$ is,
as usual in differential geometry, defined by  its action
on $f\in C^\infty(M)$ (smooth, real valued functions):  
\begin{equation}
\label{eq:FlowOfTime}
\frac{\partial}{\partial t}\Big\vert_{\Emb_s(q)}f
 =\frac{df(\Emb_{s'}(q))}{ds'}\Big\vert_{s'=s}\,.
\end{equation}   
At each point this vector field can be decomposed into its 
horizontal component that is tangential to the leaves of the
given foliation 
\index{foliation!of spacetime (by spacelike hypersurfaces)}
and its normal component. We write 
\begin{equation}
\label{eq:DecompositionLapseShift}
\frac{1}{c}\frac{\partial}{\partial t}=\alpha\,n+\beta\,,
\end{equation}   
where $\beta$ is the tangential part; 
see Figure\,\ref{fig:LapseShift}. 
The real-valued function $\alpha$ is called the \emph{lapse} (function) 
\index{lapse function}
and the horizontal vector field $\beta$ is called the \emph{shift}
(vector-field)
\index{shift vector-field}.

\subsection{Decomposition of the metric}
\label{sec:DecompMetric}
Let $\{e_0,e_1,e_2,e_3\}$ be a locally defined orthonormal frame
with dual frame $\{\theta^0,\theta^1,\theta^2,\theta^3\}$.
We call them \emph{adapted} to the foliation if $e_0=n$ and 
$\theta^0=n^\flat$.
\index{adapted frames}%
A local coordinate system $\{x^0,x^1,x^2,x^3\}$ is called \emph{adapted}
if $\partial/\partial x^a$ are horizontal for $a=1,2,3$. Note that in 
the latter case $\partial/\partial x^0$ is not required to be 
orthogonal to the leaves (i.e. it need not be parallel to $n$).
For example, we may take $x^0$ to be proportional to $t$;
say $x^0=ct$. 

In the orthonormal co-frame the spacetime metric, i.e. the field 
of signature $(\signature,+,+,+)$ metrics in the tangent spaces, has 
the simple form 
\begin{equation}
\label{eq:SpacetimeMetric}
g=\signature\theta^0\otimes\theta^0+\sum_{a=1}^3\theta^a\otimes\theta^a\,. 
\end{equation}
The inverse spacetime metric, i.e. the field 
of signature $(\signature,+,+,+)$ metrics in the co-tangent spaces,
has the form  
\begin{equation}
\label{eq:InverseSpacetimeMetric}
g^{-1}=\signature e_0\otimes e_0+\sum_{a=1}^3e_a\otimes e_a\,. 
\end{equation}

The relation that expresses the coordinate basis in terms of the 
orthonormal basis is of the form (in a self-explanatory matrix notation)
\begin{equation}
\label{eq:BasisRelations}
\begin{pmatrix}
\partial/\partial x^0\\
\partial/\partial x^m
\end{pmatrix}
=
\begin{pmatrix}
\alpha&\beta^a\\
0       & A^a_m
\end{pmatrix}
\begin{pmatrix}
e_0\\
e_a
\end{pmatrix}\,,
\end{equation}  
where $\beta^a$ are the components of $\beta$ with respect 
to the horizontal frame basis $\{e_a\}$. The inverse of 
\eqref{eq:BasisRelations} is 
\begin{equation}
\label{eq:InverseBasisRelations}
\begin{pmatrix}
e_0\\
e_a
\end{pmatrix}
=
\begin{pmatrix}
\alpha^{-1}&-\alpha^{-1}\beta^m\\
0       & [A^{-1}]^m_a
\end{pmatrix}
\begin{pmatrix}
\partial/\partial x^0\\
\partial/\partial x^m
\end{pmatrix}\,,
\end{equation}  
where $\beta^m$ are the components of $\beta$ with respect 
to the horizontal coordinate-induced frame basis 
$\{\partial/\partial x^m\}$. 

The relation for the co-bases dual to those in 
\eqref{eq:BasisRelations} is given by the transposed 
of \eqref{eq:BasisRelations}, which we write as: 
\begin{equation}
\label{eq:CoBasisRelations}
\begin{pmatrix}
\theta^0 & \theta^a
\end{pmatrix}
=
\begin{pmatrix}
dx^0&dx^m
\end{pmatrix}
\begin{pmatrix}
\alpha&\beta^a\\
0       & A^a_m
\end{pmatrix}\,.
\end{equation}  
The inverse of that is the transposed of \eqref{eq:InverseBasisRelations}:
\begin{equation}
\label{eq:InverseCoBasisRelations}
\begin{pmatrix}
dx^0&dx^m
\end{pmatrix}
=
\begin{pmatrix}
\theta^0&\theta^a
\end{pmatrix}
\begin{pmatrix}
\alpha^{-1}&-\alpha^{-1}\beta^m\\
0       & [A^{-1}]^m_a
\end{pmatrix}\,.
\end{equation}  

Orthogonality of the $e_a$ implies for the chart 
components of the spatial metric \eqref{eq:ProjMetric}
\begin{equation}
\label{eq:OrthSpatialVec}
h_{mn}:
=h\bigl(\partial/\partial x^m,\partial/\partial x^n\bigr)
=\sum_{a=1}^3 A^a_mA^a_n\,,
\end{equation}
and its inverse
\begin{equation}
\label{eq:InverseOrthSpatialVec}
h^{mn}:
=h^{-1}\bigl(dx^m,dx^n\bigr)
=\sum_{a=1}^3 [A^{-1}]^m_a[A^{-1}]^n_a\,.
\end{equation}

Inserting \eqref{eq:CoBasisRelations} into \eqref{eq:SpacetimeMetric} 
and using \eqref{eq:OrthSpatialVec} leads to the (3+1)-form of the 
metric in adapted coordinates
\begin{equation}
\label{eq:ThreePlusOneOfMetric}
\begin{split}
g&\,=\,\bigl(\signature\alpha^2+h(\beta,\beta)\bigr)c^2\,dt\otimes dt\\
 &\,+\,c\beta_m\bigl(dt\otimes dx^m+dx^m\otimes dt\bigr)\\
 &\,+\,h_{mn}\,dx^m\otimes dx^n\,,
\end{split}
\end{equation}
where $\beta_m:=h_{mn}\beta^n$ are the components of 
$\beta^\flat:=g(\beta,\,\cdot\,)=h(\beta,\,\cdot\,)$ 
with respect to the coordinate basis $\{\partial/\partial x^m\}$.
Likewise, inserting \eqref{eq:InverseCoBasisRelations} into 
\eqref{eq:InverseSpacetimeMetric} and using 
\eqref{eq:InverseOrthSpatialVec} leads to the (3+1)-form 
of the inverse metric in adapted coordinates (we write 
$\partial_t:=\partial/\partial t$ and 
$\partial_m:=\partial/\partial x^m$ for convenience)
\begin{equation}
\label{eq:ThreePlusOneOfInverseMetric}
\begin{split}
g^{-1}
&\,=\,\signature c^{-2}\alpha^{-2}\,\partial_t\otimes\partial_t\\
&\,-\,\signature c^{-1}\alpha^{-2}\,\beta^m
\bigl(\partial_t\otimes\partial_m+\partial_m\otimes\partial_t\bigr)\\
&\,+\,\bigl(h^{mn}+\signature\beta^m\beta^n\bigr)\partial_m\otimes\partial_n\,.
\end{split}
\end{equation}

Finally we note that the volume form on spacetime 
also easily follows from \eqref{eq:CoBasisRelations}
\begin{equation}
\label{eq:VolumeForm}
\begin{split}
d\mu_g&=\theta^0\wedge\theta^1\wedge\theta^2\wedge\theta^3\\
&=\alpha\sqrt{\det\{h_{mn}\}}\,cdt\wedge d^3x\,,
\end{split}
\end{equation}
where we use the standard shorthand 
$d^3x=dx^1\wedge dx^2\wedge dx^3$. 

\subsection{Decomposition of the \\ covariant derivative}
\label{sec:DecompCovDer}
Given horizontal vector fields $X$ and $Y$, the covariant 
derivative of $Y$ with respect to $X$ need not be 
horizontal. Its decomposition is written as 
\begin{equation}
\label{eq:DecompCovDer-1}
\nabla_XY=D_XY+nK(X,Y)\,,
\end{equation}
where 
\begin{alignat}{1}
\label{eq:DecompCovDer-2}
D_XY&:=P^\Vert\nabla_XY\,,\\
\label{eq:DecompCovDer-3}
K(X,Y)&:=\signature\,g(n,\nabla_XY)\,.
\end{alignat}
The map $D$ defines a covariant derivative (in the sense 
of Kozul; compare \cite{Spivak:DiffGeom-5Vols}, Vol\,2)
\index{covariant!derivative}%
for horizontal vector fields, as a trivial check of the 
axioms reveals. Moreover, since the commutator $[X,Y]$ 
of two horizontal vector fields is always horizontal (since 
the horizontal distribution is integrable by construction), 
we have 
\begin{equation}
\label{eq:Torsion-free}
\begin{split}
T^D(X,Y)
&=D_XY-D_YX-[X,Y]\\
&=P^\Vert\bigl(\nabla_XY-\nabla_YX-[X,Y]\bigr)\\
&=0
\end{split}
\end{equation}
due to $\nabla$ being torsion free. We recall that torsion 
\index{torsion}
is a tensor field $T\in\Gamma T^1_2M$ associated 
to each covariant derivative $\nabla$ via 
\begin{equation}
\label{eq:DefTorsion}
T^\nabla(X,Y)=\nabla_XY-\nabla_YX-[X,Y]\,.
\end{equation}
We have $T(X,Y)=-T(Y,X)$. As usual, even though 
the operations on the right hand side of \eqref{eq:DefTorsion}
involve tensor \emph{fields} (we need to differentiate),
the result of the operation just depends on $X$ and $Y$
pointwise. This one proves by simply checking the validity 
of $T(fX,Y)=fT(X,Y)$ for all smooth functions $f$. Hence 
\eqref{eq:Torsion-free} shows that $D$ is torsion free
because $\nabla$ is torsion free. 

Finally, we can uniquely extend $D$ to all horizontal 
tensor fields by requiring the Leibniz rule. Then, for 
$X,Y,Z$ horizontal 
\begin{equation}
\label{eq:MetricityD-1}
\begin{split}
&(D_Xh)(Y,Z)\\
&=X\bigl(h(Y,Z)\bigr)-h(D_XY,Z)-h(Y,D_XZ)\\
&=X\bigl(g(Y,Z)\bigr)-g(\nabla_XY,Z)-g(Y,\nabla_XZ)\\
&=(\nabla_Xg)(Y,Z)=0
\end{split}
\end{equation}
due to the metricity, $\nabla g=0$, of $\nabla$. 
Hence $D$ is metric in the sense 
\begin{equation}
\label{eq:MetricityD-2}
Dh=0\,.
\end{equation}

The map $K$ from pairs of horizontal vector fields 
$(X,Y)$ into functions define a symmetric tensor 
field. Symmetry follows from the vanishing torsion 
of $\nabla$, since then  
\begin{equation}
\begin{split}
\label{eq:SymmetryExtCurv}
K(X,Y)
&=\signature\,g(n,\nabla_XY)\\
&=\signature\,g(n,\nabla_YX+[X,Y])\\
&=\signature\,g(n,\nabla_YX)\\
&=K(Y,X)
\end{split}
\end{equation}
for horizontal $X,Y$. From \eqref{eq:DecompCovDer-3} one 
sees that $K(fX,Y)=fK(X,Y)$ for any smooth function $f$. 
Hence $K$ defines a unique symmetric tensor field on $M$ 
by stipulating that it be horizontal, i.e. $K(n,\cdot)=0$.
It is called the \emph{extrinsic curvature} of the foliation
\index{foliation!of spacetime (by spacelike hypersurfaces)}
\index{extrinsic curvature}
\index{curvature!extrinsic}
or \emph{second fundamental form},
\index{second fundamental form}
the first fundamental form 
\index{first fundamental form}
\index{fundamental!form (first and second)}
being the metric. From \eqref{eq:DecompCovDer-3} and the symmetry 
just shown one immediately infers the alternative expressions 
\begin{equation}
\label{eq:ExtCurvature}
K(X,Y)
=-\signature\,g(\nabla_Xn,Y)
=-\signature\,g(\nabla_Yn,X)\,.
\end{equation}
This shows the relation between the extrinsic curvature and the 
\emph{Weingarten map}, $\Wein$, also called the \emph{shape operator},
\index{Weingarten map}%
\index{shape operator}%
which sends horizontal vectors to horizontal vectors according to 
\begin{equation}
\label{eq:WeingartenMap}
X\mapsto\Wein(X):=\nabla_Xn\,.
\end{equation}
Horizontality of $\nabla_Xn$ immediately follows from $n$ being 
normalized: $g(n,\nabla_Xn)=\tfrac{1}{2}X\bigl(g(n,n)\bigr)=0$.
Hence \eqref{eq:ExtCurvature} simply becomes 
\begin{equation}
\label{eq:RelWeingartenMapExtCurv}
\begin{split}
K(X,Y)
&=-\signature\,h\bigl(\Wein(X),Y\bigr)\\
&=-\signature\,h\bigl(X,\Wein(Y)\bigr)\,,
\end{split}
\end{equation}
where we replaced $g$ with $h$---defined in \eqref{eq:ProjMetric}---since 
both entries are horizontal. It says that $K$ is $(-\signature)$ times 
the covariant tensor corresponding to the Weingarten map, and that the 
symmetry of $K$ is equivalent to the self-adjointness of 
the Weingarten map with respect to $h$. The Weingarten map characterizes 
the bending of the embedded hypersurface in the ambient space by 
answering the following question: In what direction and by what amount 
does the normal to the hypersurface tilt if, starting at point $p$, 
you progress within the hypersurface by the vector $X$. The answer is 
just $\Wein_p(X)$. Self adjointness of $\Wein$ then means that there always exist 
three ($n-1$ in general) perpendicular directions in the hypersurface  
along which the normal tilts in the same direction. These are the 
\emph{principal curvature directions} 
\index{principal!curvature directions}
mentioned above. The principal curvatures 
\index{principal!curvatures}
are the corresponding eigenvalues of $\Wein$.
  
Finally we note that the covariant derivative 
of the normal field $n$ can be written in terms 
of the acceleration and the Weingarten map as follows
\begin{equation}
\label{eq:CovDerNormalField-1}
\nabla n=\signature n^\flat\otimes a+\Wein\,.
\end{equation}
Recalling \eqref{eq:RelWeingartenMapExtCurv}, the purely 
covariant version of this is
\begin{equation}
\label{eq:CovDerNormalField-2}
\nabla n^\flat=
-\signature\bigl(K-n^\flat\otimes a^\flat\bigr)\,.
\end{equation}
From \eqref{eq:ProjMetric} and \eqref{eq:CovDerNormalField-2}
we derive by standard manipulation, using vanishing torsion,
\begin{equation}
\label{eq:LieNormal-h} 
L_nh=-2\signature K\,.
\end{equation}
In presence of torsion there would be an additional term 
$+2(i_nT)^\flat_s$, where the subscript $s$ denotes symmetrization; 
in coordinates 
$[(i_nT)^\flat_s]_{\mu\nu}=n^\lambda T^\alpha_{\lambda(\mu}g_{\nu)\alpha}$.

\section{Curvature tensors}
\label{sec:CurvatureTensors}
We wish to calculate the (intrinsic) curvature tensor of $\nabla$ and 
express it in terms of the curvature tensor of $D$, the extrinsic 
curvature $K$, and the spatial and normal derivatives of $n$ and $K$. 
Before we do this, we wish to say a few words on the definition 
of the curvature measures in general. 

All notions of curvature eventually reduce to that of curves. 
For a surface  $S$ embedded in $\reals^3$ we have the notion 
of Gaussian curvature
\index{Gaussian curvature}
\index{curvature!Gaussian}
which comes about as follows: Consider a point $p\in S$ and 
a unit vector $v$ at $p$ tangent to $S$. Consider all smooth 
curves passing through $p$ with unit tangent $v$. It is easy 
to see that the curvatures at $p$ of all such curves is not 
bounded from above  (due to the possibility to bend within the 
surface), but there will be a lower bound, $k(p,v)$, 
which just depends on the chosen point $p$ and the tangent 
direction represented by $v$. Now consider $k(p,v)$ as 
function of $v$. As $v$ varies over all tangent directions 
$k(p,v)$ will assume a minimal and a maximal value, denoted 
by $k_{min}(p)=k(p,v_{min})$ and  $k_{max}(p)=k(p,v_{max})$ 
respectively. These are 
called the principal curvatures 
\index{principal!curvatures}
of $S$ at $p$ and their 
reciprocals are called the principal radii. 
\index{principal!radii}
It is clear that the principal directions $v_{min}$ and $v_{max}$
just span the eigenspaces of the Weingarten map 
\index{Weingarten map}
discussed above. In particular, $v_{min}$ and  $v_{max}$ 
are orthogonal. The Gaussian curvature 
\index{Gaussian curvature}
\index{curvature!Gaussian}
$K(p)$ of $S$ at $p$ is then defined to be the product of the 
principal curvatures: 
\begin{equation}
\label{eq:DefGaussCurvature}
K(p)=k_{min}(p)\cdot k_{max}(p)\,.
\end{equation}
This definition is extrinsic in the sense that essential use is  
made of the ambient $\reals^3$ in which $S$ is embedded. However, 
Gauss' \emph{theorema egregium} 
\index{theorem!of Gauss (theorema egregium)} 
states that this notion of curvature can also be defined intrinsically,
in the sense that the value $K(p)$ can be obtained from geometric 
operations entirely carried out \emph{within} the surface
$S$. More precisely, it is a function of the first 
fundamental form (the metric) only, which encodes the 
intrinsic geometry of $S$, and does not involve the second 
fundamental form (the extrinsic curvature), which encodes how
$S$ is embedded into $\reals^3$.

Let us briefly state Gauss' theorem in mathematical terms.
Let 
\begin{equation}
\label{eq:FirstFundForm}
g=g_{ab}\,dx^a\otimes dx^b
\end{equation} 
be the metric of the surface in some coordinates,
and 
\begin{equation}
\label{eq:ChristoffelSymbols}
\Gamma^c_{ab}=\tfrac{1}{2}g^{cd}
\bigl(
- \partial_dg_{ab}
+ \partial_ag_{bd}
+ \partial_bg_{da}\bigr)\,,
\end{equation} 
certain combinations of first derivatives of the metric 
coefficients, known under the name of \emph{Christoffel symbols}
\index{Christoffel symbols}. Note that $\Gamma^c_{ab}$ has as many
independent components as $\partial_ag_{bc}$ and that we can 
calculate the latter from the former via  
\begin{equation}
\label{eq:FirstderFromChristoffelSymbols}
\partial_cg_{ab}=g_{an}\Gamma^n_{bc}+g_{bn}\Gamma^n_{ac}\,.
\end{equation} 
Next we form even more complicated 
combinations of first and second derivatives of the metric
coefficients, namely 
\begin{equation}
\label{eq:RiemannTensor}
R^a_{\phantom{a}b\,cd}=
 \partial_c\Gamma_{db}^a
-\partial_d\Gamma_{cb}^a
+\Gamma^a_{cn}\Gamma^n_{db}
-\Gamma^a_{dn}\Gamma^n_{cb}\,,
\end{equation}
which are now known as components of the \emph{Riemann curvature tensor}.
\index{curvature!Riemann}\index{Riemann!curvature}
\index{tensor!Riemann}\index{Riemann!tensor}
From them we form the totally covariant 
(all indices down) components: 
\begin{equation}
\label{eq:RiemannTensorCov}
R_{ab\,cd}=g_{an}R^n_{\phantom{n}b\,cd}\,.
\end{equation}
They are antisymmetric in the first and second index pair:
$R_{ab\,cd}=-R_{ba\,cd}=-R_{ab\,dc}$, so that $R_{12\,12}$ is the 
only independent component. Gauss' theorem now states that 
at each point on $S$ we have 
\begin{equation}
\label{eq:TheormaEgregium}
K=\frac{R_{12\,12}}{g_{11}g_{22}-g_{12}^2}\,.
\end{equation}
An important part of the theorem is to show that the right-hand
side of \eqref{eq:TheormaEgregium} actually makes good geometric 
sense, i.e. that it is independent of the coordinate system that 
we use to express the coefficients. 
This is easy to check once one knows that $R_{abcd}$ are the 
coefficients of a tensor with the symmetries just stated. 
In this way the curvature of a surface, which was primarily 
defined in terms of curvatures of certain curves on the surface, 
can be understood intrinsically. In what follows we will 
see that the various measures of intrinsic curvatures of $n$-dimensional 
manifolds can be reduced to that of 2-dimensional 
submanifolds, which will be called \emph{sectional curvatures}.
\index{sectional curvature}
\index{curvature!sectional}

Back to the general setting, we start from the notion 
of a covariant derivative $\nabla$. Its associated 
curvature tensor is defined by 
\begin{equation}
\label{eq:DefRiemannTensor}
R(X,Y)Z=
\bigl(\nabla_X\nabla_Y-\nabla_Y\nabla_X-\nabla_{[X,Y]}\bigr)Z\,.
\end{equation}
For each point $p\in M$ it should be thought of as a map that assigns 
to each pair $X,Y\in T_pM$ of tangent vectors at $p$ a linear map 
$R(X,Y):T_pM\rightarrow T_pM$. This assignment is antisymmetric, i.e. 
$R(X,Y)=-R(Y,X)$. If $R(X,Y)$ is applied to $Z$ the result is given by 
the right-hand side of \eqref{eq:DefRiemannTensor}. Despite first 
appearance, the right-hand side of \eqref{eq:DefRiemannTensor} at
a point $p\in M$ only depends on the values of $X,Y$, and $Z$ at 
that point and hence defines a tensor field. This one again proves by 
showing the validity of $R(fX,Y)Z=R(X,fY)Z=R(X,Y)fZ=fR(X,Y)Z$
for all smooth real-valued functions $f$ on M. In other words:
All terms involving derivatives of $f$ cancel. 

From \eqref{eq:DefRiemannTensor} and using \eqref{eq:DefTorsion}
one may show that the Riemann tensor always obeys the first and 
second \emph{Bianchi identities}\index{Bianchi identities}: 
\begin{subequations}
\label{eq:BianchiIdentities}
\begin{alignat}{1}
\label{eq:BianchiIdentity-first}
  &\sum_{(XYZ)} R(X,Y)Z \nonumber \\
 &=\sum_{(XYZ)}\Bigl\{(\nabla_XT)(Y,Z)-T\bigl(X,T(Y,Z)\bigr)\Big\}\,,\\
\label{eq:BianchiIdentity-second}
&\sum_{(XYZ)} (\nabla_XR)(Y,Z)\nonumber\\
&=\sum_{(XYZ)}R\bigl(X,T(Y,Z)\bigr)\,,
\end{alignat}
\end{subequations}
where the sums are over the three cyclic permutations of $X$, $Y$,
and $Z$. For zero torsion these identities read in component form:
\begin{subequations}
\label{eq:BianchiIdentitiesComp}
\begin{alignat}{1}
\label{eq:BianchiIdentityComp-first}
&\sum_{(\lambda\mu\nu)} R^\alpha_{\phantom{\alpha}\lambda\,\mu\nu}=0\,,\\
\label{eq:BianchiIdentityComp-second}
&\sum_{(\lambda\mu\nu)} \nabla_\lambda R^\alpha_{\phantom{\alpha}\beta\,\mu\nu}=0\,.
\end{alignat}
\end{subequations}
The second traced on $(\alpha,\mu)$ and contracted with $g^{\beta\nu}$ 
yields $(-2)$ times \eqref{eq:Bianchi2Contracted}. 

The covariant Riemann tensor is defined by 
\index{curvature!Riemann}\index{Riemann!curvature}
\index{tensor!Riemann}\index{Riemann!tensor}
\begin{equation}
\label{eq:DefCovRiemannTensor}
\Riem(W,Z,X,Y):=g\bigl(W,R(X,Y)Z\bigr)\,.
\end{equation}
For general covariant derivatives its only symmetry is 
the antisymmetry in the last pair. But for special choices it 
acquires more. In standard GR we assume the 
covariant derivative to be \emph{metric compatible}
\index{metric!connection compatible with}
and \emph{torsion free}:
\index{torsion free} 
\begin{alignat}{1}
\label{eq:MetricCompatible}
\nabla g&=0\,,\\
T&=0\,.
\end{alignat}
In that case the Riemann tensor has the symmetries   
\index{curvature!Riemann}\index{Riemann!curvature}
\index{tensor!Riemann}\index{Riemann!tensor}  
\begin{subequations}
\label{eq:SymmCovRiemannTensor}
\begin{alignat}{1}
\label{eq:SymmCovRiemannTensor-a}
 \Riem(W,Z,X,Y)&=-\Riem(W,Z,Y,X)\,,\\
\label{eq:SymmCovRiemannTensor-b}
\Riem(W,Z,X,Y)&=-\Riem(Z,W,X,Y)\,,\\
\label{eq:SymmCovRiemannTensor-c}
\Riem(W,X,Y,Z)&+\Riem(W,Y,Z,X)\,+\nonumber\\
               \Riem(W,Z,Y,X)&=0\,,\\
\label{eq:SymmCovRiemannTensor-d}
\Riem(W,Z,X,Y)&=\Riem(X,Y,W,Z)\,.
\end{alignat}
\end{subequations}
Equation \eqref{eq:SymmCovRiemannTensor-a} is true by definition  
\eqref{eq:DefRiemannTensor}, \eqref{eq:SymmCovRiemannTensor-b} is
equivalent to metricity of $\nabla$, and \eqref{eq:SymmCovRiemannTensor-c} 
is the first Bianchi identity in case of zero torsion. 
The last symmetry \eqref{eq:SymmCovRiemannTensor-d} is a 
consequence of the preceding three. Together 
\eqref{eq:SymmCovRiemannTensor-a},
\eqref{eq:SymmCovRiemannTensor-b}, and 
\eqref{eq:SymmCovRiemannTensor-d}
say that, at each point $p\in M$, $\Riem$ can be thought of as 
symmetric bilinear form on the antisymmetric tensor product 
$T_pM\wedge T_pM$. The latter has dimension $N=\tfrac{1}{2}n(n-1)$ 
if $M$ has dimension $n$, and the space of symmetric bilinear forms 
has dimension $\tfrac{1}{2}N(N+1)$. From that number we have to subtract 
the number of independent conditions \eqref{eq:SymmCovRiemannTensor-c},
which is $\binom{n}{4}$ in dimensions $n\geq 4$ and zero otherwise. Indeed,
it is easy to see that \eqref{eq:SymmCovRiemannTensor-c} is identically 
satisfied as a consequence of 
\eqref{eq:SymmCovRiemannTensor-a} and \eqref{eq:SymmCovRiemannTensor-b} 
if any two vectors $W,Z,X,Y$ coincide (proportionality is sufficient). 
Hence the number $\#$ of independent components of the curvature tensor is  
\begin{equation}
\label{eq:CompRiemannTensor}
\begin{split}
&\#\Riem=\\
&\begin{cases}
\tfrac{1}{2}N(N+1)-\binom{n}{4}=\tfrac{1}{12}n^2(n^2-1) &\text{for}\quad n\geq 4\\
6 & \text{for}\quad n=3\\
1 & \text{for}\quad n=2
\end{cases}\\
&=\tfrac{1}{12}n^2(n^2-1)\quad\text{for all}\ n\geq 2\,.
\end{split}
\end{equation}

The Ricci and scalar curvatures
\index{Ricci curvature}
\index{curvature!Ricci}
\index{scalar curvature}  
\index{curvature!scalar}
are obtained by taking traces with respect to $g$: Let $\{e_1,\cdots, e_n\}$
be an orthonormal basis, $g(e_a,e_b)=\delta_{ab}\varepsilon_a$ (no summation)
with $\varepsilon_a=\pm 1$, then 
\begin{alignat}{1}
\label{eq:DefRicciCurv}
\Ricci(X,Y)&=\sum_{a=1}^n\varepsilon_a\,\Riem(e_a,X,e_a,Y)\\
\label{eq:DefScalarCurv}
\Scalar&=\sum_{a=1}^n\varepsilon_a\,\Ricci(e_a,e_a)\,.
\end{alignat}    
The Einstein tensor is
\begin{equation}
\label{eq:DefEinsteinTensor}
\Ein=\Ricci-\tfrac{1}{2}\Scalar\,g\,.
\end{equation}

The sectional curvature is defined by   
\begin{equation}
\label{eq:DefSectCurv} 
\Sec(X,Y)=\frac{\Riem(X,Y,X,Y)}%
{g(X,X)g(Y,Y)-\bigl[g(X,Y)\bigr]^2}\,,
\end{equation}
Here $X,Y$ is a pair of linearly independent tangent vectors
that span a 2-dimensional tangent subspace restricted to 
which $g$ is non-degenerate. We will say that 
$\text{span}\{X,Y\}$ is non-degenerate. This is the necessary 
and sufficient condition for the denominator on the right-hand 
side to be non zero. The quantity $\Sec(X,Y)$ is called the 
\emph{sectional curvature}
\index{sectional curvature}
of the manifold $(M,g)$ at point $p$ tangent to 
$\Span\{X,Y\}$. From the symmetries of $\Riem$ it 
is easy to see that the right-hand side of 
\eqref{eq:DefSectCurv} does indeed only depend on the span 
of $X,Y$. That is, for any other pair $X',Y'$ such that 
$\Span\{X',Y'\}=\Span\{X,Y\}$, we have $\Sec(X',Y')=\Sec(X,Y)$. 
The geometric interpretation of $\Sec(X,Y)$ is as follows: 
Consider all geodesics of $(M,g)$ that pass through 
the considered point $p\in M$ in a direction tangential to 
$\Span\{X,Y\}$. In a neighborhood of $p$ they form an embedded 
2-surface in $M$ whose Gaussian curvature 
\index{Gaussian curvature}
\index{curvature!Gaussian}
is just $\Sec(X,Y)$. 

Now, $\Riem$ is determined by components of the form 
$\Riem(X,Y,X,Y)$, as follows from the fact that $\Riem$
is a \emph{symmetric} bilinear form on $TM\wedge TM$. 
This remains true if we restrict to those $X,Y$ whose span is 
non-degenerate, since they lie dense in $TM\wedge TM$
and $(X,Y)\mapsto\Riem(X,Y,X,Y)$ is continuous. This shows 
that the full information of the Riemann tensor can be 
reduced to certain Gaussian curvatures.
\index{Gaussian curvature}
\index{curvature!Gaussian}

This also provides a simple geometric interpretation of the 
scalar and Einstein curvatures in terms of sectional curvatures. 
Let $\{X_1,\cdots, X_n\}$ be any set of pairwise orthogonal 
non-null vectors. The $\tfrac{1}{2}n(n-1)$ 2-planes 
$\Span\{X_a,X_b\}$ are non-degenerate and also pairwise 
orthogonal. It then follows from \eqref{eq:DefScalarCurv} and
\eqref{eq:DefSectCurv} that the scalar curvature is twice the 
sum of all sectional curvatures: 
\begin{equation}
\label{eq:ScalarSectCurv}
\Scalar=2\sum_{\genfrac{}{}{0pt}{}{a,b=1}{a<b}}^n \Sec(X_a,X_b)\,.
\end{equation}
The sum on the right-hand side of \eqref{eq:ScalarSectCurv} is
the same for any set of  $\tfrac{1}{2}n(n-1)$ non-degenerate and 
pairwise orthogonal 2-planes. Hence the scalar curvature can be 
said to be twice the sum of mutually orthogonal sectional curvatures,
or $n(n-1)$ times the mean sectional curvature.  Similarly for the 
Ricci and Einstein curvatures. The symmetry of the Ricci and Einstein 
tensors imply that they are fully determined by their components 
$\Ricci(W,W)$ and $\Ein(W,W)$. Again this remains true if we  
restrict to the dense set of non-null $W$, i.e. $g(W,W)\ne 0$. Let 
now $\{X_1,\cdots, X_{n-1}\}$ be any set of mutually orthogonal vectors 
(again they need not be normalized) in the orthogonal complement of $W$. 
As before the $\tfrac{1}{2}(n-1)(n-2)$  planes $\Span\{X_a,X_b\}$ are 
non degenerate and pairwise orthogonal. From \eqref{eq:DefRicciCurv},
\eqref{eq:DefEinsteinTensor}, and \eqref{eq:DefSectCurv} it follows 
that  
\begin{equation}
\label{eq:RicciTensorGeomMeaning}
\Ricci(W,W)=g(W,W)
\sum_{a=1}^{n-1}\Sec(W,X_a)
\end{equation}
and
\begin{equation}
\label{eq:EinsteinTensorGeomMeaning}
\Ein(W,W)=-g(W,W)
\sum_{\genfrac{}{}{0pt}{}{a,b=1}{a<b}}^{n-1}\Sec(X_a,X_b)\,.
\end{equation}
Again the right-hand sides will be the same for any set  
$\{X_1,\cdots, X_{n-1}\}$ of $n-1$ mutually orthogonal 
vectors in the orthogonal complement of $W$. Note that 
$\Ricci(W,W)$ involves all sectional curvatures involving 
$W$ whereas $\Ein(W,W)$ involves all sectional curvatures 
orthogonal to $W$. For normalized $W$, where $g(W,W)=\sigma=\pm 1$, 
we can say that $-\sigma G(W,W)$ is the sum of sectional 
curvatures orthogonal to $W$, or  $\tfrac{1}{2}(n-1)(n-2)$ 
times their mean. Note that for timelike $W$ we have 
$\sigma=-1$ and $G(W,W)$ is just the sum of spatial sectional 
curvatures. 

Finally we mention the Weyl curvature tensor, 
\index{curvature!Weyl}\index{Weyl curvature}%
which contains that part of the information in the curvature 
tensor not captured by the Ricci (or Einstein-) tensor. To 
state its form in a compact form, we introduce the 
\emph{Kulkarni-Nomizu product},
\index{Kulkarni-Nomizu product} 
denoted by an encircled wedge, $\kuno$, which is a bilinear symmetric 
product on the space of covariant symmetric rank-two tensors with 
values in the covariant rank-four tensors that have the symmetries 
\eqref{eq:SymmCovRiemannTensor} of the Riemann tensor.  Let $k$ and 
$\ell$  be two symmetric covariant second-rank tensors, then their 
Kulkarni-Nomizu product is defined by  
\begin{equation}
\label{eq:KulkarniNomizuProduct}
\begin{split}
k\kuno \ell(X_1,X_2,X_3,X_4)\,:=\,&
    k(X_1,X_3)\,\ell(X_2,X_4)\\
+\,&k(X_2,X_4)\,\ell(X_1,X_3)\\
-\,&k(X_1,X_4)\,\ell(X_2,X_3)\\
-\,&k(X_2,X_3)\,\ell(X_1,X_4)\,,
\end{split}
\end{equation}
or in components 
\begin{equation}
\label{eq:KulkarniNomizuProduktKomp}
(k\kuno \ell)_{abcd}=
k_{ac}\ell_{bd}+
k_{bd}\ell_{ac}-
k_{ad}\ell_{bc}-
k_{bc}\ell_{ad}\,.
\end{equation}
The Weyl tensor, $\Weyl$, is of the same type as $\Riem$ but in addition 
totally trace-free. It is obtained from $\Riem$ by a projection map, 
$P_W$, given by  
\index{curvature!Weyl}\index{Weyl curvature}
\begin{equation}
\label{eq:DefWeylCurv}
\begin{split}
\Weyl
&:=P_W(\Riem)\\
&:=\Riem
-\tfrac{1}{n-2}\Bigl(\Ricci-\tfrac{1}{2(n-1)}\Scalar\, g\Bigr)\kuno g\,.
\end{split}
\end{equation}
$P_W$ is a linear map from the space of rank-four tensors 
with Riemann symmetries to itself. It is easy to check 
that its image is given by the totally trace-free such 
tensors and that the kernel consists of all tensors of 
the form $g\kuno K$, where $K$ is a symmetric rank-two 
tensor. The latter clearly implies $P_W\circ P_W=P_W$.
The dimension of the image corresponds to the number of 
independent components of the Weyl tensor, which is given 
by \eqref{eq:CompRiemannTensor} minus the dimension 
$\tfrac{1}{2}n(n+1)$ of the kernel. This gives for 
$n\geq 3$
\begin{equation}
\label{eq:CompWeylTensor}
\#\Weyl=
\tfrac{1}{12}n(n+1)\bigl[n(n-1)-6\bigr]
\end{equation}
and zero for $n=2$. Note that in $n=3$ dimensions the Weyl tensor 
also always vanishes, so that \eqref{eq:DefWeylCurv} can be used 
to express the Riemann tensor in terms of the Ricci and scalar 
curvature  
\begin{equation}
\label{eq:RiemInThreeDim}
\Riem=\bigl(\Ricci-\tfrac{1}{4} \Scalar\,g\bigr)\kuno g\quad (\text{for}\ n=3)\,.
\end{equation} 

A metric manifold $(M,g)$ is said to be of constant curvature
if 
\begin{equation}
\label{eq:DefConstCurv}
\Riem=k\,g\kuno g
\end{equation}
for some function $k$. Then $\Ricci=2k(n-1)g$ and $\Ein=-k(n-1)(n-2)g$.
We recall that manifolds $(M,g)$ for which the Einstein tensor 
(equivalently, the Ricci tensor) is pointwise proportional to the 
metric are called \emph{Einstein spaces}. 
\index{Einstein!spaces} 
The twice contracted second Bianchi identity~\eqref{eq:Bianchi2Contracted}
shows that $k$ must be a constant unless $n=2$. For $n=3$
equation \eqref{eq:RiemInThreeDim} shows that Einstein spaces are 
of constant curvature.  

\subsection{Comparing curvature tensors}
Sometimes one wants to compare two different curvature tensors
belonging to two different covariant derivatives $\hat\nabla$
and $\nabla$. In what follows, all quantities referring to 
$\hat\nabla$ carry a hat. Recall that a covariant derivative 
can be considered as a map 
$\nabla: \STM{1}{0}\times\STM{1}{0}\rightarrow \STM{1}{0}$, 
$(X,Y)\mapsto\nabla_XY$, which is $C^\infty(M)$-linear in the 
first and a derivation in the second argument. That is, for 
$f\in C^\infty(M)$ have $\nabla_{fX+Y}Z=f\nabla_XZ+\nabla_YZ$
and $\nabla_{X}(fY+Z)=X(f)Y+f\nabla_XY+\nabla_XZ$. This implies 
that the difference of two covariant derivatives is $C^\infty(M)$-
linear also in the second argument and hence a tensor field: 
\begin{equation}
\label{eq:ConnectionDifference}
\hat\nabla-\nabla=:\Delta\in\STM{1}{2}\,.
\end{equation}
Replacing $\hat\nabla$ with $\nabla+\Delta$ in the 
definition of the curvature tensor for $\hat\nabla$ according 
to \eqref{eq:DefRiemannTensor} directly leads to  
\begin{equation}
\label{eq:CurvatureDifference}
\begin{split}
\hat R(X,Y)Z
&= R(X,Y)Z\\
&+(\nabla_X\Delta)(Y,Z)
-(\nabla_Y\Delta)(X,Z)\\
&+\Delta\bigl(X,\Delta(Y,Z)\bigr)
-\Delta\bigl(Y,\Delta(X,Z)\bigr)\\
&+\Delta\bigl(T(X,Y),Z)\bigr)\,.
\end{split}
\end{equation}
Note that so far no assumptions have been made concerning 
torsion or metricity of $\hat\nabla$ and $\nabla$. 
This formula is generally valid. In the special case where 
$\hat\nabla$ and $\nabla$ are the Levi-Civita covariant
derivatives with respect to two metrics $\hat g$ and $g$, we 
set 
\begin{equation}
\label{eq:MetricDifference}
h:=\hat g-g\,,
\end{equation}
which is a symmetric covariant tensor field. Note that 
here, and for the rest of this subsection, $h$ has a 
different meaning from that given to it in 
\eqref{eq:ProjMetric}. We recall that the Levi-Civita 
covariant derivative is uniquely determined by the 
metric. For $\nabla$ this reads 
\begin{equation}
\label{eq:CovDerDetermined}
\begin{split}
&2\, g(\nabla_XY,Z)\\
&=X\bigl(g(Y,Z)\bigr)+Y\bigl(g(Z,X)\bigr)-Z\bigl(g(X,Y)\bigr)\\
&-g\bigl(X,[Y,Z])]\bigr)+g\bigl(Y,[Z,X])]\bigr)+g\bigl(Z,[X,Y])]\bigr).
\end{split}
\end{equation}
Subtracting \eqref{eq:CovDerDetermined} from the corresponding formula 
with $\nabla$ and $g$ replaced by $\hat \nabla$ and $\hat g$ 
yields, using $T=0$,  
\begin{equation}
\label{eq:LeviCivitaDifference}
\begin{split}
& 2\,\hat g\bigl(\Delta(X,Y),Z\bigr)= \\
& -(\nabla_Z h)(X,Y)
+(\nabla_X h)(Y,Z)
+(\nabla_Y h)(Z,X).
\end{split}
\end{equation}
This formula expresses $\Delta$ as functional of $g$ and $\hat g$.
There are various equivalent forms of it. We have chosen a 
representation that somehow minimizes the appearance of $\hat g$. 
Note that $g$ enters in $h$ as well as $\nabla$, whereas $\hat g$ 
enters in $h$ and via the scalar product on the left-hand side. 
The latter obstructs expressing $\Delta$ as functional of $g$
and $h$ alone. In components \eqref{eq:LeviCivitaDifference} reads
\begin{equation}
\label{eq:LeviCivitaDifferenceComponents}
\Delta^a_{bc}=
\tfrac{1}{2}{\hat g}^{an}
\bigl(
-\nabla_n h_{bc}
+\nabla_b h_{cn}
+\nabla_c h_{nb}
\bigr)\,.
\end{equation}
Note that one could replace the components of $h$ with those
of $\hat g=g+h$ in the bracket on the right-hand side, since 
the covariant derivatives of $g$ vanish. 
 
Now suppose we consider $h$ and its first and second derivatives 
to be small and we wanted to know the difference in the covariant 
derivatives and curvature only to leading (linear) order in $h$.
To that order we may replace $\hat g$ with $g$ on the left-hand side 
of \eqref{eq:LeviCivitaDifference} and the right-hand side of 
\eqref{eq:LeviCivitaDifferenceComponents}. Moreover we may neglect 
the $\Delta$-squared terms in \eqref{eq:CurvatureDifference} and 
obtain, writing $\delta R$ for the first order contribution to 
$\hat R-R$, 
\begin{equation}
\label{eq:CurvatureDifferenceFirstOrderComp}
\delta R^a_{\phantom{a}bcd}
=\nabla_c\Delta^a_{db}
-\nabla_d\Delta^a_{cb}\,.
\end{equation}
From this the first-order variation of the Ricci tensor follows,
writing $h_{ab}=:\delta g_{ab}$,  
\begin{equation}
\label{eq:RicciDifferenceFirstOrderComp}
\begin{split}
\delta R_{ab}
&=\nabla_n\Delta^n_{ab}-\nabla_{b}\Delta^n_{na}\\
&=\tfrac{1}{2}\bigl(
-\Delta_g\,\delta g_{ab}
-\nabla_a\nabla_b\,\delta g\\
&\qquad+\nabla^n\nabla_a\,\delta g_{nb}
+\nabla^n\nabla_b\,\delta g_{na}\bigr)\,,
\end{split}
\end{equation}
where $\Delta_g:=g^{ab}\nabla_a\nabla_b$ and $\delta g=g^{ab}\,\delta g_{ab}$.
Finally, the variation of the scalar curvature is 
(note $\delta g^{ab}=-g^{ac}g^{bd}\delta g_{cd}=-h^{ab}$)
\begin{subequations}
\label{eq:ScalarDifferenceFirstOrder}
\begin{equation}
\label{eq:ScalarDifferenceFirstOrderComp}
\delta R=R_{ab}\,\delta g^{ab}+\nabla_aU^a\,,
\end{equation}
where 
\begin{equation}
\label{eq:ScalarDifferenceFirstOrderVF}
\begin{split}
U^a
&=g^{nm}\Delta^a_{nm}-g^{an}\Delta^m_{mn}\\
&=G^{abcd}\nabla_b\,\delta g_{cd}\,.
\end{split}
\end{equation}
\end{subequations}
Here we made use of the \emph{De\,Witt metric},
\index{metric!De\,Witt}
\index{De\,Witt metric} 
which defines a symmetric non-degenerate bilinear form on the
space of symmetric covariant rank-two tensors and which in 
components reads:
\begin{equation}
\label{eq:Def-DeWittMetric}
G^{abcd}=\tfrac{1}{2}\bigl(g^{ac}g^{bd}+g^{ad}g^{bc}-2g^{ab}g^{cd}\bigr)\,.
\end{equation}
We will later have to say more about it. 

We also wish to state a useful formula that compares the 
curvature tensors for conformally related metrics, i.e. 
\begin{equation}
\label{eq:DefConfEqivalence}
\hat g=e^{2\phi}\,g\,,
\end{equation}
where $\phi:M\rightarrow\reals$ is smooth. Then 
\begin{subequations}
\label{eq:ConfTransRiem}
\begin{equation}
\label{eq:ConfTransRiem-a}
\Riem_{\hat g}=
e^{2\phi}
\Bigl[
\Riem_g+g\kuno K
\Bigr]\,,
\end{equation}
with 
\begin{equation}
\label{eq:ConfTransRiem-b}
K:=
-\nabla^2\phi
+d\phi\otimes d\phi
-\tfrac{1}{2}g^{-1}(d\phi,d\phi)\,g\,.
\end{equation}
\end{subequations}
(This can be proven by straightforward calculations using 
either \eqref{eq:DefRiemannTensor} and \eqref{eq:CovDerDetermined},
or Cartan's structure equations, or, most conveniently, normal 
coordinates.) From \eqref{eq:ConfTransRiem-a} and the fact 
that the kernel of the map $P_W$ in \eqref{eq:DefWeylCurv} is 
given by tensors of the form $g\kuno K$ it follows immediately 
that 
\begin{equation}
\label{eq:ConfTransWeyl}
\Weyl_{\hat g}=e^{2\phi}\Weyl_g\,.
\end{equation}
This is equivalently expressed by the conformal invariance of 
the contravariant version of the Weyl tensor, which is 
related to the covariant form, $\Weyl$, in the same way  \eqref{eq:DefCovRiemannTensor} as the curvature tensor $R$ 
is related to $\Riem$ (i.e., by raising the first index of 
the latter). 

From \eqref{eq:ConfTransRiem} we also deduce the 
transformation properties of the Ricci tensor:
\begin{equation}
\label{eq:ConfTransRicci}
\begin{split}
\Ricci_{\hat g}=
&\Ricci_g\\
-\ &\bigl(\Delta_g\phi+(n-2)g^{-1}(d\phi,d\phi)\bigr)\,g\\
-\ &(n-2)(\nabla\nabla\phi-d\phi\otimes d\phi)\,.
\end{split}
\end{equation}
where, as above, $\Delta_g$ denotes again the Laplacian/d'Alembertian 
for $g$. Finally, for the scalar curvature we get    
\begin{equation}
\label{eq:ConfTransScalar}
\begin{split}
\Scalar_{\hat g}=e^{-2\phi}\biggl(
&\Scalar_g\\
-\ &2(n-1)\Delta_g\phi\\
-\ &(n-1)(n-2)g^{-1}(d\phi,d\phi)\biggr)\,.
\end{split}
\end{equation}
This law has a linear dependence on the second and a quadratic 
dependence on first derivatives of $\phi$. If the conformal 
factor is written as an appropriate power of some positive 
function $\Omega:M\rightarrow\reals_+$ we can eliminate 
all dependence on first and just retain the second derivatives. 
In $n>2$ dimensions it is easy to check that the rule is this: 
\begin{equation}
\label{eq:ConfFactorPower}
e^{2\phi}=\Omega^{\frac{4}{n-2}}\,,
\end{equation}
then~\eqref{eq:ConfTransScalar} becomes
\begin{subequations}
\label{eq:ConfTransScal}
\begin{equation}
\label{eq:ConfTransScal-a}
\Scalar_{\hat g}=-\frac{4(n-1)}{n-2}\Omega^{-\frac{n+2}{n-2}}\mathcal{D}_g\Omega\,,
\end{equation}
where
\begin{equation}
\label{eq:ConfTransScal-c}
\mathcal{D}_g=\Delta_g-\frac{n-2}{4(n-1)}\Scalar_g\,.
\end{equation}
\end{subequations}
$\mathcal{D}_g$ is a linear differential operator which is 
elliptic for Riemannian and hyperbolic for Lorentzian metrics $g$.  
If we set $\Omega=\Omega_1\Omega_2$ and apply 
\eqref{eq:ConfTransScal} twice, one time to the pair 
$(\hat g,g)$, the other time to $(\hat g,\Omega_2g)$, 
we obtain by direct comparison (and renaming $\Omega_2$ to 
$\Omega$ thereafter) the conformal transformation 
property for the operator $\mathcal{D}_g$:
\begin{equation}
\label{eq:ConfEquivarianceOp}
\mathcal{D}_{\Omega^{\frac{4}{n-2}}g}=M\left(\Omega^{-\frac{n+2}{n-2}}\right)
\circ\mathcal{D}_g\circ M(\Omega)\,,
\end{equation}
where $M(\Omega)$ is the linear operator of multiplication with 
$\Omega$. This is the reason why $\mathcal{D}_g$ is called the 
\emph{conformally covariant Laplacian} (for Riemannian $g$) or the
\emph{conformally covariant wave operator} (for Lorentzian $g$). 
\index{Laplacian, conformally covariant}
\index{wave operator, conformally covariant}
\index{d'Alembertian, see wave operator}
As we will see, it has useful applications to the initial-data 
problem in GR.

\subsection{Curvature decomposition}
Using \eqref{eq:DecompCovDer-1} we can decompose the various 
curvature tensors. First we let $X,Y,Z$ be horizontal vector 
fields. We use \eqref{eq:DecompCovDer-1} in 
\eqref{eq:DefRiemannTensor} and get the general formula
(i.e. not yet making use of the fact that $\nabla$ and $D$ 
are metric and torsion free)  
\begin{equation}
\begin{split}
\label{eq:DecompCurv-1}
 R(X,Y)Z&=R^D(X,Y)Z\\
       &+(\nabla_Xn)\,K(Y,Z)-(\nabla_Yn)\,K(X,Z)\\
       &+n\bigl[(D_XK)(Y,Z)-(D_YK)(X,Z)\bigr]\\
       &+n\,K\bigl(T^D(X,Y),Z\bigr)\,,\qquad
\end{split}
\end{equation} 
where 
\begin{equation}
\label{eq:DefRiemannTensor-D}
R^D(X,Y,)Z:=\bigl(D_XD_Y-D_YD_X-D_{[X,Y]}\bigr)Z\
\end{equation}
is the horizontal curvature tensor associated to the Levi-Civita 
covariant derivative $D$ of $h$. This formula is general in the 
sense that it is valid for any covariant derivative. No assumptions 
have been made so far concerning metricity or torsion, and this 
is why the torsion $T^D$ of $D$ (defined in~\eqref{eq:Torsion-free})
makes an explicit appearance.  From now on we shall restrict to 
vanishing torsion. We observe that the first two lines on the 
right-hand side of \eqref{eq:DecompCurv-1} are horizontal whereas 
the last two lines are proportional to $n$. Decomposition into 
horizontal and normal components, respectively, leads to (where 
$T^D=0$ and $X,Y,Z$, and $W$ are horizontal), 
\begin{equation}
\begin{split}
\label{eq:DecompCurv-2}
&\Riem(W,Z,X,Y)
=\Riem^D(W,Z,X,Y)\\
&-\signature\bigl[K(W,X)K(Z,Y)-K(W,Y)K(Z,X)\bigr]\,.
\end{split}
\end{equation}
Here we used $h(W,\nabla_Xn)=-\epsilon K(W,X)$ from 
\eqref{eq:ExtCurvature}, and 
\begin{equation}
\label{eq:DecompCurv-3}
\begin{split}
&\Riem(n,Z,X,Y)\\
&=\signature\bigl[(D_XK)(Y,Z)-(D_YK)(X,Z)\bigr]\,.
\end{split}
\end{equation}
Here and in the sequel we return to the meaning of 
$h$ given by \eqref{eq:ProjMetric}.
In differential geometry \eqref{eq:DecompCurv-2} 
is referred to as \emph{Gauss equation} and  
\eqref{eq:DecompCurv-3} as \emph{Codazzi-Mainardi equation}.
\index{Gauss equation}
\index{Codazzi-Mainardi equation}

The remaining curvature components are those involving 
two entries in $n$ direction. Using \eqref{eq:CovDerNormalField-2}
we obtain via standard manipulations (now using metricity 
and vanishing torsion) 
\begin{equation}
\label{eq:DecompCurv-4}
\begin{split}
&\Riem(X,n,Y,n)\\
&=i_X\bigl(\nabla_Y\nabla_n-\nabla_n\nabla_Y-\nabla_{[Y,n]}\bigr)n^\flat\\
&=i_Xi_Y\bigl(\signature L_nK+K\circ K+Da^\flat-\signature a^\flat\otimes 
a^\flat\bigr)\,.
\end{split}
\end{equation}
Here $K\circ K\,(X,Y):=h^{-1}(i_XK,i_YK)=i_XK\bigl((i_YK)^\sharp\bigr)$
and we used the following relation between covariant and Lie
derivative (which will have additional terms in case of 
non-vanishing torsion): 
\begin{equation}
\label{eq:CovLie-K}
\nabla_nK=L_nK+2\signature\,K\circ K\,.
\end{equation}
Note also that the left-hand side of \eqref{eq:DecompCurv-4}
is symmetric as consequence of \eqref{eq:SymmCovRiemannTensor-d}.
On the right-hand side only $Da^\flat$ is not immediately seen to be 
symmetric, but that follows from \eqref{eq:AccelerationCurl-b}.
Unlike \eqref{eq:DecompCurv-2} and \eqref{eq:DecompCurv-3}, 
equation \eqref{eq:DecompCurv-4} does not seem to have a 
standard name in differential geometry.

Equations \eqref{eq:DecompCurv-1}, \eqref{eq:DecompCurv-2},
and \eqref{eq:DecompCurv-3} express all components of the 
spacetime curvature in terms of horizontal quantities 
and their Lie derivatives $L_n$ in normal direction. 
According to \eqref{eq:DecompositionLapseShift} the latter 
can be replaced by a combination of Lie derivatives along 
the time vector-field $\partial/\partial t$ and the shift 
$\beta$.  From  \eqref{eq:HorCovFields-2} we infer that 
$L_{\alpha n}=\alpha L_n$ on horizontal covariant tensor fields, 
therefore we may replace  
\begin{equation}
\label{eq:Lie-Normal-Time}
L_n
\rightarrow 
\alpha^{-1}\bigl(L_{\frac{\partial}{c\partial t}}-L_\beta\bigr)  
\rightarrow 
\alpha^{-1}\bigl(L^\Vert_{\frac{\partial}{c\partial t}}-L^\Vert_\beta\bigr)  
\end{equation}
on horizontal covariant tensor fields. Here we set $L^\Vert=P^\Vert\circ L$,
i.e. Lie derivative (as operation in the ambient spacetime) followed 
by horizontal projection.  
Moreover, using \eqref{eq:CovAcceleration}, one easily sees that 
the acceleration 1-form $a^\flat$ can be expressed in terms of the 
spatial derivative of the lapse function: 
\begin{equation}
\label{eq:AccLapseFuct-1}
a^\flat=-\signature \alpha^{-1}D\alpha\,.
\end{equation}
Hence the combination of accelerations appearing in~\eqref{eq:DecompCurv-3}
may be written as 
\begin{equation}
\label{eq:AccLapseFuct-2}
Da^\flat-\signature a^\flat\otimes a^\flat=-\signature \alpha^{-1}D^2\alpha\,.
\end{equation} 
Note that $D^2\alpha:=DD\alpha$ is just the horizontal covariant 
Hessian of $\alpha$ with respect to $h$.

\section{Decomposing Einstein's equations}
\label{sec:DecEinstEq}
The curvature decomposition of the previous section can now be 
used to decompose Einstein's equations. For this we decompose the 
Einstein tensor $\Ein$ into the normal-normal, normal-tangential,
and tangential-tangential parts. Let $\{e_0,e_1,e_2,e_3\}$ be 
an orthonormal frame with $e_0=n$, i.e. adapted to the 
foliation as in Section~\ref{sec:DecompMetric}. 
Then \eqref{eq:EinsteinTensorGeomMeaning} together with 
\eqref{eq:DecompCurv-2}  immediately lead to 
\begin{equation}
\label{eq:EinsteinTensorDecomp-00}
2\,\Ein(e_0,e_0)
=-\bigl[K_{ab}K^{ab}-(K_a^a)^2\bigr]
-\signature\Scalar^D\,,
\end{equation}
where $\Scalar^D$ is the scalar curvature of $D$, i.e. of the 
spacelike leaves in the metric $h$. Similarly we obtain 
from \eqref{eq:DecompCurv-3},
\begin{equation}
\label{eq:EinsteinTensorDecomp-0a}
\Ein(e_0,e_a)=\Ricci(e_0,e_a)=-\signature\bigl[
D^bK_{ab}-D_aK^b_b\bigr]\,.
\end{equation}

The normal-normal component of the Ricci tensor cannot likewise 
be expressed simply in terms of horizontal quantities, 
the geometric reason being that, unlike the Einstein tensor, 
it involves non-horizontal sectional curvatures (compare 
\eqref{eq:RicciTensorGeomMeaning} and \eqref{eq:EinsteinTensorGeomMeaning}).
A useful expression follows from taking the trace of 
\eqref{eq:DecompCurv-4}, considered as symmetric bilinear form 
in $X$ and $Y$. The result is :
\begin{equation}
\label{eq:RicciTensorDecomp-00}
\Ricci(e_0,e_0)=-K_{ab}K^{ab}+(K^c_c)^2+\signature \nabla\cdot V\,,
\end{equation}
where $\nabla\cdot$ denotes the divergence with respect to $\nabla$ 
and $V$ is a vector field on $M$ whose normal 
component is the trace of the extrinsic curvature and 
whose horizontal component is $\signature$ times the 
acceleration on $n$:  
\begin{equation}
\label{eq:RicciVectorField}
V=nK^c_c+\signature a\,.
\end{equation}
For the horizontal-horizontal components of Einstein's 
equation it turns 
out to be simpler to use their alternative form 
\eqref{eq:EinsteinRicciRelation-b} with the Ricci tensor 
on the left hand side. For that we need the horizontal 
components of the Ricci tensor, which we easily get from 
\eqref{eq:DecompCurv-2} and \eqref{eq:DecompCurv-4}: 
\begin{equation}
\label{eq:RicciTensorDecomp-ab}
\begin{split}
\Ricci(e_a,e_b)
&=\Ricci^D(e_a,e_b)\\
&+L_nK_{ab}+2\signature K_{ac}K^c_b-\signature K_{ab}K_c^c\\
&+\signature D_aa_b-a_aa_b\,.
\end{split}
\end{equation}
For later applications we also note the expression for the scalar 
curvature. It follows, e.g.,  from adding the horizontal trace of 
\eqref{eq:RicciTensorDecomp-ab} to $\signature$ times 
\eqref{eq:RicciTensorDecomp-00}. This leads to  
\begin{equation}
\label{eq:RicciScalarDecomp}
\Scalar
=\Scalar^D-\signature\,\bigl[K_{ab}K^{ab}-(K_a^a)^2\bigr]+2\nabla\cdot V\,.
\end{equation}
Here we made use of the relation between the $\nabla$ and $D$ 
derivative for the acceleration 1-form:
\begin{equation}
\label{eq:NablaDelta-a}
\nabla a^\flat=
Da^\flat+\signature\,n^\flat\otimes\nabla_na^\flat+i_aK\otimes n^\flat\,,
\end{equation}
whose trace gives the following relation between the 
$\nabla$ and $D$ divergences of $a$:
\begin{equation}
\label{eq:NablaDeltaDivergence-a}
\nabla\cdot a=D\cdot a-\signature h(a,a)\,.
\end{equation}
Another possibility would have been to use \eqref{eq:EinsteinTensorDecomp-00}
and \eqref{eq:RicciTensorDecomp-00} in 
$\Scalar=-2\signature(\Ein(e_0,e_0)-\Ricci(e_0,e_0))$.

Using \eqref{eq:EinsteinTensorDecomp-00} and 
\eqref{eq:EinsteinTensorDecomp-0a}, and also using the De\,Witt
metric \eqref{eq:Def-DeWittMetric} for notational ease, we can 
immediately write down the normal-normal and normal-tangential 
components of Einstein's equations \eqref{eq:EinsteinTensorEq}:
\begin{subequations}
\label{eq:Constraints-1}
\begin{alignat}{1}
\label{eq:EinsteinEquation-00}
&G^{abcd}K_{ab}K_{cd}+\signature\Scalar^D=-2\kappa\,\EMT(n,n)\,,\\
\label{eq:EinsteinEquation-0a}
&G^{abcd}D_bK_{cd}=-\signature\kappa h^{ab}\EMT(n,e_b)\,. 
\end{alignat}
\end{subequations}
From \eqref{eq:RelWeingartenMapExtCurv} and \eqref{eq:Def-DeWittMetric}
we notice that the bilinear form on the left-hand side of \eqref{eq:EinsteinEquation-00} 
can be written as 
\begin{equation}
\label{eq:KineticTerm-Weingarten}
\begin{split}
G(K,K):&=G^{abcd}K_{ab}K_{cd}\\
&=\Trace(\Wein\circ\Wein)-\bigl(\Trace(\Wein)\bigr)^2\,.
\end{split}
\end{equation}
Here the trace is natural (needs no metric for its definition) since 
$\Wein$ is an endomorphism. In a local frame in which $\Wein$ is diagonal 
with entries $\vec k:=(k_1,k_2,k_3)$ we have 
\begin{equation}
\label{eq:KineticTerm-WeingartenEigenvalues}
G(K,K):=(\delta^{ab}-3n^an^b)k_ak_b\,,
\end{equation}
where $n^a$ are the components of the normalized vector $(1,1,1)/\sqrt{3}$ in 
eigenvalue-space, which we identify with $\reals^3$ endowed with the standard 
Euclidean inner product. Hence, denoting by $\theta$ the angle between $\vec n$
and $\vec k$, we have
\begin{equation}
\label{eq:KineticTerm-Sign1}
G(K,K)=
\begin{cases}
0 &\quad\text{if}\quad\vert\cos\theta\vert=\sqrt{1/3}\\
>0&\quad\text{if}\quad\vert\cos\theta\vert<\sqrt{1/3}\\
<0&\quad\text{if}\quad\vert\cos\theta\vert>\sqrt{1/3}\,.
\end{cases}
\end{equation}
Note that $\vert\cos\theta\vert=\sqrt{1/3}$ describes a double 
cone around the symmetry axis generated by $\vec n$ and vertex 
at the origin, whose opening angle just is right so as to contain 
all three axes of $\reals^3$. For eigenvalue-vectors inside this 
cone the bilinear form is negative, outside this cone positive. 
Positive $G(K,K)$ require sufficiently anisotropic Weingarten 
maps, or, in other words, sufficiently large deviations from 
being umbilical points.  

The horizontal-horizontal component of Einstein's equations 
in the form  \eqref{eq:EinsteinTensorEq-Alt} immediately 
follows from \eqref{eq:RicciTensorDecomp-ab}. In the ensuing 
formula we use \eqref{eq:Lie-Normal-Time} to  explicitly solve 
for the horizontal Lie derivative of $K$ with respect to 
$\partial/c\partial t$ and also \eqref{eq:AccLapseFuct-2} to
simplify the last two terms in \eqref{eq:RicciTensorDecomp-ab}.
This results in 
\begin{equation}
\label{eq:K-Dot}
\begin{split}   
&\dot K_{ab}:=\bigl(L^\Vert_{\frac{\partial}{c\partial t}}K\bigr)_{ab}\\
&=\bigl(L^\Vert_\beta K\bigr)_{ab}+D_aD_b\alpha\\
&+\alpha\bigl[-2\signature K_{ac}K^c_b+\signature K_{ab}K^c_c-\Ricci^D(e_a,e_b)\bigr]\\
&-\alpha\signature\tfrac{\kappa}{n-2}h_{ab}\EMT(n,n)\\
&+\alpha\kappa\bigl(\EMT-\tfrac{1}{n-2}\Trace_h(\EMT)\,h\bigr)(e_a,e_b)\,.
\end{split}
\end{equation}
Note that in the last term the trace of $\EMT$ is taken with respect to 
$h$ and not $g$. The relation is 
$\Trace_h(\EMT)=\Trace_g(\EMT)-\signature\EMT(n,n)$. 

The only remaining equation that needs to be added here is that which 
relates the time derivative of $h$ with $K$. This we get from 
\eqref{eq:LieNormal-h} and \eqref{eq:Lie-Normal-Time}:
\begin{equation}
\label{eq:h-Dot}
\dot h_{ab}:=\bigl(L^\Vert_{\frac{\partial}{c\partial t}}h\bigr)_{ab}
=\bigl(L^\Vert_\beta h\bigr)_{ab}-2\alpha\epsilon K_{ab}\,.
\end{equation}

Equations \eqref{eq:h-Dot} and \eqref{eq:K-Dot} are six first-order in time 
evolution equations for the pair $(h,K)$. This pair cannot be freely 
specified but has to obey the four equations \eqref{eq:EinsteinEquation-00}
and \eqref{eq:EinsteinEquation-0a} which do not contain any time derivatives 
of $h$ or $K$. Equations \eqref{eq:EinsteinEquation-00} and 
\eqref{eq:EinsteinEquation-0a} are therefore referred to as \emph{constraints},
\index{constraints!in GR}
more specifically \eqref{eq:EinsteinEquation-00} as \emph{scalar constraint}
(also \emph{Hamiltonian constraint}) and \eqref{eq:EinsteinEquation-0a} as 
\emph{vector constraint} (also \emph{diffeomorphism constraint}).
\index{constraints!scalar}
\index{constraints!vector}
\index{constraints!Hamiltonian}
\index{constraints!diffeomorphism}

We derived these equations from the 3+1 split of a spacetime that we 
considered to be given. Despite having expressed all equations in terms 
of horizontal quantities, there is still a relic of the ambient space in 
our equations, namely the Lie derivative with respect to $\partial/\partial ct$.
We now erase this last relic by interpreting this Lie derivative as ordinary 
partial derivative of some $t$-dependent tensor field on a genuine 3-dimensional 
manifold $\Sigma$, which is not thought of as being embedded into a spacetime. 
The horizontal projection $L^\Vert_\beta$ of the spacetime Lie derivative 
that appears on the right-hand sides of the evolution equations above then 
translates to the ordinary intrinsic Lie derivative on $\Sigma$ with respect 
to $\beta$. This is how from now on we shall read the above equations. 
Spacetime does not yet exist. Rather, it has to be constructed from the 
evolution of the fields according to the equations, usually complemented 
by the equations  that govern the evolution of the matter fields. In these 
evolution equations $\alpha$ and $\beta$ are freely specifiable functions, 
the choice of which is subject to mathematical/computational convenience. 
Once $\alpha$ and $\beta$ are specified and $h$ as a function of 
parameter-time has been determined, we can form the expression 
\eqref{eq:ThreePlusOneOfMetric} for the spacetime metric and know that, 
by construction, it will satisfy Einstein's equations.

To sum up, the initial-value problem consists in the following steps: 
\begin{enumerate}
\item
Choose a 3-manifold $\Sigma$.
\item
Choose a time-parameter dependent lapse function $\alpha$ and 
a time-parameter dependent shift vector-field $\beta$.
\item
Find a Riemannian metric $h\in\STSigma{0}{2}$ and a 
symmetric covariant rank-2 tensor field $K\in\STSigma{0}{2}$
that satisfy equations \eqref{eq:EinsteinEquation-00} and 
\eqref{eq:EinsteinEquation-0a} either in vacuum ($\EMT=\EMT_\Lambda$; cf. 
\eqref{eq:T-Vacuum}), or after specifying some matter model.
\item
Evolve these data via \eqref{eq:h-Dot} and \eqref{eq:K-Dot},
possibly complemented by the evolution equations for the matter variables.
\item
Construct from the solution the spacetime metric $g$ via 
\eqref{eq:ThreePlusOneOfMetric}.
\end{enumerate}
For this to be consistent we need to check that the evolution 
according to  \eqref{eq:h-Dot} and \eqref{eq:K-Dot} will 
preserve the constraints \eqref{eq:EinsteinEquation-00} and 
\eqref{eq:EinsteinEquation-0a}. At this stage this could be 
checked directly, at least in the vacuum case. The easiest way 
to do this is to use the equivalence of these equations with 
Einstein's equations and then employ the twice contracted 
2nd~Bianchi identity \eqref{eq:Bianchi2Contracted}.
It follows that $\nabla_\mu E^{\mu\nu}\equiv 0$, where $E^{\mu\nu}=
G^{\mu\nu}+\lambda g^{\mu\nu}$. The four constraints 
\eqref{eq:EinsteinEquation-00} and \eqref{eq:EinsteinEquation-0a}
are equivalent to $E^{00}=0$ and $E^{0m}=0$, and the six second-order 
equations $E^{mn}=0$ to the twelve first-order evolution equations 
\eqref{eq:h-Dot} and \eqref{eq:K-Dot}. In coordinates the identity 
$\nabla_\mu E^{\mu\nu}\equiv 0$ reads 
\begin{equation}
\label{eq:Bianchi2Contracted-Coord}
\partial_0E^{0\nu}=
-\partial_mE^{m\nu}
-\Gamma^\mu_{\mu\lambda}E^{\lambda\nu}
-\Gamma^\nu_{\mu\lambda}E^{\mu\lambda}\,,
\end{equation}
which shows immediately that the time derivatives of the 
constraint functions are zero if the constraints vanished 
initially. This suffices for analytic data, but in the general 
case one has to do more work. Fortunately the equations for 
the evolution of the constraint functions can be put into an
equivalent form which is manifestly symmetric hyperbolic%
~\cite{Frittelli:1997}.
\index{symmetric hyperbolicity}
\index{constraints!preservation}
That suffices to conclude the preservation of the constraints 
in general. In fact, symmetric hyperbolicity implies more than 
that. It ensures the well-posedness 
\index{well posedness!of initial-value problem for constraints}
of the initial-value problem for the constraints, which not 
only says that they stay zero if they are zero initially, but 
also that they stay small if they are small initially. This is 
of paramount importance in numerical evolution schemes, in 
which small initial violations of the constraints must be 
allowed for and hence the consequences of these violations 
need to be controlled.  For a recent and mathematically more 
thorough discussion of the Cauchy problem \index{Cauchy problem} 
we refer to James Isenberg's survey 
\cite{Isenberg:SpringerHandbookSpacetime2014}. 

Finally we wish to substantiate our earlier claim that any 
$\Sigma$  can carry some initial data. Let us show this for 
closed $\Sigma$. To this end we choose a matter model such 
that the right-hand side of \eqref{eq:EinsteinEquation-0a} 
vanishes. Note that this still allows for arbitrary cosmological
constants since $\EMT_\Lambda(n,e_a)\propto g(n,e_a)=0$. Next we 
restrict to those pairs $(h,K)$ were $K=\lambda h$ for some 
constant $\lambda$. Geometrically this means that, in the 
spacetime to be developed, the Cauchy surface will be totally 
umbilical (isotropic Weingarten map). Due to this proportionality 
and the previous assumption the vector constraint  
\eqref{eq:EinsteinEquation-0a} will be satisfied. In the 
scalar constraint we have $G(K,K)=G(\lambda h,\lambda h)=-6\lambda^2$
so that it will be satisfied provided that  
\begin{equation}
\label{eq:KazdanWarner-1}
-\signature\Scalar^D=2\kappa\EMT(n,n)-6\lambda^2\,.
\end{equation}
For the following argument the Lorentzian signature, $\signature =-1$, 
will matter. For physical reasons we assume the weak energy 
condition so that $\kappa\EMT(n,n)\geq 0$, which makes a positive 
contribution to the right-hand side of  \eqref{eq:KazdanWarner-1}. 
However, if we choose the modulus of $\lambda$ sufficiently large we 
can make the right-hand side negative somewhere (or everywhere, since 
$\Sigma$ is compact). Now, in dimensions 3 or higher the following 
\emph{theorem of Kazdan \& Warner}
\index{theorem!of Kazdan-Warner} 
holds (\cite{Kazdan.Warner:1975}, Theorem\,1.1): Any smooth function 
on a compact manifold which is negative somewhere is the scalar curvature 
for some smooth Riemannian metric. Hence a smooth $h$ exists which 
solves \eqref{eq:KazdanWarner-1} for any given $\EMT(n,n)\geq 0$, 
provided we choose $\lambda^2>\vert\lambda\vert$  sufficiently large.
If $\Sigma$ is not closed a corresponding theorem may also be shown 
\cite{Witt:1986a}.  

The above argument crucially depends on the signs. There is no 
corresponding statement for positive scalar curvatures. In fact, 
there is a strong topological obstruction against Riemannian metrics 
of strictly positive scalar curvature. It follows from the 
\emph{theorem of Gromov \& Lawson}
\index{theorem!of Gromov-Lawson} (\cite{Gromov.Lawson:1983}, 
Theorem\,8.1) that a 3-dimensional closed orientable 
$\Sigma$ allows for Riemannian metrics with positive scalar 
curvature iff its prime decomposition consists of prime-manifolds with 
finite fundamental group or ``handles'' $S^1\times S^2$. All manifolds 
whose prime list contains at least one so-called $K(\pi,1)$-factor 
(a 3-manifold whose only non-trivial homotopy group is the first)
are excluded. See, e.g., \cite{Giulini:1994a} for more explanation 
of these notions. We conclude that the given argument crucially depends 
on $\signature =-1$.

\subsection{A note on slicing conditions}
\label{sec:Slicing conditons}
The freedom in choosing the lapse and shift functions can be of 
much importance, theoretically and in numerical evolution schemes.
This is particularly true for the lapse function $\alpha$, which 
determines the amount of proper length by which the Cauchy slice 
advances in normal direction per unit parameter interval.
If a singularity is to form in spacetime due to the collapse of 
matter within a bounded spatial region, it would clearly be advantageous 
to not let the slices run into the singularity before the outer parts of
it have had any chance to develop a sufficiently large portion of 
spacetime that one might be interested in, e.g. for the study of gravitational 
waves produced in the past. This means that one would like to slow down 
$\alpha$ in regions which are likely to develop a singularity and speed up 
$\alpha$ in those regions where it seems affordable. Take as an example 
the ``equal-speed'' gauge $\alpha=1$ and $\beta=0$, so that 
$g=-c^2\,dt^2+h$. This means that $n=\partial/\partial ct$ is geodesic.
Taking such a gauge from the $t=0$ slice in the Schwarzschild/Kruskal 
spacetime would let the slices run into the singularity after a 
proper-time of $t=\pi GM/c^3$, where $M$ is the mass of the black hole. 
In that short period of time the slices had no chance to explore a 
significant portion of spacetime outside the black hole.  

A gauge condition that one may anticipate to have singularity-avoiding 
character is that where $\alpha$ is chosen such that the divergence of 
the normal field $n$ is zero. This condition just means that the locally 
co-moving infinitesimal volume elements do not change volume, for 
$L_nd\mu=(\nabla\cdot n)\,d\mu$, where $d\mu=\det\{h_{ab}\}d^3x$ is the 
volume element of $\Sigma$. From \eqref{eq:CovDerNormalField-2} we see that 
$n$ has zero divergence iff $K$ has zero trace, i.e. the slices are of zero
mean-curvature. The condition on $\alpha$ for this to be preserved under 
evolution follows from 
\begin{equation}
\label{eq:YorkSlicing-1}
0=L_n(h^{ab}K_{ab})=-K^{ab}L_nh_{ab}+h^{ab}L_nK_{ab}\,.
\end{equation}
Here we use \eqref{eq:LieNormal-h} to eliminate $L_nh_{ab}$ in the 
first term and \eqref{eq:DecompCurv-4} to eliminate $L_nK_{ab}$ in the 
second term, also making use of \eqref{eq:AccLapseFuct-2}. This leads to 
the following equivalent of \eqref{eq:YorkSlicing-1}:
\begin{equation}
\label{eq:YorkSlicing-2}
\Delta_h\alpha+\signature\bigl(\Ricci(n,n)+K^{ab}K_{ab}\bigr)\alpha=0\,.
\end{equation}
This is a linear elliptic equation for $\alpha$.
The case of interest to us in GR is $\signature=-1$. In the closed case 
we immediately deduce by standard arguments that $\alpha=0$ is the only 
solution, provided the strong energy-condition holds (which implies 
$\Ricci(n,n)\geq 0$). In the open case, where we might impose 
$\alpha\rightarrow 1$ as asymptotic condition, we deduce existence and 
uniqueness again under the assumption of the strong energy condition. 
Hence we may indeed impose the condition $h^{ab}K_{ab}=0$, or 
$\Trace(\Wein)=0$, for non-closed $\Sigma$. It is called the 
\emph{maximal slicing condition}
\index{maximal slicing}\index{slicing!maximal}
or \emph{York gauge}~\cite{York:1979}.
\index{York!gauge}\index{gauge!York}

Whereas this gauge condition has indeed the desired singularity-avoiding 
character it is also not easy to implement due to the fact that at each 
new stage of the evolution one has to solve the elliptic equation 
\eqref{eq:YorkSlicing-2}. For numerical studies it is easier to implement 
evolution equations for $\alpha$. Such an equation is, e.g., obtained 
by asking the time function \eqref{eq:TimeFunction} to be harmonic, in the 
sense that 
\begin{equation}
\label{eq:HarmonicGauge-1}
\begin{split}
&0=\Box_gt:
=g^{\mu\nu}\nabla_{\mu}\nabla_{\nu} t\\
&=\vert\det\{g_{\alpha\beta}\}\vert^{-\frac{1}{2}}\partial_\mu
\Bigl(\vert\det\{g_{\alpha\beta}\}\vert^{\frac{1}{2}}g^{\mu\nu}\partial_\nu\Bigr) t\,.
\end{split}
\end{equation}
This is clearly just equivalent to   
\begin{equation}
\label{eq:HarmonicGauge-2}
\partial_\mu
\Bigl(\vert\det\{g_{\mu\nu}\}\vert^{\frac{1}{2}}g^{\mu 0}\Bigr)=0\,,
\end{equation}
which can be rewritten using \eqref{eq:ThreePlusOneOfInverseMetric} 
and \eqref{eq:VolumeForm} to give
\begin{equation}
\label{eq:HarmonicGauge-3}
\begin{split}
\dot\alpha:&=\frac{\partial\alpha}{c\partial t}=
L_\beta\alpha-\signature K^a_a\alpha^2\\
&=L_\beta\alpha+\Trace(\Wein)\ \alpha^2\,.
\end{split}
\end{equation}
This is called the \emph{harmonic slicing condition}.
\index{harmonic slicing}
\index{slicing!harmonic}
Note that we can still choose $\beta=0$ and try to determine 
$\alpha$ as function of the trace of $\Wein$. There also exist 
generalizations to this condition where $\alpha^2$ on the 
right-hand side is replaced with other functions $f(\alpha)$.

\subsection{A note on the De\,Witt metric}
\label{sec:DeWittMetric-Note}
\index{metric!De\,Witt}
\index{De\,Witt metric}
At each point $p$ on $\Sigma$ the De\,Witt metric \eqref{eq:Def-DeWittMetric} 
can be regarded as a symmetric bilinear form on the space of positive-definite 
inner products $h$ of $T_p\Sigma$. The latter is an open convex cone in 
$T^*_P\Sigma\otimes T^*_P\Sigma$. We wish to explore its properties a little 
further. 

A frame in $T_p\Sigma$ induces a frame in $T^*_p\Sigma\otimes T^*_p\Sigma$
(tensor product of the dual frame). If $h_{ab}$ are the components of $h$
then we have the following representation of the generalized De\,Witt 
metric
\begin{subequations}
\label{eq:DeWittMetric-1}
\begin{equation}
\label{eq:DeWittMetric-1a}
G_{(\lambda)}=
G_{(\lambda)}^{abcd}\,dh_{ab}\otimes dh_{cd}\,,
\end{equation}
where
\begin{equation}
\label{eq:DeWittMetric-1b}
G_{(\lambda)}^{abcd}=\frac{1}{2}\bigl(
 h^{ac}h^{bd}
+h^{ad}h^{bc}
-2\lambda\,h^{ab}h^{cd}
\bigr)\,.
\end{equation}
\end{subequations}
Here we introduced a factor $\lambda$ in order to parametrize the 
impact of the negative trace term. We also consider $\Sigma$ to be 
of general dimension $n$.

The inverse metric to  \eqref{eq:DeWittMetric-1} is given by 
\begin{subequations}
\label{eq:DeWittMetric-2}
\begin{equation}
\label{eq:DeWittMetric-2a}
G^{-1}_{(\lambda)}= 
G^{-1}_{(\lambda)\,abcd}\,\frac{\partial}{\partial h_{ab}}\otimes\frac{\partial}{\partial h_{cd}}\,,
\end{equation}
where
\begin{equation}
\label{eq:DeWittMetric-2b}
G^{-1}_{(\lambda)\,abcd}=\frac{1}{2}\bigl(
 h_{ac}h_{bd}
+h_{ad}h_{bc}
-2\mu\,h_{ab}h_{cd}
\bigr)\,.
\end{equation}
\end{subequations}
The relation between $\lambda$ and $\mu$ is 
\begin{equation}
\label{eq:DeWittMetric-3}
\lambda+\mu=n\lambda\mu\,,
\end{equation}
so that 
\begin{equation}
\label{eq:DeWittMetric-4}
G_{(\lambda)}^{abnm}G^{-1}_{(\lambda)\,nmcd}
=\tfrac{1}{2}\bigl(\delta^a_c\delta^b_d+\delta^a_d\delta^b_c\bigr)\,.
\end{equation}
In ordinary GR $n=3$, $\lambda=1$, and $\mu=1/2$. 
Note that there are good reasons the keep the superscript $-1$
even in component notation, that is, to write $G^{-1}_{(\lambda)\,abcd}$
rather than just $G_{(\lambda)\,abcd}$, since $G^{-1}_{(\lambda)\,abcd}$
does not equal $h_{ak}h_{bl}h_{cm}h_{dn}G_{(\lambda)}^{klmn}$ unless
$\lambda=2/n$, in which case $\lambda=\mu$.   
 
If we change coordinates according to 
\begin{equation}
\label{eq:DeWittMetric-NewCoord}
\begin{split}
\tau:&=\ln\Bigl(\bigl[\det\{h_{ab}\}\bigr]^{\frac{1}{n}}\Bigr)\,,\\
r_{ab}:&=h_{ab}/\bigl[\det\{h_{ab}\}\bigr]^{\frac{1}{n}}\,,
\end{split}
\end{equation}
where $\tau$ parametrizes conformal changes and $r_{ab}$ 
the conformally invariant ones, the metric \eqref{eq:DeWittMetric-1} 
reads
\begin{equation}
\label{eq:DeWittMetric-5}
G_{(\lambda)}=n(1-\lambda n)\,d\tau\otimes d\tau+
r^{ac}r^{bd}\,dr_{ab}\otimes dr_{cd}\,,
\end{equation}
where $r^{an}r_{nb}=\delta^a_b$. Since $h$ is positive definite, so is $r$.
Hence the second part is positive definite on the $\bigl(\tfrac{1}{2}n(n+1)-1\bigr)$-- 
dimensional vector space of trace-free symmetric tensors. Hence the De\,Witt 
metric is positive definite for $\lambda <1/n$, Lorentzian for $\lambda >1/n$,
\index{Lorentz!signature of De\,Witt metric}
and simply degenerate (one-dimensional null space) for the critical value 
$\lambda =1/n$. In the GR case we have $\lambda =1$ and $n=3$, so that the 
De\,Witt metric is Lorentzian of signature $(-,+,+,+,+,+)$. Note that this 
Lorentzian signature is independent of $\signature$, i.e. it has nothing to 
do with the Lorentzian signature of the spacetime metric. 

In the Hamiltonian formulation it is not $G$ but rather a conformally related 
metric that is important, the conformal factor being $\sqrt{\det\{h_{ab}\}}$.
If we set 
\begin{equation}
\label{eq:ConfDeWittMetric-1}
{\hat G}_{(\lambda)}:=\bigl[\det\{h_{ab}\}\bigr]^{1/2}\,G_{(\lambda)}
\end{equation}
and correspondingly 
\begin{equation}
\label{eq:ConfDeWittMetric-2}
{\hat G}^{-1}_{(\lambda)}:=\bigl[\det\{h_{ab}\}\bigr]^{-1/2}\,G^{-1}_{(\lambda)}\,,
\end{equation}
we can again write ${\hat G}_{(\lambda)}$ in terms of $(\tau,r_{ab})$. In fact, 
the conformal rescaling clearly just corresponds to multiplying 
\eqref{eq:DeWittMetric-5} with $\sqrt{\det\{h_{ab}\}}=e^{n\tau/2}$. 
Setting 
\begin{equation}
\label{eq:DefHatTau}
T:=4\bigl[(1-n\lambda)/n\bigr]^{1/2}\,e^{n\tau/4}
\end{equation}
we get, excluding the degenerate case $\lambda=1/n$,
\begin{equation}
\label{eq:ConfDeWittMetric-3}
\begin{split}
{\hat G}_{(\lambda)}
&=\mathrm{sign}(1-n\lambda)\,dT\otimes dT\\
&+T^2\,C\,r^{ac}r^{bd}\,dr_{ab}\otimes dr_{cd}\,,
\end{split}
\end{equation}
where $C=n/(16\vert 1-n\lambda\vert)$ ($=3/32$ in GR).
This is a simple warped product metric of $\reals_+$ 
with the left-invariant metric on the homogeneous space 
$GL(3,\reals)/SO(3)\times\reals_+$ of symmetric positive 
definite forms modulo overall scale, the warping function 
being just $T^2$ if $T$ is the coordinate on $\reals_+$. 
Now, generally, quadratic warped-product metrics of the form  
$\pm dT\otimes dT+T^2g$, where $g$ is independent of $T$, are 
non-singular for $T\searrow 0$ iff $g$ is a metric of constant 
curvature $\pm 1$ (like for a unit sphere in $\reals^n$, with 
$T$ being the radius coordinate, or the unit spacelike hyperboloid 
in $n$-dimensional Minkowski space, respectively). This is not the 
case for \eqref{eq:ConfDeWittMetric-3}, which therefore has a curvature 
singularity for small $T$, i.e. small $\det\{h_{ab}\}$. Note that 
this is a singularity in the space of metrics (here at a fixed 
space point), which has nothing to do with spacetime singularities. 
In the early days of Canonical Quantum Gravity this has led to 
speculations concerning ``natural'' boundary conditions for the wave 
function, whose domain is the space of 
metrics~\cite{DeWittQTGI:1967}. The intention was to pose conditions 
such that the wave function should stay away from such singular 
regions in the space of metrics; see also \cite{Kiefer:QuantumGravity}
for a more recent discussion. 

We stress once more that the signature of the De\,Witt metric is 
\emph{not} related to the signature of spacetime, i.e. independent 
of $\signature$. For example, for the GR values $\lambda=1$ and 
$n=3$, it is Lorentzian even if spacetime were given a Riemannian 
metric. Moreover, by integrating over $\Sigma$, the pointwise metric 
\eqref{eq:ConfDeWittMetric-3} defines a bilinear form on the 
infinite dimensional space of Riemannian structures on $\Sigma$, 
the geometry of which may be investigated to some limited 
extent~\cite{Giulini:1995c}\cite{Giulini:2009a}.

\section{Constrained Hamiltonian systems}
\label{sec:ConstrainedHamiltonianSystems}
In this section we wish to display some characteristic 
features of Hamiltonian dynamical systems with constraints. 
We restrict attention to finite-dimensional systems in order
to not overload the discussion with analytical subtleties.

Let $Q$ be the $n$-dimensional configuration manifold 
of a dynamical system that we locally coordinatize by 
$(q^1,\cdots ,q^n)$. By $TQ$ we denote its tangent bundle, 
which we coordinatize by $(q^1,\cdots,q^n\,,\,v^1,\cdots,v^n)$, 
so that a tangent vector $X\in TQ$ is given by 
$X=v^a\partial/\partial q^a$. The dynamics of the system is 
described by a \emph{Lagrangian}
\index{Lagrangian (Lagrange function)}
\begin{equation}
\label{eq:Lagrangian}
L:TQ\rightarrow\reals\,,
\end{equation}
which selects the dynamically possible trajectories in $TQ$
as follows: Let $\reals\ni t\mapsto x(t)\in Q$ be a (at least 
twice continuously differentiable) curve, then it is dynamically 
possible iff the following 
\emph{Euler Lagrange equations} hold (we set $dx/dt=:\dot x$): 
\index{Euler Lagrange equations}
\begin{equation}
\label{eq:EulerLagrangeEquations-1}
\frac{\partial L}{\partial q^a}\bigg\vert_{\genfrac{}{}{0pt}{}{q=x(t)}{v=\dot x(t)}}
-\frac{d}{dt}\left[\frac{\partial L}{\partial v^a}%
\bigg\vert_{\genfrac{}{}{0pt}{}{q=x(t)}{v=\dot x(t)}}\right]=0\,.
\end{equation}
Performing the $t$-differentiation on the second term, 
this is equivalent to 
\begin{equation}
\label{eq:EulerLagrangeEquations-2}
H_{ab}\bigl(x(t),\dot x(t)\bigr){\ddot x}^b=V_a\bigl(x(t),\dot x(t)\bigr)\,,
\end{equation}
where
\begin{equation}
\label{eq:EulerLagrangeEquations-3}
H_{ab}(q,v):=
\frac{\partial^2L(q,v)}{\partial v^a\partial v^b}\,,
\end{equation}
and 
\begin{equation}
\label{eq:EulerLagrangeEquations-4}
V_a(q,v):=
\frac{\partial L(q,v)}{\partial q^a}
-\frac{\partial^2 L(q,v)}{\partial v^a\partial q^b}\ v^b\,.
\end{equation}
Here we regard $H$ and $V$ as function on $TQ$ with values in 
the symmetric $n\times n$ matrices and $\reals^n$ respectively. 
In order to be able to solve \eqref{eq:EulerLagrangeEquations-2} 
for the second derivative $\ddot x$ the matrix $H$ has to be 
invertible, that is, it must have rank $n$. That is the case 
usually encountered in mechanics. On the other hand, 
\emph{constrained systems}
\index{constrained Hamiltonian system}%
\index{systems, constrained}%
are those where the rank of $H$ is not maximal.  This is the case 
we are interested in. 

We assume $H$ to be of constant rank $r<n$. Then, for each point 
on $TQ$,  there exist $s=(n-r)$ linearly independent kernel elements 
$K_{(\alpha)}(q,v)$, 
$\alpha=1,\cdots, s$, such that $K^a_{(\alpha)}(q,v)H_{ab}(q,v)=0$. 
Hence any solution $x(t)$ to \eqref{eq:EulerLagrangeEquations-2}
must be such that the curve $t\mapsto\bigl(x(t),\dot x(t)\bigr)$ 
in $TQ$ stays on the subset
\begin{subequations}
\label{eq:ConstrainedSurfaceTQ}
\begin{equation}
\label{eq:ConstrainedSurfaceTQ-a}
\Constraint:=
\bigl\{
(q,v)\in TQ :  \psi_{\alpha}(q,v)=0\,,\quad
\alpha=1,\cdots,s\bigr\},
\end{equation} 
where  
\begin{equation}
\label{eq:ConstrainedSurfaceTQ-b}
\psi_\alpha(q,v)=K^a_{(\alpha)}(q,v)V_a(q,v)\,.
\end{equation}
\end{subequations}
We assume $\Constraint\subset TQ$ to be a smooth closed 
submanifold of co-dimension $s$, i.e. of dimension $2n-s=n+r$. 

Now we consider the cotangent bundle $T^*Q$ over $Q$. 
On $T^*Q$ we will use so-called 
\emph{canonical coordinates},
\index{canonical!coordinates}
\index{coordinates, canonical}
denoted by $\{q^1,\cdots,q^n,p_1,\cdots,p_n\}$, the precise 
definition of which we will give below.  The Lagrangian 
defines a map $\FL: TQ\rightarrow T^*Q$, which in these 
coordinates reads 
\begin{equation}
\label{eq::FL-Map}
\FL(q,v)=\left(q,p:=\frac{\partial L(q,v)}{\partial v}\right)\,.
\end{equation}
From what has been said above it follows that the Jacobian of 
that map has constant rank $n+r$. Given sufficient regularity,
we may further assume that 
\begin{equation}
\label{eq:ConstrainedSurfaceTQStar-1}
\Constraint^*:=\FL\bigl(\Constraint\bigr)\subset T^*Q
\end{equation}
is a smoothly embedded closed submanifold in phase space $T^*Q$ 
of co-dimension $s$. Hence there are $s$ functions $\phi_\alpha$, 
$\alpha=1,\cdots, s$ such that 
\begin{equation}
\label{eq:ConstrainedSurfaceTQStar-2}
\Constraint^*:=
\bigl\{
(q,p)\in T^*Q :  \phi_{\alpha}(q,p)=0\,,\
\alpha=1,\cdots,s\bigr\}.
\end{equation} 
This is called the \emph{constraint surface}
\index{constraints!surface}
\index{surface, constraint} 
in phase space. It is given as the intersection of the 
zero-level sets of $s$ independent functions. Independence 
means that at each $p\in\Constraint^*$ the $s$ 
one-forms $d\phi_1\vert_p\,\cdots,d\phi_s\vert_p$ are linearly 
independent elements of $T^*_pT^*Q$. 

The dynamical trajectories of our system will stay entirely 
on $\Constraint^*$. The trajectories themselves are 
integral lines of a Hamiltonian flow. But what is the 
Hamiltonian function that generates this flow?  To explain 
this we first recall the definition of the \emph{energy function} 
\index{energy!function}
\index{function!energy}
for the Lagrangian $L$. It is a function $E:TQ\rightarrow\reals$ 
defined through 
\begin{equation}
\label{eq:EnergyFunction}
E(q,v):=\frac{\partial L(q,v)}{\partial v^a}\, v^a-L(q,v)\,.
\end{equation}
At first sight this function cannot be defined on phase space, 
for we cannot invert $\FL$ to express $v$ as function of $q$ and
$p$ which we could insert into $E(q,v)$ in order to get $E(q,v(q,p))$.
However, one may prove the following: There exists a function 
\begin{subequations}
\label{eq:ZeroHamiltonian}
\begin{equation}
\label{eq:ZeroHamiltonian-a}
H_{\Constraint^*}:\Constraint^*\rightarrow\reals\,,
\end{equation}
so that 
\begin{equation}
\label{eq:ZeroHamiltonian-b}
E=H_{\Constraint^*}\circ\FL\,.
\end{equation}
\end{subequations}
A local version of this is seen directly from taking the differential 
of \eqref{eq:EnergyFunction}, which yields 
$dE=v^ad(\partial L/\partial v^a)-(\partial L/\partial q^a)dq^a$,
expressing the fact that $dE_{(q,v)}(X)=0$ if $\FL_{*(q,v)}(X)=0$
for $X\in T_{(q,v)}TQ$, or in simple terms: $E$ does not vary if 
$q$ and $p$ do not vary. 

So far the function $H_{\Constraint^*}$ is only defined on $\Constraint^*$.
By our regularity assumptions there exists a smooth extension of 
it to $T^*Q$, that is a function $H_0:T^*Q\rightarrow\reals$ such 
that $H_0\vert_{\Constraint^*}=H_{\Constraint^*}$. This is clearly not unique. 
But we can state the following: Let $H_0$ and $H$ both be smooth 
(at least continuously differentiable) extensions of $H_{\Constraint^*}$ to 
$T^*Q$, then there exist $s$ smooth functions 
$\lambda^\alpha:T^*Q\rightarrow\reals$ such that 
\begin{equation}
\label{eq:RelationHamiltonians-1}
H=H_0+\lambda^\alpha\phi_\alpha\,.
\end{equation}
Locally a proof is simple: Let $f:T^*Q\rightarrow\reals$ be 
continuously differentiable and such that 
$f\vert_{\Constraint^*}\equiv 0$. Consider a point 
$p\in\Constraint^*$ and coordinates $(x^1,\cdots,x^{2n-s}, y^1,\cdots y^s)$
in a neighborhood $U\subset T^*Q$ of $p$, where the $x$'s are 
coordinates on the constraint surface and the $y$'s are just the 
functions $\phi$. In $U$ the constraint surface is clearly just 
given by $y^1=\cdots =y^s=0$. Then 
\begin{subequations}
\label{eq:ProofVanishingFunctionLemma}
\begin{equation}
\label{eq:ProofVanishingFunctionLemma-a}
\begin{split}
f\vert_U(x,y)
&=\int_0^1dt\frac{d}{dt}f(x,ty)\\
&=\int_0^1dt\frac{\partial f}{\partial y^\alpha}(x,ty)\ y^\alpha
=\lambda_\alpha(x,y)\,y^\alpha\,,
\end{split}
\end{equation}
where 
\begin{equation}
\label{eq:ProofVanishingFunctionLemma-b}
\lambda_\alpha(x,y):=\int_0^1dt\frac{\partial f}{\partial y^\alpha}(x,ty)\,.
\end{equation}
\end{subequations}
For a global discussion see \cite{HenneauxTeitelboim:QGS}.

As \emph{Hamiltonian}
\index{Hamiltonian}
\index{function!Hamiltonian} 
for our constraint system we address any smooth (at least continuously 
differentiable) extension $H$ of $H_{\Constraint^*}$. So if $H_0$ is a somehow 
given one, any other can be written as  
\begin{equation}
\label{eq:GeneralHamiltonianFunction}
H=H_0+\lambda^\alpha\phi_\alpha
\end{equation}
for some (at least continuously differentiable) real-valued  
functions $\lambda^\alpha$ on $T^*Q$. 

Here we have been implicitly assuming that the Hamiltonian 
dynamics does not leave the constraint surface
\eqref{eq:ConstrainedSurfaceTQStar-1}. If this were not the 
case we would have to restrict further to proper submanifolds 
of $\Constraint^*$ such that the Hamiltonian vector fields 
evaluated on them lie tangentially. (If no such submanifold can 
be found the theory is simply empty). This is sometimes expressed 
by saying that the \emph{primary constraints}
\index{primary constraints}
\index{constraints!primary}
(those encountered first in the Lagrangian/Hamiltonian analysis) 
are completed by \emph{secondary}, \emph{tertiary}, etc.
constraints for consistency.
\index{secondary constraints}
\index{constraints!secondary}

Here we assume that our system is already dynamically consistent. 
This entails that the Hamiltonian vector-fields $X_{\phi_\alpha}$ for 
the $\phi_\alpha$ are tangential to the constraint surface.  
This is equivalent to $X_{\phi_\alpha}(\phi_\beta)\vert_{\Constraint^*}=0$,
or expressed in Poisson brackets: 
\begin{equation}
\label{eq:FirstClassPissonBracket-1}
\{\phi_\alpha,\phi_\beta\}\big\vert_{\Constraint^*}=0\,,
\end{equation} 
for all $\alpha,\beta\in\{1,\cdots,s\}$. Following Dirac~\cite{Dirac:LQM}, 
constraints which satisfy this condition are said to be of \emph{first class}.
\index{first class constraints}
\index{constraints!first class}  
By the result shown (locally) above in \eqref{eq:ProofVanishingFunctionLemma} this 
is equivalent to the existence of $\tfrac{1}{2}s^2(s-1)$ (at least continuously 
differentiable) real-valued functions $C^\gamma_{\alpha\beta}=-C^\gamma_{\beta\alpha}$ 
on $T^*Q$, such that 
\begin{equation}
\label{eq:FirstClassPissonBracket-2}
\{\phi_\alpha,\phi_\beta\}=C^\gamma_{\alpha\beta}\,\phi_\gamma\,.
\end{equation} 
Note that as far as the intrinsic geometric properties of the 
constraint surface are concerned  \eqref{eq:FirstClassPissonBracket-1}
and  \eqref{eq:FirstClassPissonBracket-2} are equivalent. 

The indeterminacy of the Hamiltonian due to the freedom to choose 
any set of $\lambda^\alpha$ seems to imply an $s$-dimension
worth of indeterminacy in the dynamically allowed motions. 
But the difference in these motions is that generated by the 
constraint functions on the constraint surface. In order to 
actually tell apart two such motions requires observables 
(phase-space functions) whose Poisson brackets with the 
constraints do \emph{not} vanish on the constraint surface. 
The general attitude is to assume that this is not possible, 
i.e. to assume that \emph{physical observables}
\index{physical!observables}
\index{observables, physical}
correspond exclusively to phase-space functions whose Poisson  
bracket with all constraints vanish on the constraint surface.
This is expressed by saying that all motions generated by the 
constraints are \emph{gauge transformations}.
\index{gauge!transformations}%
This entails that they are 
undetectable in principle and merely correspond to a 
mathematical redundancy in the description rather than to  
any physical degrees of freedom. It is therefore more 
correct to speak of gauge \emph{redundancies} rather than of 
gauge \emph{symmetries}, 
\index{gauge!redundancies versus symmetries}
\index{symmetries!gauge (versus redundancies)}
\index{redundancy!gauge (versus symmetries)}
as it is sometimes done, for the word 
``symmetry'' is usually used for a physically meaningful 
operation that does change the object to which it is applied 
in at least some aspects (otherwise the operation is the 
identity). Only some ``relevant'' aspects, in the context of
which one speaks of symmetry, are not changed.

\subsection{Geometric theory}
\label{sec:GeometricTheory}
Being first class has an interpretation in terms of symplectic 
geometry. To see this, we first recall a few facts and notation 
from elementary symplectic geometry of cotangent bundles. Here 
some sign conventions enter and the reader is advised to compare 
carefully with other texts. 

A \emph{symplectic structure}
\index{symplectic!structure} on a manifold is a non-degenerate 
closed two-form.  Such structures always exist in a natural way 
on cotangent bundles, where they even derive from a 
\emph{symplectic potential}. 
\index{symplectic!potential}
The latter is a one-form field $\theta$ on $T^*Q$ whose general 
geometric definition is as follows: Let $\pi: T^*Q\rightarrow Q$ 
be the natural projection from the co-tangent bundle of 
$Q$ (phase space) to $Q$ itself. Then, for each $p\in T^*Q$, 
we define
\begin{equation}
\label{eq:DefSymplPot-Inv}
\theta_p:=p\circ\pi_{*p}\,.
\end{equation}
So in order to apply $\theta_p$ to a vector $X\in T_pT^*Q$,
we do the following: Take the differential $\pi_*$ of the projection 
map $\pi$, evaluate it at point $p$ and apply it to $X\in T_pT^*Q$
in order to push it forward to the tangent space $T_{\pi(p)}Q$ 
at point $\pi(p)\in Q$. Then apply $p$ to it, which makes sense since 
$p$ is, by definition, an element of the co-tangent space at 
$\pi(p)\in Q$.

The symplectic structure, 
\index{symplectic!structure}
$\omega$, is now given by 
\begin{equation}
\label{eq:DefSymplStructure-Inv}
\omega=\,-\,d\theta\,.
\end{equation}
The minus sign on the right-hand side has no significance 
other than to comply with standard conventions. 
Let us stress that $\theta$, and hence $\omega$, is globally defined. 
This is obvious from the global definition \eqref{eq:DefSymplPot-Inv}.
Therefore $\omega$ is not only closed, $d\omega=0$, but even globally 
exact for any $Q$. Non-degeneracy of $\omega$ will be immediate from 
the expression in canonical coordinates to be discussed below 
(cf.~\eqref{eq:SympStructureCoordinates-b}).

A diffeomorphism $F:T^*Q\rightarrow T^*Q$ is called a 
\emph{canonical transformation}
\index{canonical!transformation}
\index{transformation, canonical}
or \emph{symplectic morphism}
\index{symplectic!morphism}
if it preserves $\omega$, that is, if $F^*\omega=\omega$. 
We explicitly mention two kinds of canonical transformations,
which in some sense are complementary to each other. 

The first set of canonical transformations are fibre-preserving 
ones.  This means that, for each $q\in Q$,  points in the fibre 
$\pi^{-1}(q)$ are moved to points in the same fibre $\pi^{-1}(q)$. 
This is equivalent to the simple equation  
\begin{equation}
\label{eq:FibrePreservingTrans}
\pi\circ F=\pi\,. 
\end{equation}
The special fibre-preserving diffeomorphisms we wish to mention 
are given by adding to each momentum $p\in T^*Q$ the value 
$\sigma\bigl(\pi(p)\bigr)$ of a section $\sigma: Q\rightarrow T^*Q$:
\begin{equation}
\label{eq:CanTransFibreShift}
F(p)=p+\sigma_{\pi(p)}\,.
\end{equation}
This transforms the symplectic potential at $p\in T^*Q$ 
into 
\begin{equation}
\label{eq:VertShiftDiffeo-1}
\begin{split}
(F^*\theta)_p
&\stackrel{\phantom{\eqref{eq:DefSymplPot-Inv}}}{=}\theta_{F(p)}\circ F_{*p}\\
&\stackrel{\eqref{eq:DefSymplPot-Inv}}{=}  F(p)\circ\bigl(\pi\circ F\bigr)_{*p}\\
&\stackrel{\eqref{eq:FibrePreservingTrans}}{=} F(p)\circ\pi_{*p} \\
&\stackrel{\eqref{eq:CanTransFibreShift}}{=}\theta_p+\sigma_{\pi(p)}\circ\pi_{*p}\\
&\stackrel{\phantom{\eqref{eq:DefSymplPot-Inv}}}{=}\theta_p+\bigl(\pi^{*}\sigma\bigr)_p\,.
\end{split}
\end{equation}
Hence 
\begin{subequations}
\label{eq:VertShiftDiffeo-2}
\begin{alignat}{2}
\label{eq:VertShiftDiffeo-2-a}
&F^*\theta&&\,=\,\theta+\pi^*\sigma\,,\\
\label{eq:VertShiftDiffeo-2-b}
&F^*\omega&&\,=\,\omega-\pi^*d\sigma\,.
\end{alignat}
\end{subequations}
This is a canonical transformation if $\sigma$ is a closed 
covector field on $Q$. By Poincar\'e's Lemma  such a $\sigma$ 
is locally exact, but this need not be the case globally. 
Obstructions to global exactness are the first De\,Rahm cohomology 
class $H^1_{\rm DR}(Q)$, which is just defined to be the 
vector space of closed modulo exact covector fields on $Q$. 
The dimension of this vector space equals the rank of the free 
part of the ordinary first homology group $H_1(Q,\integers)$ on 
$Q$ with integer coefficients. This latter group is always abelian 
and isomorphic to the abelianization of the (generally non-abelian) 
first homotopy group $\pi_1(Q)$. Hence for non-simply connected 
$Q$ the possibility of canonical transformations exist which 
change the symplectic potential by a closed yet non-exact 
covector field. 

The second set of canonical transformations that we wish 
to mention are natural extensions to $T^*Q$ of diffeomorphisms 
of $Q$. These extensions not only leave invariant the symplectic 
structure $\omega$ but also the symplectic potential $\theta$. 
To see this we note that any diffeomorphism $f:Q\rightarrow Q$ 
has a natural \emph{lift} 
\index{lift, of diffeomorphims to cotangent bundle} 
to $T^*Q$. We recall that a \emph{lift} of a diffeomorphism $f$ 
of the base manifold $Q$ is a diffeomorphism 
$F:T^*Q\rightarrow T^*Q$ such that 
\begin{equation}
\label{eq:DefLiftOfDiffeo}
\pi\circ F=f\circ\pi \,.
\end{equation}
This is equivalent to saying that the following diagram of 
maps commutes (a tailed arrow indicates injectivity and a 
double-headed arrow surjectivity) 
\begin{equation}
\label{eq:CommDiagLiftCotBundle}
\bfig
\square/ >->>`>>`>>` >->>/<800,495>[T^*Q`T^*Q`Q`Q;F`\pi`\pi`f]
\efig
\end{equation}
Here the map $F$ is just the pull-back of the inverse $f^{-1}$.
Hence the image of $p\in T^*Q$ is given by 
\begin{equation}
\label{eq:LiftCotBundle}
F(p)
=p\circ f^{-1}_{*f(\pi(p))}\,.
\end{equation}
From that it follows that the symplectic potential is 
invariant under all lifts of diffeomorphisms on $Q$:
\begin{equation}
\label{eq:DiffInvSymplPot}
\begin{split}
(F^*\theta)_p
&\stackrel{\phantom{\eqref{eq:DefSymplPot-Inv}}}{=}\theta_{F(p)}\circ{F}_{*p}\\
&\stackrel{\eqref{eq:DefSymplPot-Inv}}{=} F(p)\circ\bigl(\pi\circ F\bigr)_{*p}\\
&\stackrel{\eqref{eq:DefLiftOfDiffeo}}{=}F(p)\circ f_{*\pi(p)}\circ\pi_{*p}\\
&\stackrel{\eqref{eq:LiftCotBundle}}{=}p\circ f^{-1}_{*f(\pi(p))}\circ f_{*\pi(p)}\circ\pi_{*p}\\
&\stackrel{\phantom{\eqref{eq:DefSymplPot-Inv}}}{=}p\circ\bigl(f^{-1}\circ f\bigr)_{*\pi(p)}\circ\pi_{*p}\\
&\stackrel{\phantom{\eqref{eq:DefSymplPot-Inv}}}{=}\theta_p\,.
\end{split}
\end{equation}

So far we deliberately avoided intoducing local coordinates
in order to stress global existence of the quantities in question.
We now introduce convenient coordinates in which the symplectic 
potential and structure take on the familiar form.  These are 
called \emph{canonical coordinates}, which we  already mentioned 
above and the definition of which we now give. Let $(x,U)$ be a 
local chart on $Q$ such that $x:Q\supset U\rightarrow\reals^n$ 
is the chart map with component functions $x^a$. This chart 
induces a chart $(z,V)$ on $T^*Q$, where $V=\pi^{-1}(U)\subset T^*Q$ 
and $z:V\rightarrow\reals^{2n}$. We follow general tradition and 
label the first set of $n$ component functions by $z^a=q^a$ 
(for $a=1,\cdots,n$) and the second set by $z^{n+a}=p_a$ 
(for $a=1,\cdots,n$). For the first set we define
\begin{subequations}
\label{eq:DefCanCoord}
\begin{equation}
\label{eq:DefCanCoord-a}
q^a(\lambda):=x^a\bigl(\pi(\lambda)\bigr)\,,
\end{equation}
and for the second 
\begin{equation}
\label{eq:DefCanCoord-b}
p_a(\lambda):=
\lambda\left(\frac{\partial}{\partial x^a}\bigg\vert_{\pi(\lambda)}\right)\,,
\end{equation}
\end{subequations}
for any $\lambda\in V$. Note that \eqref{eq:DefCanCoord-b} just says 
that  $\lambda=p_a(\lambda)\,dx^a\vert_{\pi(\lambda)}$. 
In this way we get a ``canonical'' extension of any chart on 
$Q$ with domain $U$ to a chart on $T^*Q$ with domain 
$V=\pi^{-1}(U)$. From the definition it is clear that 
\begin{subequations}
\label{eq:CanCoordCoordProjections}
\begin{equation}
\label{eq:CanCoordCoordProjections-a}
\pi_{*\lambda}\left(\frac{\partial}{\partial q^a}\bigg\vert_\lambda\right)=
\frac{\partial}{\partial x^a}\bigg\vert_{\pi(\lambda)}
\end{equation}
and
\begin{equation}
\label{eq:CanCoordCoordProjections-b}
\pi_{*\lambda}\left(\frac{\partial}{\partial p_a}\bigg\vert_\lambda\right)=0\,.
\end{equation}
\end{subequations}
It immediately follows from the definition \eqref{eq:DefSymplPot-Inv}
that 
\begin{subequations}
\label{eq:SymplPotComponents}
\begin{equation}
\label{eq:SymplPotComponents-a}
\theta_\lambda\left(\frac{\partial}{\partial q^a}\bigg\vert_\lambda\right)=
p_a(\lambda)
\end{equation}
and 
\begin{equation}
\label{eq:SymplPotComponents-b}
\theta_\lambda\left(\frac{\partial}{\partial p_a}\bigg\vert_\lambda\right)=
0\,.
\end{equation}
\end{subequations}
Hence, in canonical coordinates, the symplectic potential and 
structure take on the form  
\begin{subequations}
\label{eq:SympStructureCoordinates}
\begin{alignat}{2}
\label{eq:SympStructureCoordinates-a}
&\theta\vert_V=p_a\,dq^a\,,\\
\label{eq:SympStructureCoordinates-b}
&\omega\vert_V=dq^a\wedge dp_a\,.
\end{alignat}
\end{subequations}

Note again that \eqref{eq:SympStructureCoordinates} is valid in 
any canonical completion of a chart on $Q$. As advertised above, 
it is immediate from \eqref{eq:SympStructureCoordinates-b} that 
$\omega\vert_V$ is non-degenerate at any point $p\in V$. Since 
non-degeneracy is a pointwise property and valid in any canonical 
chart, it follows that $\omega$ is non-degenerate everywhere. 
In the sequel we shall drop the explicit mention of the chart domain 
$V$.

The non-degeneracy of $\omega$ allows to uniquely associate a 
vector field $X_f$ to any real-valued function $f$ on $T^*Q$
through  
\begin{equation}
\label{eq:DefHamVF}
i_{X_f}\omega=df\,.
\end{equation}
It is called the \emph{Hamiltonian vector field}
\index{Hamiltonian vector field}
\index{vector field, Hamiltonian} 
of $f$. An immediate consequence of \eqref{eq:DefHamVF} and 
$d\omega=0$ is that $\omega$ has vanishing Lie derivative with 
respect to any Hamiltonian vector field: 
\begin{equation}
\label{eq:OmegaVanLieDer}
L_{X_f}\omega=(i_{X_f}\circ d+d\circ i_{X_f})\omega=0\,.
\end{equation}
In coordinates $X_f$ looks like this: 
\begin{equation}
\label{eq:HamVFCoordinates}
X_f=
\frac{\partial f}{\partial p_a}\frac{\partial}{\partial q^a}-
\frac{\partial f}{\partial q^a}\frac{\partial}{\partial p_a}\,.
\end{equation}
The \emph{Poisson bracket} 
\index{Poisson!bracket}
between two functions $f$ and $g$ is defined as 
\begin{subequations}
\label{eq:DefPoissonBracket}
\begin{alignat}{1}
\label{eq:DefPoissonBracket-a}
\{f,g\}
:&=\omega(X_f,X_g)=X_g(f)=-X_f(g)\\
\label{eq:DefPoissonBracket-b}
&=\frac{\partial f}{\partial q^a}\frac{\partial g}{\partial p_a}-
 \frac{\partial f}{\partial p_a}\frac{\partial g}{\partial q^a}\,.
\end{alignat}
\end{subequations}
It provides $C^\infty(T^*Q)$ with a structure of a Lie algebra, 
\index{Lie!algebra}\index{algebra!Lie}
which means that for all $f,g,h\in C^\infty(T^*Q)$ and all 
$a\in\reals$ we have,
\begin{subequations}
\label{eq:PoissonIsLieAlg}
\begin{alignat}{1}
\label{eq:PoissonIsLieAlg-a}
&\{f,g\}=-\{g,f\}\,,\\
\label{eq:PoissonIsLieAlg-b}
&\{af+g,h\}=a\{f,h\}+\{g,h\}\,,\qquad\\
\label{eq:PoissonIsLieAlg-c}
&\{f,\{g,h\}\}+
\{g,\{h,f\}\}+
\{h,\{f,g\}\}=0\,.
\end{alignat} 
\end{subequations}
Antisymmetry and bi-linearity are obvious from 
\eqref{eq:DefPoissonBracket}. The third property
\eqref{eq:PoissonIsLieAlg-c}, called the 
\emph{Jacobi identity},
\index{Jacobi!identity} 
can of course be directly checked using the coordinate expression~\eqref{eq:DefPoissonBracket-b}, but the geometric 
proof is more instructive, which we therefore wish to present here.

The first thing we note is that the map $f\mapsto X_f$ obeys 
\begin{equation}
\label{eq:HamVF-AntiLieHoemo}
X_{\{f,g\}}=-[X_f,X_g]\,.
\end{equation}
This follows from 
\begin{equation}
\label{eq:HamVF-PB}
\begin{split}
d\{f,g\}
&=d\bigl(\omega(X_f,X_g)\bigr)\\
&=di_{X_g}i_{X_f}\omega\\
&=L_{X_g}i_{X_f}\omega-i_{X_g}di_{X_f}\omega\\
&=i_{[X_g,X_f]}\omega=-i_{[X_f,X_g]}\omega\,,
\end{split}
\end{equation}
where in the last step we used $di_{X_f}\omega=0$ due to 
$d\omega=0$ and $L_{X_f}\omega=0$ and once more $L_{X_g}\omega=0$
in the first term. As $\omega$ is non-degenerate comparison 
with \eqref{eq:DefHamVF} leads to \eqref{eq:HamVF-AntiLieHoemo}. 
Next we recall that the exterior differential of a general 
$k$-form field $\alpha$, applied to the $k+1$ vectors 
$X_0,X_1,\cdots,X_k$, can be written as
\begin{equation}
\label{eq:ExtDerivGlobalExpression}
\begin{split}
&d\alpha(X_0,\cdots,X_k)\\
=&\sum_{0\leq i\leq k} 
(-1)^i\ X_i\bigl(\alpha(X_0,\cdots,\hat X_i,\cdots,X_k)\bigr)\\
+&\sum_{0\leq i<j\leq k}
(-1)^{i+j}\ \alpha\bigl([X_i,X_j],X_0,\cdots,\\
&\hspace{3.0cm}\hat X_i,\cdots, \hat X_j,\cdots,X_k\bigr)\,.\\
\end{split}
\end{equation}
Here the hatted entries are omitted. This we apply to 
$d\omega(X_f,X_g,X_h)$ and make use of, e.g.,  $X_f\bigl(\omega(X_g,X_h)\bigr)=(L_{X_f}\omega)(X_g,X_h)+\omega([X_f,X_g],X_h)+
\omega(X_g,[X_f,X_h])$, as well as \eqref{eq:OmegaVanLieDer}.
Then we get, writing $\sum_{(fgh)}$ for the cyclig sum over $f,g,h$: 
\begin{equation}
\label{eq:JacobiIdentity-GeometricProof}
\begin{split}
d\omega(X_f,X_g,X_h)
&=-\sum_{(fgh)}\omega\bigl(X_f,[X_g,X_h]\bigr)\\
&\stackrel{\eqref{eq:HamVF-AntiLieHoemo}}{=}
\sum_{(fgh)}\omega\bigl(X_f,X_{\{g,h\}}\bigr)\\
&\stackrel{\eqref{eq:DefPoissonBracket-a}}{=}
\sum_{(fgh)}\{f,\{g,h\}\}\,.
\end{split}
\end{equation} 
Hence we see that the Jacobi identity \eqref{eq:PoissonIsLieAlg-c}
follows from $d\omega=0$.

Equations \eqref{eq:PoissonIsLieAlg} state that $C^\infty (T^*Q)$ 
is a Lie algebra 
\index{Lie!algebra}\index{algebra!Lie}
with $\{\cdot\,,\,\cdot\}$ as Lie product. With 
respect to the ordinary pointwise product of functions 
$C^\infty (T^*Q)$ is already a commutative and associative 
algebra. Both structures are linked via 
\begin{equation}
\label{eq:LinkAssLie}
\{f\cdot g,h\}=f\cdot\{g,h\}+\{f,h\}\cdot g\,,
\end{equation}
which immediately follows from the Leibniz rule and 
\eqref{eq:DefPoissonBracket} (which implies that 
$\{f\cdot g,h\}$ is the derivative of $f\cdot g$ 
along $X_h$). All this is expressed by saying that  
$C^\infty (T^*Q)$ is a 
\emph{Poisson algebra},
\index{Poisson!algebra (general)}
\index{algebra!Poisson}
meaning the coexistence of two algebraic structures, 
one of a commutative and associative algebra, one of 
a Lie algebra,
\index{Lie!algebra}\index{algebra!Lie}
and their compatibility via \eqref{eq:LinkAssLie}. 
Note that the liner space $\ST(T^*Q)$ of vector fields is also 
a Lie algebra, 
\index{Lie!algebra}\index{algebra!Lie}
whose Lie product is the commutator of vector fields. 
Equation \eqref{eq:HamVF-AntiLieHoemo} then shows that the map 
from $C^\infty (T^*Q)$ to $\ST(T^*Q)$ that sends $f\mapsto X_f$ is 
an anti homomorphism of Lie algebras. 
\index{Lie!homomorphism}

After this brief digression we now return to the geometric 
interpretation of first-class constraints. For any $p\in\Constraint^*$ 
we define 
\begin{equation}
\label{eq:DefKernelSyplForm}
\begin{split}
&T_p^\perp(T^*Q):=\\
&\bigl\{X\in T_p(T^*Q) : 
\omega(X,Y)=0\,,\forall Y\in T_p\Constraint^*\bigr\}\,.
\end{split}
\end{equation}

The non-degeneracy of $\omega$ implies that the dimension 
of $T_p^\perp(T^*Q)$ equals $s$, the co-dimension of $\Constraint^*$
in $T^*Q$. But note that as $\omega$ is skew, $T_p^\perp(T^*Q)$
might well have a non-trivial intersection with $T_p\Constraint^*$.   
This gives rise to the following characterizations for the 
submanifold $\Constraint^*\subset T^*Q$ (understood to
hold at each point $p\in\Constraint^*$): $\Constraint^*$ is 
called
\index{isotropic submanifold}
\index{submanifold!isotropic}
\index{co-isotropic submanifold}
\index{submanifold!co-isotropic}
\index{Lagrangian submanifold}
\index{submanifold!Lagrangian}
\begin{itemize}
\item
\emph{isotropic}
\index{isotropic submanifold} 
iff $T_p\Constraint^*\subset T_p^\perp(T^*Q)$;
\item
\emph{co-isotropic}
\index{isotropic submanifold} 
iff $T_p\Constraint^*\supset T_p^\perp(T^*Q)$;
\item
\emph{Lagrangian}
\index{Lagrangian submanifold} 
iff $T_p\Constraint^*=T_p^\perp(T^*Q)$.
\end{itemize}

Since $\{\phi_\alpha,\phi_\beta\}=d\phi_{\alpha}(X_{\phi_{\beta}})$
we see that condition \eqref{eq:FirstClassPissonBracket-1} is 
equivalent to the statement that the Hamiltonian vector-fields 
for the constraint functions $\phi_\alpha$ are tangent to the 
constraint hypersurface:
\begin{equation}
\label{eq:HamVFTangentCStar}
X_{\phi_{\alpha}}\vert_{\Constraint^*}\in\Gamma T\Constraint^*\,.
\end{equation}
Our assumption that the $s$ differentials $d\phi_\alpha$ be linearly 
independent at each $p\in\Constraint^*$ now implies that the 
$s$ vectors $X_{\phi_\alpha}(p)$ span an $s$-dimensional subspace 
of $T_p\Constraint^*$. But they are also elements of
$T_p^\perp(T^*Q)$ since $\omega(X_{\phi_\alpha},Y)=d\phi_\alpha(Y)=0$
for all $Y$ tangent to $\Constraint^*$. As the dimension of 
$T_p^\perp(T^*Q)$ is $s$, this shows 
\begin{equation}
\label{eq:IntersectionConstOrth} 
T_p^\perp(T^*Q)=
\mathrm{span}\bigl\{X_{\phi_1}\,\cdots,X_{\phi_s}\bigr\}
\subset T_p\Constraint^*\,,
\end{equation}
that is, co-isotropy of $\Constraint^*$. 
First-class constraints are precisely those which give rise to 
co-isotropic constraint surfaces. 

The significance of this lies in the following result, 
which we state in an entirely intrinsic geometric fashion. 
Let $\Constraint^*\subset T^*Q$ be co-isotropic of co-dimension 
$s$ and  let $e:\Constraint^*\rightarrow T^*Q$ be its embedding. 
We write 
\begin{equation}
\label{eq:PullBackSymForm}
\hat\omega:=e^*\omega
\end{equation}
for the pull back of $\omega$ to the constraint surface 
(i.e. essentially the  restriction of $\omega$ to the 
tangent bundle of the constraint surface). $\hat\omega$
is now $s$-fold degenerate, its kernel at $p\in\Constraint^*$
being just $T_p^\perp(T^*Q)\subset T_p\Constraint^*$. We have the 
smooth assignment of subspaces    
\begin{equation}
\label{eq:CoIsotropicDistribution}
\Constraint^*\ni p\mapsto \text{kernel}_p(\hat\omega)=T_p^\perp(T^*Q)\,,
\end{equation}
which forms a sub-bundle of $T\Constraint^*$ called the 
\emph{kernel distribution} of $\hat\omega$ . Now, the crucial result 
is that this sub-bundle is \emph{integrable}, i.e tangent to locally 
embedded submanifolds $\gamma^*\subset\Constraint^*$ of co-dimension 
$s$ in $\Constraint^*$, or co-dimension $2s$ in $T^*Q$. Indeed, 
in order to show this we only need to show that whenever two 
vector fields $X$ and $Y$ on $\Constraint^*$ take values in the 
kernel distribution their commutator $[X,Y]$ also takes values in 
the kernel distribution. That this suffices for local 
integrability is known as \emph{Frobenius' theorem}
\index{Frobenius theorem}
\index{theorem!of Frobenius} 
in differential geometry. Writing  
\begin{equation}
\label{eq:KernelIntegrability-1}
i_{[X,Y]}\hat\omega=
L_X\bigl(i_Y\hat\omega\bigr)
-i_Y\bigl(L_X\hat\omega\bigr)
\end{equation}
we infer that the first term on the right-hand side vanishes 
because $Y$ is in $\hat\omega$'s kernel and $L_X\hat\omega$ 
vanishes because $L_X=d\circ i_X+i_X\circ d$  on forms, where  
$i_X\hat\omega =0$ again due to $X$ being in the kernel
and $d\hat\omega=de^*\omega=e^*d\omega=0$ due to $\omega$
being closed. 

The program of \emph{symplectic reduction}
\index{symplectic!reduction}
\index{reduction!symplectic} 
is now to form the $(2n-2s)$-dimensional quotient space 
$\Constraint^*/\!\!\sim$, where $\sim$ is the equivalence relation 
whose equivalence classes are the maximal integral submanifolds 
of the kernel distribution of $\hat\omega$. $\Constraint^*/\!\!\sim$ 
is called the \emph{physical phase-space}
\index{physical!phase space}
or \emph{reduced space of states}.
\index{reduced space!of states}%

We stress that this geometric formulation of the reduction program, 
\index{reduction!program (geometric)}and the characterization of 
$\Constraint^*/\!\!\sim$ in
particular, does not refer to any set of functions $\phi_\alpha$ 
that one might use in order to characterize $\Constraint^*$. 
If one uses such functions, it is understood that they obey 
the above-mentioned regularity conditions of being at least 
continuously differentiable in a neighborhood of $\Constraint^*$ 
and giving rise to a set of $s$ linearly independent differentials $d\phi_\alpha$ at any point of $\Constraint^*$. Hence redefinitions 
of constraint functions like $\phi\mapsto\sqrt{\vert\phi\vert}$ or $\phi\mapsto\phi^2$, albeit leading to the same surface $\Constraint^*$, 
are a priori not allowed. 

The reduced phase space can be identified with the set of 
physical states.
\index{physical!states} Smooth functions on this space then 
correspond to physical observables.
\index{physical!observables} 
But how can we characterize the latter without explicitly 
constructing $\Constraint^*/\!\!\sim$? This we shall explain in
the remaining part of this section.
 
We define the \emph{gauge algebra} 
\index{gauge!algebra}\index{algebra!gauge} 
as the set of smooth functions on unreduced phase space that 
vanish on the constraint surface:
\begin{equation}
\label{eq:DefGaugeAlgebra}
\Gau:=\bigl\{f\in C^\infty(T^*Q) :  f\vert_{\Constraint^*}=0\bigr\}\,.
\end{equation}
This set is clearly an associative ideal 
\index{ideal!associative}
with respect to the 
pointwise product. But it is also a Lie sub-algebra with respect 
to the Poisson bracket. To see this, we first remark that 
$X_f\vert_{\Constraint^*}$ is $T^\perp(T^*Q)$-valued if $f\in\Gau$. 
Indeed, $f\in\Gau$ implies that $\text{kernel}(df_p)$ includes 
$T_p\Constraint^*$ for $p\in\Constraint^*$. But then (\ref{eq:DefHamVF})
shows $X_f(p)\in T_p^\perp(T^*Q)$. Now, this immediately implies 
that $\Gau$ is a Lie algebra 
\index{Lie!algebra} \index{algebra!Lie}
for co-isotropic $\Constraint^*$,
for then (\ref{eq:DefPoissonBracket}) implies $\{f,g\}\vert_{\Constraint^*}=\omega(X_f,X_g)\vert_{\Constraint^*}=0$.
Hence $\Gau$ is a Poisson sub-algebra of $C^\infty(T^*Q)$ and also 
an associative ideal 
\index{ideal!associative}
with respect to pointwise multiplication. 
However, it is \emph{not} a Lie-ideal 
\index{Lie!ideal}
\index{ideal!Lie}
with respect to 
$\{\cdot,\cdot\}$. Indeed, if $f\in\Gau$ we just need to take 
a $g\in C^\infty(T^*Q)$ which is not constant along the flow of $X_f\vert_{\Constraint^*}$; then $\{f,g\}\vert_{\Constraint^*}\ne 0$.    
This means that we cannot define physical observables 
\index{physical!observables}
by the quotient $C^\infty(T^*Q)/\Gau$, since this will not result 
in a Poisson algebra.
\index{Poisson!algebra (general)}
As we insist that all elements in $\Gau$ generate gauge 
transformations, we have no choice but to reduce the size of    
$C^\infty(T^*Q)$ in order to render the quotient a Poisson algebra. 
Economically the most effective possibility is to take the 
Lie-idealizer
\index{Lie!idealizer}%
\index{idealizer, Lie}%
of $\Gau$ in  $C^\infty(T^*Q)$, which is defined as follows
\begin{equation}
\label{eq:DefLieIdealizer}
I_{\Gau}:=\{f\in C^{\infty}(T^*Q)
: \{f,g\}\big\vert_{\Constraint^*}=0\ \forall g\in\Gau\}\,.
\end{equation} 
Note that $I_{\Gau}$ is the set of smooth functions that,
to use a terminology introduced by Dirac\cite{Dirac:LQM}, 
\emph{weakly} (Poisson) commute with the constraints (i.e. 
with $\Gau$). Here \emph{weak (Poisson) commutativity} 
\index{commutativity (Poisson)!weak}
means that the Poisson brackets of observables and constraints 
need not vanish globally, i.e. on $T^*Q$, but only after 
restriction to $\Constraint^*$. We will briefly come back to 
the  case of  \emph{strong (Poisson) commutativity} 
\index{commutativity (Poisson)!strong}
below. 

Now, if $I_{\Gau}/\Gau$ is to make sense as Poisson algebra 
of physical observables $I_{\Gau}$ must be a Poisson algebra 
containing $\Gau$ as an Poisson ideal. 
\index{ideal!Poisson}
\index{Poisson!ideal}
That $I_{\Gau}$ is an associative 
algebra under pointwise multiplication immediately follows from 
\eqref{eq:LinkAssLie}. That it is also a Lie subalgebra follows 
from the Jacobi identity 
\index{Jacobi!identity}
\eqref{eq:PoissonIsLieAlg-c}. Indeed, 
let $f,g\in I_{\Gau}$ and $h\in\Gau$; then
\eqref{eq:PoissonIsLieAlg-c} immediately gives
\begin{equation}
\label{eq:LieIdealizerSubLie}
\begin{split}
&\{\{f,g\},h\}\big\vert_{\Constraint^*}\\
=-&\{\underbrace{\{g,h\}}_{\in\Gau},f\}\big\vert_{\Constraint^*}
  -\{\underbrace{\{h,f\}}_{\in\Gau},g\}\big\vert_{\Constraint^*}=0\,.
\end{split}
\end{equation}

Hence we have shown that $I_{\Gau}$ is a Poisson subalgebra of 
$C^\infty(T^*M)$ which contains $\Gau$ as a Poisson ideal.
\index{ideal!Poisson}
\index{Poisson!ideal}
By construction $I_{\Gau}$ is the largest Poisson subalgebra of 
$C^\infty(T^*M)$ with that property. Hence we may identify the 
Poisson algebra of physical observables, or 
\emph{reduced space of observables}
\index{reduced space!of observables} 
\index{Poisson!algebra (of physical observables)}
with the quotient
\begin{equation}
\label{eq:PoissonAlgPhysObs}
\mathcal{O}_{\rm phys}:=I_{\Gau}/\Gau\,.
\end{equation}
This complements the definition of the reduced space of 
physical states. Again we stress that the definition given 
here does not refer to any set of functions $\phi_\alpha$ 
that one might use in order to characterize $\Constraint^*$.

We also stress that instead of the Lie idealizer
\index{Lie!idealizer}
\index{idealizer, Lie} 
\eqref{eq:DefLieIdealizer} we could \emph{not} have taken the 
Lie centralizer 
\index{Lie!centralizer}
\begin{equation}
\label{eq:DefLieCentralizer}
C_{\Gau}:=\{f\in C^{\infty}(T^*Q)
: \{f,g\}=0\ \forall g\in\Gau\}\,.
\end{equation} 
Note that here the only difference to \eqref{eq:DefLieIdealizer}
is that  $\{f,g\}$ is required to vanish strongly (i.e. 
on all of $T^*Q$) and not only weakly (i.e. merely on $\Constraint^*$). 
This makes a big difference and the quotient $I_{\Gau}/\Gau$ 
will now generally be far too small. In fact, it is intuitively 
clear and also easy to prove (see, e.g., Lemma\,5 of 
\cite{Giulini:2003}) that the Hamiltonian vector fields $X_f$ 
corresponding to functions $f\in\Gau$, which span $T_p^\perp(T^*\Constraint)\subset T_p\Constraint^*$ at each  
$p\in\Constraint$, span all of $T_p(T^*Q)$ for each point $p$ 
off $\Constraint^*$. This implies that smooth functions in 
$C^\infty(T^*Q)$ which strongly (Poisson) commute with $\Gau$
are locally constant outside $\Constraint^*$. Sometimes strong 
(Poisson) commutativity is required not with respect to $\Gau$ 
but with respect to a complete set $\{\phi_1,\cdots,\phi_s\}$
of functions in $\Gau$ defining $\Constraint^*$; for example 
the component functions of the momentum map 
\index{momentum!map}(see next subsection).  But even then 
strong commutativity is too strong, as a smooth function commuting
with all $\phi_\alpha$ on $\Constraint^*$ need generally not 
extend to a smooth function defined in a neighborhood of 
$\Constraint^*$ in $T^*Q$ which still commutes with all 
$\phi_\alpha$. The reason is that the leaves of the foliation 
\index{foliation!leaves of}
defined by the $\phi_\alpha$ may become `wild' off 
$\Constraint^*$. Compare, e.g., the discussion in 
\cite{Bordemann-etal:2000} (including the example on 
p.\,116).

\subsection{First-class constraints from zero momentum-maps}
First class constraints often arise from group actions (see Appendix
for group actions). This is also true in GR, at least partially. 
So let us explain this in more detail. Let a Lie group $G$ act 
\index{Lie!group}
on the left on $T^*Q$. This means that there is a map 
$G\times T^*Q\rightarrow T^*Q$, here denoted simply by 
$(g,p)\mapsto g\cdot p$, so that $g_1\cdot(g_2\cdot p)=(g_1g_2)\cdot p$
and $e\cdot p=p$ if $e\in G$ is the neutral element. 
As already seen earlier in \eqref{eq:LieAntiHomo} 
and explained in detail in the Appendix, there is then an 
anti-homomorphism 
\index{Lie!anti-homomorphism}
from $\Lie(G)$, the Lie algebra of $G$, 
\index{Lie!algebra}\index{algebra!Lie}
to the Lie algebra of vector fields on $T^*Q$. Recall that the vector 
field $V^X$ corresponding to $X\in\Lie(G)$, evaluated at 
point $p\in T^*Q$, is given by    
\begin{equation}
\label{eq:LeftActionVectorField}
V^X(p):=\frac{d}{dt}\Big\vert_{t=0}\exp(tX)\cdot p\,.
\end{equation}
Then  
\begin{equation}
\label{eq:LeftActionCommutator}
\bigl[V^X,V^Y\bigr]=-V^{[X,Y]}\,.
\end{equation}
Let us further suppose that the group action on $T^*Q$ is of a special 
type, namely it arises from a group action on $Q$ by a canonical 
lift. (Every diffeomorphism $f$ of $Q$ can be lifted to a 
diffeomorphism $F$ of $T^*Q$ given by the pull back of the 
inverse $f^{-1}$.) Then it is easy to see from the geometric 
definition \eqref{eq:DefSymplPot-Inv} that the symplectic 
potential $\theta$ is invariant under this group action and 
consequently the group acts by symplectomorphisms 
($\omega$ preserving diffeomorphisms). The infinitesimal version
of this statement is that, for all $X\in\Lie(G)$,
\begin{equation}
\label{eq:LieThetaZero}
L_{V^X}\theta=0\,.
\end{equation}
Since $L_{V^X}=i_{V^X}\circ d+d\circ i_{V^X}$ this is 
equivalent to 
\begin{equation}
\label{eq:LieTetaZero}
i_{V^X}\omega=d\bigl(\theta(V^X)\bigr)
\end{equation}
which says that $V^X$ is the Hamiltonian vector field of the function 
$\theta(V^X)$. We call the map
\begin{equation}
\label{eq:MomentumMap}  
\Lie(G)\ni X\mapsto P(X):=\theta(V^X)\in C^{\infty}(T^*Q)
\end{equation}
the \emph{momentum map}
\index{momentum!map}
 for the action of $G$. It is a linear map from 
$\Lie(G)$ to $C^{\infty}(T^*Q)$ and satisfies 
\begin{equation}
\label{eq:MomentumMap-Homeomorphism}  
\begin{split}
\bigl\{P(X),P(Y)\bigr\}
&=V^Y\bigl(\theta(V^X)\bigr)\\
&=\bigl(L_{V^Y}\theta\bigr)(V^X)+\theta\bigl(L_{V^Y}V^X\bigr)\\
&=\theta\bigl(V^{[X,Y]}\bigr)\\
&=P\bigl([X,Y]\bigr)\,,
\end{split}
\end{equation}
where we used \eqref{eq:LieThetaZero} and \eqref{eq:LeftActionCommutator}
for the third equality. Hence we see that the map \eqref{eq:MomentumMap}
is a Lie homomorphism from $\Lie(G)$ into the Lie algebra of smooth, 
real-valued functions on $T^*Q$ (whose Lie product is the Poisson bracket).
\index{Lie!algebra}\index{algebra!Lie}

Now, first class constraints are often given by the condition of 
\emph{zero momentum mappings}, i.e., by $P(X)=0$ for all $X\in\Lie(G)$.
By linearity in $X$, this is equivalent to the set of $s:=\dim(G)$
conditions     
\begin{equation}
\label{eq:ZeroMomentumMap-1}
\phi_\alpha:=P(e_\alpha)=0\,,
\end{equation}
where $e_\alpha=\{e_1,\cdots,e_s\}$ is a basis of $\Lie(G)$.
Let the structure constants for this basis be $C^\gamma_{\alpha\beta}$, 
i.e. $[e_\alpha,e_\beta]=C^\gamma_{\alpha\beta}e_\gamma$, then 
\eqref{eq:MomentumMap-Homeomorphism} becomes
\begin{equation}
\label{eq:ZeroMomentumMap-2}
\{\phi_\alpha,\phi_\beta\}=C^\gamma_{\alpha\beta}\phi_\gamma\,.
\end{equation}

Constraints in gauge theories will typically arise as zero momentum 
maps in the fashion described here, the only necessary generalization 
being the extension to infinite-dimensional groups and Lie algebras.
\index{Lie!algebra}\index{algebra!Lie}
In fact, for gauge theories our $G$ will correspond to the 
infinite-dimensional \emph{group of gauge transformations},
\index{group!of gauge transformations}%
which is not to be confused with the finite-dimensional gauge group.
\index{gauge!group}%
\index{group!gauge}%
The former consists of functions, or sections in bundles, with 
values in the latter. 
On the other hand, the constraints in GR will only partially 
be of this type. More precisely, those constraints arising from 
3-dimensional diffeomorphisms (called the vector or diffeomorphism 
constraints) will be of this type, those from non-tangential 
hypersurface deformations (scalar or Hamiltonian constraint)
will not fit into this picture. For the former $G$ will
correspond to $\Diff(\Sigma)$, or some appropriate subgroup 
thereof, and $\Lie\bigl(\Diff(\Sigma)\bigr)$ to the 
infinite-dimensional Lie algebra of vector fields on $\Sigma$
\index{Lie!algebra}\index{algebra!Lie}
(possibly with special support and/or fall-off conditions). 
The different nature of the latter constraint will be signaled 
by structure \emph{functions} $C^\gamma_{\alpha\beta}(q,p)$ 
appearing on the right-hand side rather than constants. 
\index{structure!functions}
\index{structure!constants}  
This has recently given rise to attempts to generalize the 
group-theoretic setting described above to that of \emph{groupoids}
and \emph{Lie algebroids}, 
\index{groupoids}
\index{Lie!algebroids} 
in which the more general structure of GR can be 
accommodated~\cite{Blohmann.etal:2013}.

\section{Hamiltonian GR}
\label{sec:HamiltonianGR}
The Hamiltonian formulation of GR proceed along the lines outlined
in the previous section. For this we write down the action in a 
(3+1)-split form, read off the Lagrangian density, define the 
conjugate momenta as derivatives of the latter with respect to the 
velocities, and finally express the energy function 
\eqref{eq:EnergyFunction} in terms of momenta. The constraint 
functions will not be determined on the Lagrangian level, but 
rather directly on the Hamiltonian level as primary and secondary 
constraints (there will be no tertiary ones), the primary ones 
being just the vanishing of the momenta for lapse and shift.

The Lagrangian density for GR is essentially just the scalar
curvature of spacetime. However, upon variation of this quantity, 
which contains second derivatives in the metric, we will pick up
boundary terms from partial integrations which need not vanish by 
just keeping the metric on the boundary fixed. Hence we will 
need to subtract these boundary terms which will otherwise obstruct 
functional differentiability. Note that this is not just a matter 
of aesthetics: Solutions to differential equations (like Einstein's
equation) will not be stationary points of the action if the latter 
is not differentiable at these points. Typically, Euler Lagrange 
equations will allow for solutions outside the domain of 
differentiability of the action they are derived from. Including 
some such solutions will generally need to adapt the action by 
boundary terms. This clearly matters if one is interested in the 
values of the action, energies, etc. for these solutions and 
also, of course, in the path-integral formulations of the 
corresponding quantum theories. 

The Einstein-Hilbert action of GR is
\index{principle!of least action (in GR)}
\index{Einstein-Hilbert action} 
\begin{equation}
\label{eq:ActionGR}
S_{\rm GR}[\Omega,g]=-\frac{\signature}{2\kappa}\int_\Omega\Scalar\,
d\mu_g+\text{boundary terms}
\end{equation}
where in local coordinates $x^\mu=(x^0=ct,x^1,x^2,x^3)$,
\begin{equation}
\label{eq:ActionGR-Measure}
d\mu_g=\sqrt{\signature\det\{g_{\mu\nu}\}}\,cdt\wedge dx^1\wedge dx^2\wedge dx^3\,.
\end{equation}
The sign convention behind the prefactor $-\signature$ in 
\eqref{eq:ActionGR} is such that in the Lorentzian as well as 
the Riemannian case the Lagrangian density contains the bilinear 
De\,Witt inner product of the extrinsic curvatures (compare 
\eqref{eq:RicciScalarDecomp}) with a positive sign, i.e. 
transverse traceless modes have positive kinetic energy. 

The boundary term can be read off \eqref{eq:RicciScalarDecomp}
and \eqref{eq:RicciVectorField}. If the integration domain 
$\Omega\subset M$ is such that the spacelike boundaries are 
contained in two hypersurfaces $\Sigma_s$, i.e. two 
$t=\text{const.}$ surfaces, say 
$\partial\Omega_i:=\partial\Omega\cap\Sigma_{\rm initial}$ and 
$\partial\Omega_f:=\partial\Omega\cap\Sigma_{\rm final}$, we 
would have to add the two boundary terms (dependence on 
$\signature$ drops out)
\begin{equation}
\label{eq:SpacelikeBoundaryTerms}
\kappa^{-1}
 \int_{\partial\Omega_f}\Trace_h(K)\,d\mu_h
-\kappa^{-1}
\int_{\partial\Omega_i}\Trace_h(K)\,d\mu_h\,.
\end{equation}
Here we used that the second term in \eqref{eq:RicciVectorField}
does not contribute due to $a$ being orthogonal to $n$. 
$d\mu_h$ is the standard measure from the induced metric, $h$, 
on the hypersurfaces. If the cylindrical timelike boundary 
$\partial\Omega_{\rm cyl}$ is chosen such that its spacelike normal 
$m$ is orthogonal to $n$, only the second term in 
\eqref{eq:RicciVectorField} contributes and we get one 
more boundary term (again $\signature$ drops out)  
\begin{equation}
\label{eq:SpacelikeBoundaryTerm}
\kappa^{-1}\int_{\partial\Omega_{\rm cyl}}\hat K(n,n)\,d\mu_h\,.
\end{equation}
Here $\hat K$ is the extrinsic curvature of $\partial\Omega_{\rm cyl}$
in $M$, which we picked up because 
$g(a,m)=g(\nabla_nn,m)=-g(n,\nabla_nm)=\hat K(n,n)$.

Once the boundary terms are taken care of we can just read off 
the Lagrangian density from \eqref{eq:RicciScalarDecomp} also using 
\eqref{eq:VolumeForm},
\begin{equation}
\label{eq:LagrangianDensityGR}
\LDensity_{\rm GR}
=(2\kappa)^{-1}\bigl[
G(K,K)-\signature R\bigr]\,\alpha\sqrt{h}\,,
\end{equation}
where we now used the standard abbreviations 
\begin{equation}
\label{eq:Abbreviations}
\begin{split}
G(K,K):&=G^{abcd}K_{ab}K_{cd}\,,\\
R:&=\Scalar^D\,,\\
\sqrt{h}:&=\sqrt{\det\{h_{ab}\}}\,.
\end{split}
\end{equation}
Moreover, $K_{ab}$ has here to be understood as expressed in 
terms of the time and Lie derivatives of $h_{ab}$:
\begin{equation}
\label{eq:K-InTermsOfLieH}
K=-\tfrac{\signature}{2}\alpha^{-1}\bigl(\dot h-L_\beta h)\,.
\end{equation}
We keep in mind that an overdot denotes differentiation with respect 
to $ct$ (not $t$). In passing we also note that $\LDensity_{\rm GR}$ 
has the right physical dimension of an energy-density ($\alpha$ is 
dimensionless).

The Hamiltonian density is now obtained by the usual Legendre transform
with respect to all configuration variables that are varied in the 
action. These comprise all components $g_{\mu\nu}$ and hence in the 
(3+1)-split parametrization all $h_{ab}$ as well as the lapse $\alpha$ 
and the three shift components $\beta^a$. However, it is immediate 
that \eqref{eq:LagrangianDensityGR} does not contain any time 
derivatives of the latter; hence their conjugate momenta vanish:
\begin{subequations}
\label{eq:LapseShiftMomenta}
\begin{alignat}{2}
\label{eq:LapseShiftMomenta-a}
&\pi_\alpha&&:=\frac{1}{c}\frac{\partial\LDensity_{\rm GR}}{\partial\dot\alpha}=0\,,\\
\label{eq:LapseShiftMomenta-b}
&\pi_{\beta^a}&&:=
\frac{1}{c}\frac{\partial\LDensity_{\rm GR}}{\partial\dot{\beta}^a}=0\,.
\end{alignat}
\end{subequations}
This leaves us with the momenta for the metric components 
$h_{ab}$ 
\begin{equation}
\label{eq:H-Momenta}
\begin{split}
\pi^{ab}:
&=\frac{1}{c}\frac{\partial\LDensity_{\rm GR}}{\partial{\dot h}_{ab}}\\
&=\frac{(-\signature)\sqrt{h}}{2\kappa c}G^{abcd}K_{cd}\\
&=\frac{(-\signature)}{2\kappa c}{\hat G}^{abcd}K_{cd}\,.
\end{split}
\end{equation}
Here again $K$ stands for the expression \eqref{eq:K-InTermsOfLieH}.
We also made use of the conformally rescaled De\,Witt metric 
\eqref{eq:ConfDeWittMetric-1} whose significance appears here for 
the first time. Again in passing we note that the physical 
dimension of $\pi^{ab}$ is right, namely that of momentum per 
area (the dimension of $K$ is an inverse length).

In order to compute the Hamiltonian density we express $\dot h$
in terms of the momenta  
\begin{equation}
\label{eq:VelocityIn Momenta}
\begin{split}
{\dot h}_{ab}
&=(L_\beta h)_{ab}-2\signature\alpha\, K_{ab}\\
&=(D_a\beta_b+D_b\beta_a)+4\alpha\kappa c\,{\hat G}^{-1}_{abcd}\pi^{cd}
\end{split}
\end{equation}
and obtain   
\begin{equation}
\label{eq:GR-Hamiltonian-1}
\begin{split}
\HDensity_0[h,\pi]
&=\pi^{ab}\,c{\dot h}_{ab}-\LDensity_{\rm GR}\\
&=\alpha\Bigl[(2\kappa c^2){\hat G}^{-1}_{abcd}\pi^{ab}\pi^{cd}\Bigl.\\
&\quad+\Bigl.\epsilon(2\kappa)^{-1}\sqrt{h}R\Bigr]\\
&\quad+2c\pi^{ab}D_a\beta_b\,.
\end{split}
\end{equation}
The Hamiltonian, $H_0$, is just the integral of this density over 
$\Sigma$. The subscript $0$ is to indicate that this Hamiltonian 
is still to be modified by constraints according to the general 
scheme.  Also, we have to once more care about surface terms in order to 
ensure functional differentiability  without which the Hamiltonian 
flow does not exist~\cite{ReggeTeitelboim:1974}.  

The first thing to note is that we have found the primary constraints 
\eqref{eq:LapseShiftMomenta}. For them to be maintained under the 
evolution we need to impose 
\begin{subequations}
\label{eq:SecondaryConstraints1}
\begin{alignat}{4}
\label{eq:SecondaryConstraints1-a}
&c{\dot\pi}_\alpha
&&=\{\pi_{\alpha},H_0\}
&&=-\frac{\delta H_o}{\delta\alpha}
&&=0\,,\\
\label{eq:SecondaryConstraints1-b}
&c{\dot\pi}_{\beta^a}
&&=\{\pi_{\beta^a},H_0\}&&=-\frac{\delta H_o}{\delta\beta^a}&&=0\,,
\end{alignat}
\end{subequations}
giving rise to the secondary constraints
\begin{subequations}
\label{eq:SecondaryConstraints2}
\begin{alignat}{1}
\label{eq:SecondaryConstraints2-a}
&(2\kappa c^2){\hat G}^{-1}(\pi,\pi)+\signature(2\kappa)^{-1}\sqrt{h}R=0\,,\\
\label{eq:SecondaryConstraints2-b}
&-2D_a\pi^{ab}=0\,,
\end{alignat}
\end{subequations}
respectively. It may be checked directly that these equations respectively 
are equivalent to \eqref{eq:Constraints-1} for $\EMT=0$. If we had 
included a cosmological constant this would have led to the replacement 
of $R$ in \eqref{eq:SecondaryConstraints2-a} with $(R-2\Lambda)$.

In passing we make the following geometric observation regarding the 
bilinear form $G^{-1}(\pi,\pi)$, which is analogous to that 
made for $G(K,K)$ below equation \eqref{eq:KineticTerm-Weingarten}.
We can choose a local frame $\{e_a\}$ so that 
$h_{ab}=h(e_a,e_b)=\delta_{ab}$ and $\pi^{ab}=\pi(\theta^a,\theta^b)=
\text{diag}(p_1,p_2,p_3)$, where $\{\theta^a\}$ is dual to  
$\{e_a\}$. Then 
\begin{equation}
\label{eq:KineticTerm-MomentumEigenvalues}
{\hat G}^{-1}(\pi,\pi):=\bigl(\delta^{ab}-\tfrac{3}{2}n^an^b\bigr)p_ap_b\,,
\end{equation}
where, as before, $n^a$ are the components of the normalized vector 
$\vec n:=(1,1,1)/\sqrt{3}$ in eigenvalue-space, which we identify 
with $\reals^3$ endowed with the standard Euclidean inner product. 
Denoting again by $\theta$ the angle between $\vec n$ and 
$\vec p:=(p_1,p_2,p_3)$, we have
\begin{equation}
\label{eq:KineticTerm-Sign2}
{\hat G}^{-1}(\pi,\pi)=
\begin{cases}
0 &\quad\text{if}\quad\vert\cos\theta\vert=\sqrt{2/3}\\
>0&\quad\text{if}\quad\vert\cos\theta\vert<\sqrt{2/3}\\
<0&\quad\text{if}\quad\vert\cos\theta\vert>\sqrt{2/3}\,.
\end{cases}
\end{equation}
This should be compared with \eqref{eq:KineticTerm-Sign1}.
The difference is that $\sqrt{1/3}$ is replaced by  $\sqrt{2/3}$, 
which is due to $\lambda=1$ in \eqref{eq:DeWittMetric-1b}
but $\mu=1/2$ in \eqref{eq:DeWittMetric-2b} in the GR case. 
This has an interesting consequence: The 
condition $\vert\cos\theta\vert=\sqrt{2/3}$ now describes 
a double cone around the symmetry axis generated by $\vec n$ 
and vertex at the origin, whose opening angle is strictly 
smaller than that of the cone considered in  
\eqref{eq:KineticTerm-Sign1}. In fact, it is small enough to just 
touch boundaries of the positive and negative octants in 
$\reals^3$. This means that $\vert\cos\theta\vert>\sqrt{2/3}$ 
implies that either all $p_a$ are positive or all  $p_a$ are
negative. In other words ${\hat G}^{-1}(\pi,\pi)<0$ implies that 
the symmetric bilinear form $\pi$ is either positive or 
negative definite. In contrast, \eqref{eq:KineticTerm-Sign1}
did not allow to conclude definiteness of the symmetric bilinear 
form $K$ from  $G(K,K)<0$, since the interiors of the double-cone 
$\vert\cos\theta\vert>\sqrt{1/3}$ intersect the complements 
of the definite octants.

Let us now return to the constraints. We have found the primary and 
secondary constraints \eqref{eq:LapseShiftMomenta} and \eqref{eq:SecondaryConstraints2} respectively. The most important 
thing to note next is that there will be no further (tertiary, etc.) constraints. Indeed, this follows from the general argument following~\eqref{eq:Bianchi2Contracted-Coord}, which ensures the 
preservation of the secondary constraints under Hamiltonian evolution. 
The primary constraints are now taken care of by simply eliminating 
the canonical pairs $(\alpha,\pi_{\alpha})$ and $(\beta^a,\pi_{\beta^a})$
from the list of canonical variables. As we will see shortly, 
the secondary constraints \eqref{eq:SecondaryConstraints2} are of 
first-class, so that, according to the general theory outlined 
above, they should be added with arbitrary coefficients to the 
initial Hamiltonian $H_0$ to get the general Hamiltonian. This leads to 
\begin{equation}
\label{eq:GR-Hamiltonian}
H[\alpha,\beta]=C_s(\alpha)+C_v(\beta)+\text{boundary terms}\,,
\end{equation} 
where 
\begin{subequations}
\label{eq:GR-ConstraintsHamForm}
\begin{alignat}{2}
\label{eq:GR-ConstraintsHamForm-a}
C_s(\alpha):&=\int_\Sigma d^3x\ \alpha
&&\Bigl[
(2\kappa c^2){\hat G}^{-1}(\pi,\pi)\Bigr.\nonumber \\
& &&\Bigl.\quad+\ \signature(2\kappa)^{-1}\sqrt{h}R
\Bigr]\,,\\
\label{eq:GR-ConstraintsHamForm-b}
C_v(\beta):&=\int_\Sigma d^3x\ \beta^a&&\Bigl[-2ch_{ab}D_c\pi^{bc}\Bigr]\,,
\end{alignat}
\end{subequations}
where $\alpha$ and $\beta^a$ are now arbitrary coefficients  
corresponding to the $\lambda$'s in \eqref{eq:GeneralHamiltonianFunction}. 
In particular, they may depend on the remaining canonical variables 
$h$ and $\pi$. Note that up to boundary terms the Hamiltonian 
is just a sum of constraints, where $S$ stands for the scalar- 
(or Hamiltonian-) and and $V$ for the vector (or diffeomorphism-) 
constraint. 

The equations of motion generated by $H$ will clearly be 
equivalent to \eqref{eq:h-Dot} and \eqref{eq:K-Dot}. Let us 
write them down explicitly. To do this we first note that 
the functional derivatives of $C_v(\beta)$ with respect to 
$h$ and $\pi$ are easily obtained if we note that, 
modulo surface terms, the integrand can either be written as 
$c (L_\beta h)_{ab}\pi^{ab}$ or as $-c (L_\beta \pi)^{ab}h_{ab}$.
Hence, given that the surface terms are so chosen to guarantee 
functional differentiability, we have 
\begin{equation}
\label{eq:PiDerivVecConstr}
\frac{\delta C_v(\beta)}{\delta \pi^{ab}}
=c\,L_\beta h_{ab}=c\,(D_a\beta_b+D_b\beta_a)
\end{equation}
and 
\begin{equation}
\label{eq:MetricDerivVecConstr}
\begin{split}
&\frac{\delta C_v(\beta)}{\delta h_{ab}}
=-\,c\,(L_\beta\pi)^{ab}\\
=\ & 
-\,c\,\bigl[D_c(\beta^c\pi^{ab})-(D_c\beta^a)\pi^{cb}-(D_c\beta^b)\pi^{ac}\bigr]\,.
\end{split}
\end{equation}
Note that in the first term on the right-hand side of \eqref{eq:MetricDerivVecConstr} the $\beta^c$ appears under the 
differentiation $D_c$ because $\pi$ is a tensor-density of 
weight one. The functional derivative with respect to $\pi$ of 
$C_s(\alpha)$ is simply $4\alpha\kappa c^2\hat G(\pi,\,\cdot\,)$,
so that the equation of motion for $h$ is readily written down:
\begin{equation}
\label{eq:GRHamEqMotion-a}
\begin{split}
c\dot h_{ab}
&=\{h_{ab},H\}=\frac{\delta H}{\delta\pi^{ab}}\\
&=4\alpha\kappa c^2{\hat G}^{-1}_{ab\,cd}\pi^{cd}+cL_\beta h_{ab}\,.
\end{split}
\end{equation}
Using \eqref{eq:H-Momenta} this is immediately seen to be 
just \eqref{eq:h-Dot}. From the explicit $h$-dependence of 
$\hat G$, as displayed in~\eqref{eq:DeWittMetric-2b} and  \eqref{eq:ConfDeWittMetric-1}, we obtain, using 
$\partial\sqrt{h}/\partial h_{ab}=\tfrac{1}{2}\sqrt{h}h^{ab}$,
\begin{equation}
\label{eq:MetricDerivScalConstr-a}
\frac{\partial{\hat G}^{-1}(\pi,\pi)}{\partial h_{ab}}
= -\tfrac{1}{2}h^{ab}{\hat G}^{-1}(\pi,\pi)+2h^{an}\pi^{bm}{\hat G}^{-1}_{nmcd}\pi^{cd}.
\end{equation}
For the variational derivative of the second term in~\eqref{eq:GR-ConstraintsHamForm} we use the following 
standard formula for the variation of $\sqrt{h}$ times 
the scalar curvature of $h$  
\begin{equation}
\label{eq:MetricVarScalCurv}
\begin{split}
&\delta\bigl(\sqrt{h}\,R(h)\bigr)\\
&=\sqrt{h}
\bigl(
-G^{ab}(h)\delta h_{ab}+G^{nmab}D_nD_m\delta h_{ab}
\bigr)\,,
\end{split}
\end{equation}
which immediately follows from \eqref{eq:ScalarDifferenceFirstOrder},
where we recall that $g_{ab}$ there corresponds to $h_{ab}$ here and
also that $\delta h^{ab}=-h^{ac}h^{bd}\delta h_{cd}$.
$G^{ab}(h)$ denote the contravariant components of the Einstein 
tensor for $h$. Taken together we get
\begin{equation}
\label{eq:MetricDerivScalConstr}
\begin{split}
&\frac{\delta C_s(\alpha)}{\delta h_{ab}}=\\
\ \kappa c^2\alpha&\Bigl
  [-h^{ab}{\hat G}^{-1}(\pi,\pi)+4h^{an}\pi^{bm}{\hat G}^{-1}_{nmcd}\pi^{cd}\Bigr]\\
-\frac{\signature}{2\kappa}&\Bigl[\alpha G^{ab}(h)-G^{abnm}D_nD_m\alpha\Bigr]
\sqrt{h}
\end{split}
\end{equation}
With \eqref{eq:MetricDerivVecConstr} this gives the second 
Hamilton equation
\begin{equation}
\label{eq:GRHamEqMotion-b}
\begin{split}
c\dot \pi^{ab}
=&\{\pi^{ab},H\}=-\frac{\delta H}{\delta h_{ab}}\\
=&\ \kappa c^2\alpha\Bigl
  [h^{ab}{\hat G}^{-1}(\pi,\pi)-4h^{an}\pi^{bm}{\hat G}^{-1}_{nmcd}\pi^{cd}\Bigr]\\
&+\frac{\signature}{2\kappa}\Bigl[\alpha G^{ab}(h)-G^{abnm}D_nD_m\alpha\Bigr]
\sqrt{h}\\
&+cL_\beta\pi^{ab}\,.
\end{split}
\end{equation}

These are seen to be equivalent to \eqref{eq:h-Dot} and 
\eqref{eq:K-Dot} respectively. Before we discuss the boundary 
terms we  write down the Poisson brackets for the constraints. 
\begin{subequations}
\label{eq:ConstraintAlgebra}
\begin{alignat}{2}
\label{eq:ConstraintAlgebra-a}
&\bigl\{C_v(\beta),C_v(\beta')\bigr\}&&= C_v\bigl([\beta,\beta']\bigr)\,,\\
\label{eq:ConstraintAlgebra-b}
&\bigl\{C_v(\beta),C_s(\alpha)\bigr\}&&=C_s\bigl(\beta(\alpha)\bigr)\,,\\
\label{eq:ConstraintAlgebra-c}
&\bigl\{C_s(\alpha),C_s(\alpha')\bigr\}&&=
\signature C_v\bigl(\alpha (d\alpha')^\sharp-\alpha' (d\alpha)^\sharp\bigr)\,.
\end{alignat}
\end{subequations}
These may be obtained by direct computation, but are also
dictated by geometry. Before discussing the geometry behind 
them, we note the following more or less obvious points:
\begin{enumerate}
\item
The vector constrains form a Lie algebra. 
\index{Lie!algebra}\index{algebra!Lie}
The map $\beta\rightarrow V(\beta)$ is a Lie homomorphism form the 
Lie algebra of vector fields in $\Sigma$ to the Lie algebra
(within the Poisson algebra)
\index{algebra!Poisson}
of phase-space functions. 
In fact, this map is just the momentum map for the action 
of the diffeomorphism group $G=\Diff(\Sigma)$ on phase 
$T^*Q$, which is a lift of the action on $Q=\RiemMet(\Sigma)$,
the space of Riemannian metrics on $\Sigma$. Note that here 
the symplectic potential can be written in a symbolic 
infinite-dimensional notation 
(cf.\,\eqref{eq:SympStructureCoordinates-a}) 
\begin{equation}
\label{eq:GR-SymplPotential}
\theta=\int_\Sigma d^3x\
\pi^{ab}(x)\,\delta h_{ab}(x)
\end{equation}
and the vector field $V_\beta$ generated by the action of  
$G=\Diff(\Sigma)$ on $Q=\RiemMet(\Sigma)$:
\begin{equation}
\label{eq:GR-VectorFieldOnRiem}
V_\beta=\int_\Sigma d^3x\
L_\beta h_{ab}(x)\frac{\delta}{\delta h_{ab(x)}}\,.
\end{equation}
The momentum map \eqref{eq:MomentumMap} is then given by 
\begin{equation}
\label{eq:GR-MomentumMap}
\begin{split}
P_\beta
&=\theta(V_\beta)=\int_\Sigma d^3x\ \pi^{ab}L_\beta h_{ab}\\
&=c^{-1}C_v(\beta)+2\int_{\partial\Sigma}d^2x\beta_a\pi^{ab}\nu_b\,,
\end{split}
\end{equation}
where $\nu^b$ denote the components of the outward pointing 
normal of $\partial\Sigma$. This shows that for vector fields
$\beta$ for which the surface term does not contribute the 
vector constraint is just the momentum map (up to a factor of 
$c^{-1}$, which comes in because the physical dimension of 
the values of the momentum map are that of momentum whereas 
the physical dimension of the constraints are that of an 
Hamiltonian, that is, energy). The surface term will be 
discussed below. What is important here is that the 
vector constraint coincides to the zero momentum-map for 
those diffeomorphisms which are asymptotically trivial, 
i.e. for which the surface term vanishes. Only those 
are to be considered as gauge transformations! Long ranging 
diffeomorphism for which the surface term is non zero, 
i.e. for configurations of non vanishing linear and/or 
angular momentum (cf. Section~\ref{sec:AsympFlatnessCharges}),
have to be considered as proper changes in physical state.
If we required these motions to be pure gauge we would 
eliminate all states with non-zero asymptotic charges.      
Compare the closing remarks of 
Section\,\ref{sec:ConstrainedHamiltonianSystems}.
\item
Once we have understood that the vector constraint is the momentum 
map for diffeomorphisms, its Poisson bracket with any other
phase-space function $F$ that defines a \emph{geometric object} 
\index{geometric object}
on $\Sigma$  (i.e. an object with well defined transformation 
properties under diffeomorphisms) is fixed. We simply have 
\begin{equation}
\label{eq:GR-GeomObject}
\bigl\{F,V_\beta\bigr\}=L_\beta F\,.
\end{equation}
In this sense \eqref{eq:ConstraintAlgebra-b} says no more than 
that the expression \eqref{eq:SecondaryConstraints2-a} is a scalar
density of weight one. Recall that if $F$ is a scalar density 
of weight one then $L_\beta F=D_a(\beta^aF)$. If we multiply 
$F$ by $\alpha$ and integrate over $\Sigma$ we get after 
partial integration and assuming the boundary term to give no
contribution (which for non closed $\Sigma$ requires certain  
fall-off conditions) an integral of $-F\beta(\alpha)$, which is 
just what  \eqref{eq:ConstraintAlgebra-b} expresses. 
Algebraically speaking, the fact that the Poisson bracket 
of a vector and a scalar constraint is proportional to a 
scalar rather than a vector constraint means that the vector 
constraints do \emph{not} form an \emph{ideal}. 
\index{ideal}
\index{vector constraints, no ideal}
Geometrically this means that the Hamiltonian vector fields 
for the scalar constraint, if evaluated on the hypersurface 
for the vector constraint, will generally not be tangential 
to it, except for the points where this hypersurface intersects 
that of the scalar constraint. This has very important 
consequences for algorithms of phase-space reduction, 
\index{reduction!of phase space}
i.e. algorithms that aim to ``solve'' the constraints. 
\index{constraints!solve (methods, meaning)}
It means that a reduction in steps is \emph{not} possible, 
whereby one \emph{first} solves for the vector constraint 
and \emph{then} seeks for solutions of the scalar constraint
within the class of solutions to the vector constraint. 
\item
According to \eqref{eq:ConstraintAlgebra-c} two scalar 
constraints Poisson commute into a vector constraint. Two 
facts are remarkable concerning the vector field that 
forms the argument of this vector constraint: 
First, it depends on the signature of spacetime (overall 
multiplication with $\signature$). Second, it depends on the 
phase space variable $h$ through the $\sharp$-operation of 
``index raising''; explicitly: 
\begin{equation}
\label{eq:ScalarConstH-Dep}
\begin{split}
&\alpha(d\alpha')^\sharp-\alpha'(d\alpha)^\sharp\\
&=h^{ab}\bigl(\alpha\partial_b\alpha'-\alpha'\partial_b\alpha
\bigr)\frac{\partial}{\partial x^a}\,.
\end{split}
\end{equation}
This is the fact already mentioned at the end of 
Section\,~\ref{sec:ConstrainedHamiltonianSystems}, 
that the constraints in GR are not altogether in 
the form of a vanishing momentum map. This fact has 
led to some discussion in the past and attempts 
have been made to consider different algebraic combinations 
of the constraints which define the same constraint 
hypersurfaces but display structure constants rather 
than structure functions in their Poisson brackets; e.g., 
\cite{Markopoulou:1996}. But as already discussed in 
Section\,\ref{sec:ConstrainedHamiltonianSystems} it is 
important that these redefinitions do not spoil the 
regularity properties of the functions that define the 
constraint surface.       
\end{enumerate}
This ends the immediate discussion of \eqref{eq:ConstraintAlgebra}.
But there is another aspect that is related to the last point 
just discussed and that deserves to be mentioned. 

\subsection{Hypersurface deformations and 
their representations}
\label{sec:HypersurfaceDeformations}
Even though the constraints cannot be understood in a straightforward 
fashion as zero momentum map of a group action, they nevertheless 
do furnish a representation of an algebraic object (a groupoid) 
of hypersurface motions. As a result, the relations  
\eqref{eq:ConstraintAlgebra} are \emph{universal},
\index{constraints!algebra, universality of} 
in the sense that \emph{any}  spacetime diffeomorphism invariant 
theory, whatever its field content, will give rise to the very 
same relations \eqref{eq:ConstraintAlgebra}; see 
\cite{Teitelboim:1973} and \cite{Hojman.etal:1976} for early 
and lucid discussions and \cite{Isham.Kuchar:1985a}\cite{Isham.Kuchar:1985b}
for a comprehensive account.

The idea is to regard the space of (spacelike) embeddings 
$\Embeddings(\Sigma,M)$ of $\Sigma$ into $M$ as an infinite-dimensional 
manifold, on which the diffeomorphism group of $M$ acts on the 
left by simple composition. Then there is a standard 
anti-homomorphism 
\index{Lie!anti-homomorphism}
from the Lie algebra of $\Diff(M)$ to the 
Lie algebra of vector fields on $\Emb(\Sigma,M)$, just as in 
\eqref{eq:LeftActionCommutator}. A tangent vector at a particular 
$\Emb\in\Embeddings(\Sigma,M)$ can be visualized as a vector 
field $\xi$ on $\Sigma\subset M$ with normal and tangential 
components, more precisely, as a section in the pull-back 
bundle $\Emb^*TM$ over $ \Sigma$. Its decomposition into 
normal and tangential components depends on $\Emb$. 
If we think of $M$ as being locally coordinatized by functions 
$y^\mu$ and $\Sigma$ by functions $x^a$ then $\Emb$ can be locally 
represented by four functions $y^\mu$ of three variables $x^a$.
A vector field $V_\xi$ can then be represented in a symbolic 
infinite-dimensional notation 
\begin{equation}
\label{eq:HypDefVF-1}
V_\xi=\int_\Sigma d^3x\
\xi^\mu\bigl(y(x)\bigr)\frac{\delta}{\delta y^\mu(x)}\,.
\end{equation}
In full analogy to \eqref{eq:LeftActionCommutator}, this 
immediately leads to 
\begin{equation}
\label{eq:HypDefVF-2}
[V_\xi,V_\eta]=-V_{[\xi,\eta]}\,.
\end{equation}
If we now decompose $\xi$ in an embedding dependent 
fashion into its normal component $\alpha n$ and 
tangential component $\beta$ we can rewrite \eqref{eq:HypDefVF-1}
into  
\begin{equation}
\label{eq:HypDefVF-3}
\begin{split}
&V(\alpha,\beta)=\\
&\int_\Sigma d^3x\
\Bigl(
\alpha(x)n^\mu[y](x)+\beta^a\partial_ay^\mu(x)
\Bigr)
\frac{\delta}{\delta y^\mu(x)}\,,
\end{split}
\end{equation}
where the components $n^\mu$ of the normal $n$ to the image 
$\Emb(\Sigma)\subset M$ have to be considered as functional 
of the embedding. Again we can compute the commutator 
explicitly. The only non-trivial part is the functional 
derivative of the $n^\mu$ with respect to the $y^\nu$. 
How this is done is explained in the Appendix of 
\cite{Teitelboim:1973}. The result is 
\begin{subequations}
\label{eq:HypDefAlgebra}
\begin{equation}
\label{eq:HypDefAlgebra-1}
\bigl[V(\alpha_1,\beta_1),V(\alpha_2,\beta_2)\bigr]
=-V(\alpha,\beta)\,,
\end{equation}
where 
\begin{alignat}{2}
\label{eq:HypDefAlgebra-2}
&\alpha&&=\beta_1(\alpha_2)-\beta_2(\alpha_1)\,,\\
\label{eq:HypDefAlgebra-3}
&\beta &&=[\beta_1,\beta_2]
+\signature\bigl(
\alpha_1(d\alpha_2)^\sharp-\alpha_2(d\alpha_1)^\sharp
\bigr)\,.
\end{alignat} 
\end{subequations}
This is just \eqref{eq:ConstraintAlgebra} up to a relative minus 
sign that has the same origin as that between 
\eqref{eq:LeftActionCommutator} and \eqref{eq:MomentumMap-Homeomorphism}.
We therefore see that \eqref{eq:ConstraintAlgebra} is a representation 
of a general algebraic structure which derives from the geometry of
deformations of (spacelike) hypersurfaces in spacetime.

\begin{figure}[ht]
\begin{center}
\includegraphics[width=0.50\linewidth]{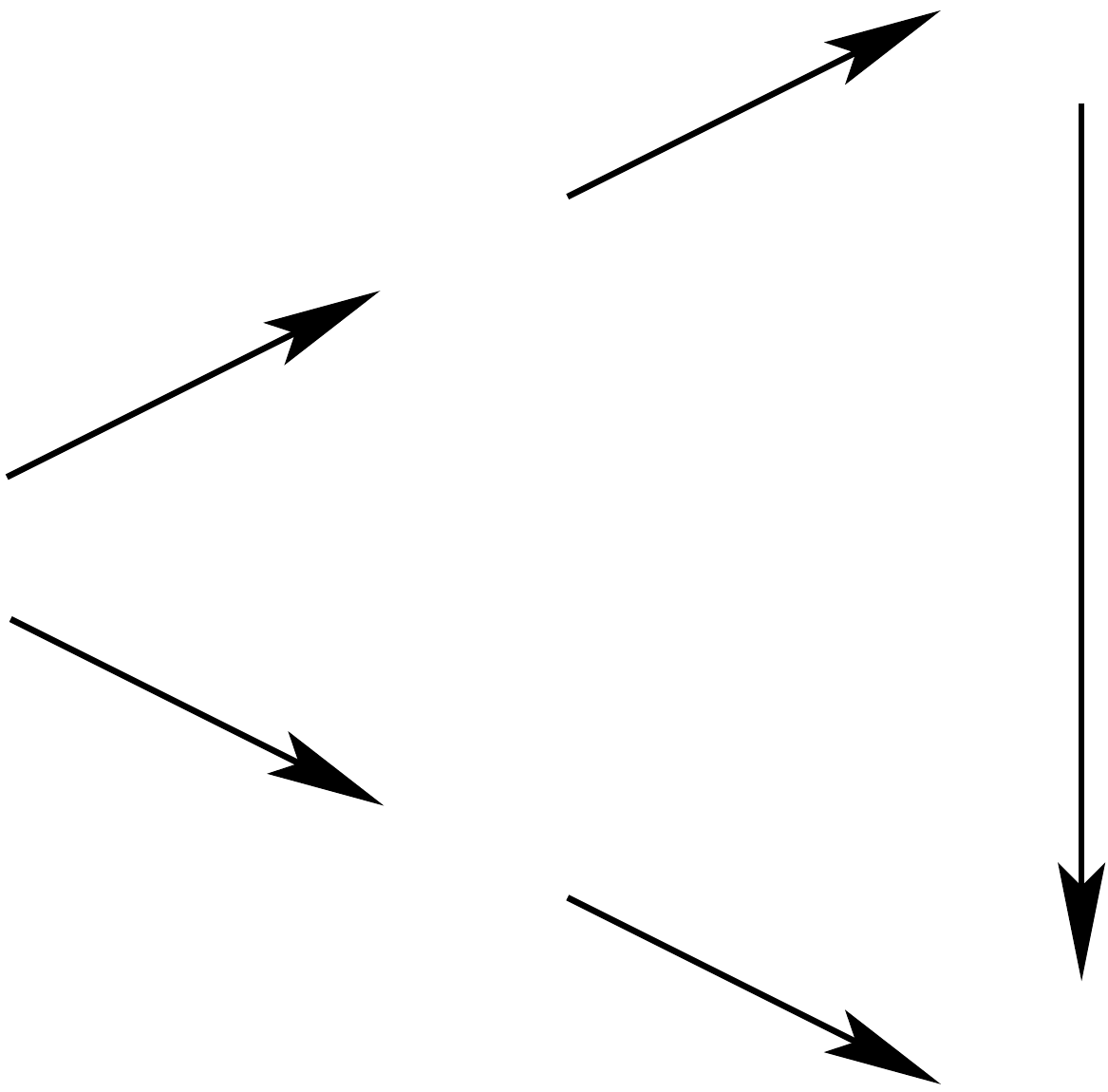}
\put(-123,51){$\Sigma$}
\put(-68,81){$\Sigma_1$}
\put(-69,21){$\Sigma_2$}
\put(-10,108){$\Sigma_{12}$}
\put(-10,-3){$\Sigma_{21}$}
\put(-113,65){\tiny
 \begin{rotate}{25}$(\alpha_1,\beta_1)$\end{rotate}}
\put(-56,94){\tiny
 \begin{rotate}{25}$(\alpha_2,\beta_2)$\end{rotate}}
\put(-113,41){\tiny 
 \begin{rotate}{-25}$(\alpha_2,\beta_2)$\end{rotate}}
\put(-56,13){\tiny 
 \begin{rotate}{-25}$(\alpha_1,\beta_1)$\end{rotate}}
\put(3,54){$(\alpha,\beta)$}
\caption{\label{fig:FigCommDia}\footnotesize%
An (infinitesimal) hypersurface deformation with parameters 
$(\alpha_1,\beta_1)$ that maps $\Sigma\mapsto\Sigma_1$,
followed by one with parameters $(\alpha_2,\beta_2)$ that 
maps $\Sigma_1\mapsto\Sigma_{12}$ differs by one with 
parameters $(\alpha,\beta)$ given by \eqref{eq:HypDefAlgebra-2}
from that in which the maps with the same parameters are 
composed in the opposite order.}
\end{center}
\end{figure}
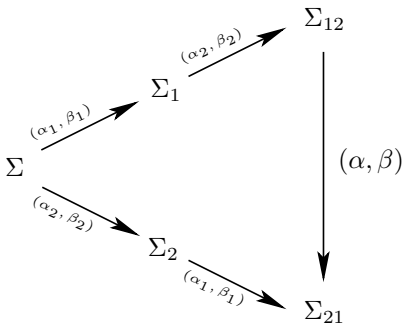

We can now address the inverse problem, namely to find all Hamiltonian 
representations  of \eqref{eq:HypDefAlgebra} on a given phase space. 
As in GR the phase space is $T^*Q$, where $Q=\RiemMet(\Sigma)$, 
That is, we may ask for the most general phase-space functions 
$H(\alpha,\beta):T^*\RiemMet(\Sigma)\rightarrow\reals$, 
parametrized by $(\alpha,\beta)$, so that  
\begin{equation}
\label{eq:H-Rep}
\bigl\{H(\alpha_1,\beta_1),H(\alpha_2,\beta_2)\bigr\}=H(\alpha,\beta)\,.
\end{equation}   
The meaning of this relation is once more explained in 
Fig.\,\ref{fig:FigCommDia}. It is also sometimes expressed as
\emph{path independence}, for it implies 
\index{paths independence}%
that the Hamiltonian flow corresponding to two different paths 
in $\Emb(\Sigma,M)$ reaching the same final hypersurface will
also result in the same physical state (phase-space point). 

To answer this question one first has to choose a phase space. 
Here we stick to the same phase space as in GR, that is $T^*Q$, 
where $Q=\RiemMet(\Sigma)$. The representation problem can be 
solved under certain additional hypotheses concerning the 
geometric interpretation of $H(\alpha=0,\beta)$ and 
$H(\alpha, \beta=0)$:
\begin{itemize}
\item[1.]
$H(0,\beta)$ should represent an infinitesimal spatial 
diffeomorphism, so that 
\begin{subequations}
\begin{equation}
\label{eq:Uniqueness-Assumpt1}
\{F,H(0,\beta)\}=L_\beta F
\end{equation}
for any phase-space function $F$. This fixes $H(0,\beta)$ to 
be the momentum map for the action of $\Diff(\Sigma)$ on 
phase space. 
\item[2.]
$H(\alpha,0)$ should represent an infinitesimal $\Diff(M)$ 
action ``normal to $\Sigma$''. In absence of $M$, which is 
not yet constructed, this phrase is taken to mean that 
\eqref{eq:LieNormal-h} must hold, i.e. 
\begin{equation}
\label{eq:Uniqueness-Assumpt2}
\{h,H(\alpha,0)\}=-2\signature\alpha K\,,
\end{equation}
\end{subequations}
where $K$ is the extrinsic curvature of $\Sigma$ in the 
ambient spacetime that is yet to be constructed. 
\end{itemize}

It has been shown that under these conditions the Hamiltonian of GR,
including a cosmological constant, provides the unique 2-parameter 
family of solutions, the parameters being $\kappa$ and $\Lambda$.
See \cite{Hojman.etal:1976} for more details and \cite{Kuchar:1973}
for the most complete proof (see below for a small topological gap). 
This result may be seen as Hamiltonian analog to Lovelock's 
uniqueness result \cite{Lovelock:1972} for Einstein's equations 
using spacetime covariance.   

A particular consequence of this result is the impossibility 
to change the parameter $\lambda$ in the De\,Witt metric 
\eqref{eq:DeWittMetric-1} to any other than the GR value 
$\lambda=1$ without violating the representation condition, 
that is, without violating covariance under spacetime 
diffeomorphisms. Such theories include those of Ho{\v r}ava-Lifshitz
type \cite{Horava:2009}, which were suggested as candidates for 
ultraviolet completions of GR.

At this point we must mention a topological subtlety which 
causes a small gap in the uniqueness proofs mentioned above 
and might have important consequences in Quantum Gravity. 
To approach this issue we recall from the symplectic framework 
that we can always perform a canonical transformation of the form
\begin{equation}
\label{eq:CanTrans}
\pi\mapsto \pi':=\pi+\Theta\,,
\end{equation}
where $\Theta$ is a closed one-form on $\RiemMet(\Sigma)$. 
Closedness ensures that all Poisson brackets remain the same 
if $\pi$ is replaced with $\pi'$. Since $\RiemMet(\Sigma)$ 
is an open positive convex cone in a vector space and hence 
contractible, it is immediate that $\Theta=df$ for some 
function $f:\RiemMet(\Sigma)\rightarrow\reals$. However, $\pi$ 
and $\pi'$ must satisfy the diffeomorphism constraint, which is 
equivalent to saying that the kernel of $\pi$ (considered as 
one-form on $\RiemMet(\Sigma)$) contains the vector fields 
generated by spatial diffeomorphisms, which implies that 
$\Theta$, too, must annihilate all those, so that $f$ is 
constant on each connected component of the $\Diff(\Sigma)$ 
orbit in $\RiemMet(\Sigma)$. But unless these orbits are 
connected this does not imply that $f$ is the pull back of a 
function on the quotient $\RiemMet(\Sigma)/\Diff(\Sigma)$, 
as assumed in \cite{Kuchar:1973}. We can only conclude that 
$\Theta$ is the pull back of a closed 
but not necessarily exact one-form on superspace. Hence there is 
an analogue of the Bohm-Aharonov-like  ambiguity that one always 
encounters if the configuration space is not simply connected. 
The quantum theory is then expected to display a sectorial 
structure labeled by the equivalence classes of unitary irreducible 
representations of the fundamental group of configuration space,
which in analogy to Yang-Mills-type gauge theories are sometimes 
referred to as \emph{$\theta$-sectors}~\cite{Isham:1982}. 
\index{theta sectors, in Quantum Gravity}
In GR the fundamental group of configuration space is 
\index{fundamental!group (of configuration space)}
isomorphic to a certain mapping-class group of the 
3-manifold $\Sigma$. The theta-structure then depends on the topology 
of $\Sigma$ and can range from `trivial' to `very complicated'. 
See \cite{Giulini:2007a} for more details on the role and
determination of these mapping-class groups and \cite{Giulini:2009a}
for a more general discussion of the configuration space in GR, 
which, roughly speaking, is the quotient 
$\RiemMet(\Sigma)/\Diff(\Sigma)$, often referred to as 
\emph{Wheeler's superspace}~\cite{Wheeler:1964-b}\cite{DeWittQTGI:1967}. 
\index{superspace, Wheeler's}

We finally note that additional theta-structures may emerge if 
the gravitational field is formulated by means of different 
field variables including more mathematical degrees of freedom 
and more constraints (so as to result in the same number of 
physical degrees of freedom upon taking the quotient). The global 
structure of the additional gauge transformations may then add to the 
non-triviality of the fundamental group of configuration space and 
hence to the complexity of the sectorial structure. Examples have 
been discussed in the context of \emph{Ashtekar variables}
\index{Ashtekar!variables} (cf. final Section) in connection 
with the \emph{CP-problem} 
\index{CP-problem, in Quantum Gravity}
in Quantum Gravity~\cite{Ashtekar.Balachandran.Jo:1989}.

\subsection{An alternative action principle}
\label{sec:AlternativeActionPrinciple}
A conceptually interesting albeit mathematically awkward 
alternative form of the action principle for GR was given
by Baierlein, Sharp, and Wheeler in 
\cite{Baierlein.Sharp.Wheeler:1962}. Its underlying idea,
as far as the initial-value problem is concerned, is as 
follows: We have seen that initial data $(h,K)$ (or $(h,\pi)$)
had to obey four constraints (per space point) but that the 
four functions $\alpha$ and $\beta$ could be specified freely. 
Could we not let the constraints determine $\alpha$ and $\beta$ 
and thereby gain full freedom in specifying the initial data?
In that case we would, for example, gain full control over the 
initial geometry, whereas, as we will see later, the standard 
\emph{conformal method}
\index{conformal method}
\index{constraints!solve (methods, meaning)}
to solve the constraints only provides control over the 
conformal equivalence class of the initial geometry, the 
representative within that class being determined by the 
solution to the scalar constraint. For black-hole collision 
data this, e.g.,  means that we cannot initially specify 
the initial distances. We will now discuss to what extent 
this can indeed be done. At the end of this subsection we will add 
some more comments regarding the conceptual issues associated with 
this alternative formulation. 
     
We start from the action 
\begin{equation}
\label{eq:BSW-1}
S_{\rm GR}[g;\Omega]=\int_\Omega d^4x\, 
\LDensity_{\rm GR}\,, 
\end{equation}
where $\LDensity_{\rm GR}$ is as in \eqref{eq:LagrangianDensityGR}. 
In it we express $K$ in terms of $\dot h$ as given in \eqref{eq:K-InTermsOfLieH}. This results in  
\begin{equation}
\label{eq:BSW-2}
\begin{split}
&S_{\rm GR}[\alpha,\beta,h,\dot h\,;\,\Omega]=\\
&\int_\Omega d^4x\frac{\sqrt{h}}{2\kappa}\left\{
\frac{1}{4\alpha}\ 
G\bigl(\dot h-L_\beta h\,,\,\dot h-L_\beta h\bigr)
-\signature R\alpha
\right\}\,.
\end{split} 
\end{equation}
For fixed domain $\Omega\subset M$ this is to be regarded 
as functional of $g$, that is of $\alpha$, $\beta$ and $h$.
Note that $\alpha$ enters in an undifferentiated form. 
Variation with respect to it gives 
\begin{equation}
\label{eq:BSW-3}
\alpha=\alpha_*(h,\dot h,\beta):=\frac{1}{2}\sqrt{
\frac{-\signature\,G\bigl(\dot h-L_\beta h\,,\,\dot h-L_\beta h\bigr)}{R}}\,,
\end{equation}
where we have chosen the positive root for $\alpha$ and introduced 
the abbreviation $\alpha_*$ for the function of $h$, $\dot h$, and 
$\beta$ defined by the expression on the right-hand side in 
\eqref{eq:BSW-3}. Note that this only makes sense if $R$ has no 
zeros and if the sign of 
$-\signature\,G\bigl(\dot h-L_\beta h\,,\,\dot h-L_\beta h\bigr)$
equals that of $R$. Hence we have to restrict to a particular 
sign for the latter expression. We set 
\begin{equation}
\label{eq:BSW-4}
\begin{split}
\sigma:
&=\text{sign}\Bigl(G\bigl(\dot h-L_\beta h\,,\,\dot h-L_\beta h\bigr)\Bigr)\\
&=-\signature\ \text{sign}\bigl(R\bigr)\,.
\end{split}
\end{equation}
Reinserting \eqref{eq:BSW-3} into \eqref{eq:BSW-1}, 
taking into account \eqref{eq:BSW-4}, gives 
\begin{equation}
\label{eq:BSW-5}
\begin{split}
&S_{\rm BSW}[\beta,h,\dot h;\Omega]
:=S_{\rm GR}[\alpha=\alpha_*\,,\,\beta,h,\dot h;\Omega]=\\
&\int_\Omega d^4x\, 
\frac{\sigma\sqrt{h}}{2\kappa}\sqrt{%
-\signature\,R\,G(\dot h-L_\beta h\,,\,\dot h-L_\beta h)}\,. 
\end{split}
\end{equation}
Here we explicitly indicated the functional dependence on the 
time derivative of $h$ to stress the independence on the time 
derivative of $\beta$. With reference to 
\cite{Baierlein.Sharp.Wheeler:1962} this form of the action is 
sometimes called the \emph{Baierlein-Sharp-Wheeler action}.
\index{Baierlein-Sharp-Wheeler action}
(Hence the subscript BSW in \eqref{eq:BSW-5}.) 
We can now try to further reduce this action so
as to only depend on $h$ and $\dot h$. For this we would have 
to proceed with $\beta$ in the same fashion as we have just 
done with $\alpha$; i.e., vary \eqref{eq:BSW-5} with respect 
to $\beta$ and then reinsert the solution $\beta[h,\dot h]$ of 
the ensuing variational equations back into \eqref{eq:BSW-5}. 
The variational equations for $\beta$ are easily obtained
(note that $\sigma$ drops out): 
\begin{equation}
\label{eq:ThinSandwichEquation}
\begin{split}
\frac{2\kappa}{\sqrt{h}}\,&\frac{\delta S_{\rm BSW}}{\delta\beta_d}=\hfill\\
\quad&D_c\left\{\frac{G^{abcd}(\dot h-L_\beta h)_{ab}}{\alpha_*(h,\dot h,\beta)}\right\}=0\,.
\end{split}
\end{equation}
These three equations are traditionally referred to as the 
\emph{thin-sandwich equations}.
\index{thin-sandwich!equations}
They are meant to determine $\beta$ for given pairs $(h,\dot h)$.

Suppose there is a unique solution to \eqref{eq:ThinSandwichEquation} 
for given $(h,\dot h)$, i.e.,
\begin{subequations}
\label{eq:BSW-6}
\begin{equation}
\label{eq:BSW-6a}
\beta=\beta_*(h,\dot h)\,.
\end{equation}
Inserting this into \eqref{eq:BSW-3} determines the lapse 
for given $(h,\dot h)$:
\begin{equation}
\label{eq:BSW-6b}
\alpha=\alpha_*\bigl(h,\dot h,\beta=\beta_*(h,\dot h)\bigr)\,.
\end{equation}
\end{subequations}
Our initial goal is then achieved if we consider $h\in\STSigma{0}{2}$
(positive definite) and $\dot h\in\STSigma{0}{2}$ (arbitrary) as 
freely specifiable initial data. As intended, $h$ represents a 
freely specifiable Riemannian geometry of $\Sigma$ and $\dot h$ its
initial rate of change with respect to some formal parameter. The 
relation between this formal time parameter and proper time is 
fully determined  by the solutions \eqref{eq:BSW-6}, in the way 
explained in Section\,\ref{sec:DecompMetric}. This means that 
the specification of two infinitesimally nearby configurations 
$h$ and $h+\dot h\,dx^0$ allows to deduce the proper time that 
separates the corresponding spatial slices in the spacetime to 
be constructed. In this sense, and subject to the solvability of
\eqref{eq:ThinSandwichEquation}, physically meaningful 
\emph{durations} can be deduced from two infinitesimally close 
instantaneous configurations. This is why Baierlein, Sharp, and 
Wheeler concluded in \cite{Baierlein.Sharp.Wheeler:1962} that ``three-dimensional geometry is a carrier of information about 
time''; or in Barbour's congenial dictum (\cite{Barbour:1994b}, p.\,2885): 
``The instant is not in time; time is in the instant''. 
Also, it should now be obvious where the term \emph{thin-sandwich} 
comes from and why the problem of finding a solution to \eqref{eq:ThinSandwichEquation} is often referred to as the  
\emph{thin-sandwich problem}. 
\index{thin-sandwich!problem} 
   
The thin-sandwich equations \eqref{eq:ThinSandwichEquation} 
were first conceived and discussed in \cite{Wheeler:1964-b} and \cite{Baierlein.Sharp.Wheeler:1962}. Existence and uniqueness 
of solutions were originally conjectured (henceforth known as 
\emph{thin-sandwich conjecture}) 
\index{thin-sandwich!conjecture}, 
e.g. in \cite{Wheeler:1964-b}, but first mathematical 
investigations soon showed that the unqualified thin-sandwich
conjecture is false; see  \cite{Belasco.Ohanian:1969}
and \cite{Christodoulou.Francaviglia:1979}. 

To see what positive statements can be made let us summarize 
the situation: At fixed time, i.e. on each 3-manifold $\Sigma$ 
with given Riemannian metric $h$ and given parameter-time 
derivative $\dot h$ (covariant symmetric second-rank tensor 
field), equations \eqref{eq:ThinSandwichEquation} form a 
system of three quasi-linear (though highly non-linear) second 
order equations for the three components 
$\beta_a$. The restriction $\alpha>0$, where $\alpha$ is given by \eqref{eq:BSW-3}, involves $\beta$ and hence implies an 
\emph{a priori bound} for the unknown $\beta$. As a consequence 
one may first of all expect only \emph{local} results (if at all), 
in the sense that if $\beta_*$ is a solution for given $(h_*,\dot h_*)$ 
then there exists some open neighborhood $U$ of $(h_*,\dot h_*)$ in 
field space (in an appropriate topology) such that existence 
and uniqueness of solutions follows for all $(h,\dot h)\in U$. 
Such a local result can only be expected if the initial velocity 
$\dot h_*$, or any of its space-point dependent reparametrizations, 
is not just of the form of an infinitesimal diffeomorphism $L_\xi h_*$. 
In other words, $(h_*,\dot h_*)$ must be chosen such that for any 
smooth function $\eta$ and vector field $\xi$ on $\Sigma$ we have 
the following implication   
\begin{equation}
\label{eq:FullyDeveloppedDirection}
\dot h_*=\eta\,L_\xi h_*
\quad\Leftrightarrow\quad
\eta=0\quad\text{and}\quad\xi=0\,.
\end{equation}
Assuming \eqref{eq:FullyDeveloppedDirection}, local results were 
first proven in \cite{Bartnik.Fodor} and subsequently in a 
more geometric form and with generalizations, also including 
matter fields, in \cite{Giulini:1999}. The idea of proof is to 
write $\beta=\beta_*+\delta\beta$ and linearize the differential 
operator \eqref{eq:ThinSandwichEquation} acting on 
$\delta\beta$ around the solution $(\beta_*;h_*,\dot h_*)$. 
The resulting linear operator turns out to be symmetric (no 
surprise, being the second functional derivative of $S_{\rm BSW}$
in $\beta$) with a principal symbol $\sigma(k)$ whose 
determinant is proportional to a power of $\Vert k\Vert$ times 
$[\pi(k,k)]^2$ (see eq.~(3.14) in \cite{Giulini:1999}), which 
vanishes for non-zero $k$ iff $\pi(k,k)=0$. Hence the linearized 
operator is elliptic iff the quadratic form $\pi$ is either 
positive or negative definite. Granted this, \eqref{eq:FullyDeveloppedDirection} then ensures that the 
elliptic operator has a trivial kernel. Together this allows 
to deduce an implicit-function theorem that immediately implies 
local existence and uniqueness.   
As regards ellipticity, we recall that from the discussion following 
\eqref{eq:KineticTerm-Sign2} that a definite $\pi$ is 
equivalent to ${\hat G}(\pi,\pi)<0$, i.e. $\sigma=-1$ (compare \eqref{eq:BSW-4}). Hence a local existence and uniqueness result 
holds provided that $(h_*,\dot h_*,\beta_*)$ satisfy $\sigma=-1$,
comprising the two equations 
\begin{equation}
\label{eq:BSW-EllipticityCond}
\begin{split}
&\epsilon R(h_*)>0\,,\\
&G(\dot h_*-L_{\beta_*}h_*\,,\,\dot h_*-L_{\beta_*}h_*\,,\,)<0\,.
\end{split}
\end{equation}
We see that we actually were not free in choosing either 
sign for $\sigma$ in \eqref{eq:BSW-4}: We are bound to 
$\sigma=-1$ in order to ensure at least local existence 
and uniquness of the thin-sandwich equation 
\eqref{eq:ThinSandwichEquation}.

In passing we note the choice of the negative sign $\sigma=-1$ 
has very different topological consequences depending on whether 
$\signature =-1$ (Lorentzian spacetime), where it implies $R<0$, or $\signature=1$ (Riemannian/Euclidean spacetime), where it implies $R>0$.
In the former case, given any orientable closed 3-manifold 
$\Sigma$, the theorem of Kazdan \& Warner
\index{theorem!of Kazdan-Warner} 
(see paragraph above Section\,\ref{sec:Slicing conditons}) 
ensures the existence of a Riemannian metric $h$ on $\Sigma$ with 
negative scalar curvature. In contrast, in the latter case (Euclidean spacetimes), the theorem of Gromov \& Lawson 
\index{theorem!of Gromov-Lawson} 
implies a severe topological obstruction against Riemannian 
metrics $h$ on $\Sigma$ with positive scalar curvature 
(as already discussed above: only connected sums of handles and 
lens spaces survive this obstruction).

The result on local \emph{uniqueness} can be generalized to 
a global argument, first given in \cite{Belasco.Ohanian:1969} 
and generalized in \cite{Giulini:1999} to also include matter 
fields. On the other hand, it is not difficult to see that global \emph{existence} cannot hold, i.e. there are 
more or less obvious data $(h,\dot h)$ for which \eqref{eq:ThinSandwichEquation} has no solution for $\beta$; 
see, e.g., \cite{Murchadha.Roszkowski:2006}.  

Finally we 
mention the \emph{conformal thin-sandwich} approach 
of York's \cite{York:1999}, which is a conceptually weaker 
but mathematically less awkward variant \index{thin-sandwich!conformal variant} of the full thin-sandwich problem, in which only the 
conformal equivalence class of the metric and its time derivative 
is initially specified, together with the lapse function and 
the extrinsic curvature. The constraints are then solved for 
the conformal factor and the shift vector field. The equation
for the conformal factor is as in the conformal method 
discussed below (York's equation \eqref{eq:YorkEquation}),
but the equation for the shift is mathematically less awkward 
than in the full thin-sandwich equation 
\eqref{eq:ThinSandwichEquation}. Recall that the latter is 
rendered complicated to the insertion of the solution 
\eqref{eq:BSW-3} for $\alpha$, which is precisely what is 
\emph{not} done in the conformal variant. But, clearly, 
the price for not solving for the lapse is that we have 
again no initial control over the full metric. 
Nevertheless, the better behaved equations of the conformal 
variant of the thin-sandwich method make it a useful 
tool in numerical investigations; see, e.g.,
\cite{Baumgarte.Shapiro:NumericalRelativity}%
\cite{Gourgoulhon:ThreePlusOneGR}.

\subsubsection*{Comparison with Jacobi's principle}
It is conceptually interesting to compare the Baierlein-Sharp-Wheeler 
action of GR with Jacobi's principle 
\index{principle!Jacobi's}
in mechanics. So let us briefly recall Jacobi's original idea 
\cite{Jacobi:VorlesungenDynamik1866}, where only the notation 
will be adapted. 

As in Section\,\ref{sec:ConstrainedHamiltonianSystems} we 
consider a (so far unconstrained) mechanical system with 
$n$-dimensional configuration space $Q$. Let it be characterized 
by a Lagrangian $L:TQ\rightarrow\reals$ of the form 
\begin{equation}
\label{eq:JabociPrinciple-1}
L(q,v)=\tfrac{1}{2}G_q(v,v)-V(q)\,. 
\end{equation}
Here $G_q$ is a positive-definite bilinear form on $T_qQ$  
called the \emph{kinetic-energy metric}
\index{kinetic-energy metric}. We already know that as 
$L$ does not explicitly depend on time, any dynamically 
possible trajectory will run entirely within a hypersurface 
of constant energy $E$ (the energy function being 
\eqref{eq:EnergyFunction}). 
Maupertuis' \emph{principle of least action}, 
\index{principle!Maupertuis'}%
\index{Maupertuis' principle} 
states that a dynamically possible trajectory 
$x:\reals\supset I\rightarrow Q$, connecting fixed 
initial and final points $q_i$ and $q_f$,  extremizes 
the ``action''
\begin{equation}
\label{eq:MaupertuisPrinciple}
\int_{q_i}^{q_f} G_{x(t)}\bigl(\dot x(t),\dot x(t)\bigr)\,dt
\end{equation}
relative to all other curves with the same endpoints and on 
the same energy hypersurface. We note in passing that since 
$\dot x\,dt=dq$ and $G(v,\cdot)=\partial L(q,v)/\partial v=p$, 
the integrals in \eqref{eq:MaupertuisPrinciple} equal the 
integrals of the canonical one-form $\theta$ (compare \eqref{eq:SympStructureCoordinates-a}) along the paths 
$q=x(t)$ and $p=G\bigl(\dot x(t),\cdot\bigr)$ in $T^*Q$ with 
fixed endpoints of the curve projected into $Q$ and the curves 
all running on a hypersurface of fixed value of the Hamiltonian 
function. This is the form the principle of least action
\index{principle!of least action (in mechanics)}
is given in modern formulations, like in Arnold's book 
\cite{Arnold:Mechanics}(Chapter 9, Section 45\,D). In this 
(modern, Hamiltonian) form, time plays no role. Indeed, on 
phase space $T^*Q$ the integral of the one-form $\theta$, as 
well as the level sets for the Hamiltonian function, are 
defined without reference to any time parameter $t$. But in 
the traditional (19th century, Lagrangian) form stated above, 
the parameter enters in an essential way. In fact, in this 
formulation $t$ is not an \emph{independent} variable because 
the energy condition expressed in terms of positions and 
velocities (measured with respect to $t$) introduces an 
implicit dependency of $t$ with coordinates on $TQ$. 
These dependencies have to be respected by variations of \eqref{eq:MaupertuisPrinciple}. This is why Jacobi complained 
in his lectures that the form given above appears incomprehensible
to him.%
\footnote{``Dieses Prinzip wird fast in allen Lehrb\"uchern, 
auch den besten, in denen von \emph{Poisson, Lagrange} und 
\emph{Laplace}, so dargestellt, dass es nach meiner Ansicht 
nicht zu verstehen ist." (\cite{Jacobi:VorlesungenDynamik1866}, p.\,44)} 
His crucial observation was that the energy condition allows to
eliminate $t$ altogether. Indeed, if we solve the energy condition 
for $dt$, 
\begin{equation}
\label{eq:JacobiEnergyCondition}
dt=\sqrt{\frac{G_q(dq,dq)}{2\bigl(E-V(q)\bigr)}}\,,
\end{equation}
and use that to eliminate $dt$ in 
\eqref{eq:MaupertuisPrinciple}, we can put the integral into the 
form      
\begin{equation}
\label{eq:JacobisPrinciple}
\int_{q_i}^{q_f} \sqrt{2\bigl(E-V(q)\bigr)\,G_q(dq,dq)}\,,
\end{equation}
which is independent of any parametrization. In fact, it has the 
simple geometric interpretation of the length functional or the 
conformally rescaled kinetic-energy metric: 
\begin{equation}
\label{eq:FirstJacobiMetric}
\hat G=2(E-V)\,G\,,
\end{equation}
where $E$ is a constant. \emph{Jacobi's principle}
\index{Jacobi!principle} then says that the dynamically 
possible trajectories of energy $E$ are the geodesics of 
$\hat G$, and that (Newtonian) time along such a geodesic 
is obtained by integrating \eqref{eq:JacobiEnergyCondition} 
along it. Note that Jacobi's principle
\index{principle!Jacobi's}
\index{Jacobi!principle}
defines in fact \emph{two} new metrics on configuration space $Q$, both 
of which are conformally equivalent to the kinetic-energy metric
$G$. The first is \eqref{eq:FirstJacobiMetric}, which determines 
the trajectory in $Q$, the other is 
\begin{equation}
\label{eq:SecondJacobiMetric}
\tilde G=\frac{G}{2(E-V)}\,,
\end{equation}
which determines Newtonian time along the trajectories
selected by the first. We call it the \emph{first} 
and \emph{second Jacobi metric} respectively.
\index{Jacobi!metric (first and second)} 
Note also that \eqref{eq:JacobiEnergyCondition} gives a 
measure for duration, $dt$, in terms of changes of mechanical 
coordinates. 

In passing we remark that a special realization of this
far reaching idea, namely to read off Newtonian time from
simultaneous configurations (i.e., generalized positions) 
of mechanical systems, assuming the systems to obey Newtonian
laws of motion, is \emph{emphemeris (or astronomical) time} in
astronomy~\cite{Clemence:1957}\cite{Barbour:AbsRelMot1},
where the relative configurations (ephemerides) of the Moon,
Sun, and planets (as seen from the Earth) are used as positions. 
Ephemeris time was first proposed as time standard in 1948
\cite{Clemence:1948} in order to establish a reference with 
respect to which non uniformities in the Earth's daily rotation
could be accounted for, though the idea goes back at least to 
1929~\cite{Clemence:1971}. Ephemeris time ideed became the 
time standard in 1952 until the 1970s, when atomic time took over
its place, though we remark that ``atomic time'' is really just 
based on a straightforward generalization of the very same 
principle to Quantum Mechanics. Here, again,  one reads off time 
from simultaneously measurable states or observables of one or 
more systems obeying known deterministic ``laws of motion'', 
like Schr\"odinger's equation for states or Heisenberg's 
equation for observables. These equations of motion 
correlate the a priori unobservable parameter $t$ with observable 
properties of the system, which render $t$ observable in a 
context dependent fashion through inverting these relations
(solving the equations of motion for $t$).

Now, coming back to gravity, \eqref{eq:JacobisPrinciple} 
should be compared to 
\eqref{eq:BSW-5}. Except for the terms involving the 
lapse-function $\beta$ the latter is (almost!) like
 \eqref{eq:JacobisPrinciple} for $E=0$ and $V=\signature R$.
The analogy is incomplete (hence ``almost'') because in 
\eqref{eq:BSW-5} the spatial integration is outside the 
square-root, so that the integrand for the 
parameter-integration along the curve in configuration 
space is a sum of square-roots rather than a single 
square-root of a sum. This difference renders the 
expression in \eqref{eq:BSW-5} different from usual 
``length functionals''. Note that sums of square roots
(involving themselves sums of squares) generally do not 
even form a Finsler metric (compare the discussion in 
\cite{Giulini:1999}).  

Taking the analogy further, \eqref{eq:JacobiEnergyCondition} 
should be compared to \eqref{eq:BSW-6b} with $\alpha_*$ 
given by the integral \eqref{eq:BSW-3}. Again, except for 
the terms involving $\beta$, they seem to closely correspond 
to each other. Equation \eqref{eq:BSW-6b} determines one 
proper time per spatially fixed (with respect to the spatial 
coordinates) observer  in the spacetime to be developed 
from the initial data. Hence there is something like a 
continuum of second Jacobi metrics, one for each space point.

An interesting observation in this connection is the 
following~\cite{Christodoulou:1975}, which we give in a 
simplified form. Suppose that we tried to define, from 
first principles, \emph{duration} by some measure of change 
in the gravitational degrees of freedom, i.e., some kind of 
gravitational \emph{ephemeris time}
\index{ephemeris time!gravitational}. In analogy to \eqref{eq:JacobiEnergyCondition} We assume the 
``measure of change'', $d\tau$,  to be given by some 
local rescaling of a pseudo-Riemannian distance measure
\begin{equation}
\label{eq:WDW-Distance}
ds^2=\int_\Sigma d^3x\ G^{ab\,cd}[h(x)] dh_{ab}(x)dh_{cd}(x)\,,
\end{equation} 
so that 
\begin{equation}
\label{eq:ChronosTime}
d\tau^2=\frac{ds^2}{\int_\Sigma d^3x\ R(x)}\,.
\end{equation}
Here $R$ must be a scalar function of the spatial metric $h$. 
The simplest non-constant such function is the scalar 
curvature, which depends on $h$ and its derivatives up to 
second order. A priori such a measure of duration seems 
to depend equally on all gravitational degrees of freedom at 
all points in space, thus giving rise to a highly non-local 
concept of time with respect to which durations of processes, 
even local ones, can be measured. However, suppose we required 
that the measure of time be compatible with arbitrarily fine 
localization $\Sigma\rightarrow U\subset\Sigma$. Following 
\cite{Christodoulou:1975} we call this the \emph{chronos principle}.
\index{chronos principle}
It implies that the numerator and denominator of \eqref{eq:ChronosTime} 
are proportional for each restriction $\Sigma\rightarrow U$. 
This is only possible if the integrands are proportional. 
Without loss of generality we can take this constant of 
proportionality (which cannot be zero) to be $1$ (this just 
fixes the overall scale of physical time) and obtain (here written 
without cosmological constant for simplicity)  
\begin{equation}
\label{eq:ChronosTimeLocProp}
G^{ab\,cd}[h(x)] \frac{dh_{ab}(x)}{d\tau}\frac{dh_{cd}(x)}{d\tau}-R[h](x)=0\,.
\end{equation}
which is just the Hamiltonian constraint. The duration in time 
is then given by 
\begin{equation}
\label{eq:TimeFormula}
\Delta\tau(h_i,h_f)=\int_{h_i}^{h_f}
\sqrt{\frac{
G\bigl(dh/d\lambda,dh/d\lambda\bigr)}
{-R\bigl[h(\lambda)\bigr]}}\ d\lambda\,,
\end{equation}
which is obviously just the analog of the integrated version of
\eqref{eq:JacobiEnergyCondition} for $E=0$ (which here amounts to 
the Hamiltonian constraint), i.e. the integrated version of 
\eqref{eq:BSW-3} for $\beta=0$ (this is the simplification we 
alluded to above) and $\signature=-1$.

\section{Asymptotic flatness and global charges}
\label{sec:AsympFlatnessCharges}
Isolated systems are described by geometries which at 
large spatial distances approach a matter-free spacetime. 
In case of vanishing cosmological constant the latter will be 
flat Minkowski spacetime. For non-zero $\Lambda$ it will either 
be de\,Sitter ($\Lambda >0$) or anti-de\,Sitter ($\Lambda <0$) 
space. Here we are interested in the case $\Lambda=0$. We refer to 
the survey \cite{Fischetti.Kelly.Marolf:SpringerHandbookSpacetime2014}
for a recent discussion of the anti-de\,Sitter case. 

An initial data set $(h,\pi)$ or $(h,K)$ on $\Sigma$ needs 
to satisfy certain asymptotic conditions in order to give 
rise to an asymptotically flat spacetime. Before going into 
this, we point out that there is also a topological condition 
on $\Sigma$ in order to sensibly talk about 
\emph{asymptotic regions}.
\index{asymptotic!regions} 
The condition is that there exists a compact set $K\subset\Sigma$
such that its complement $\Sigma-K$ is diffeomorphic to the 
disjoint union of manifolds $\reals^3 -B$, where $B$ is a closed 
ball. These pieces in which $\Sigma$ decomposes if one cuts out 
increasingly large compact sets are called \emph{ends} of $\Sigma$.
\index{end (of a manifold)}
In passing we note that the theory of `ends' for topological spaces 
and groups was developed by Freudenthal in 1931~\cite{Freudenthal:1931}.
Now, the first condition we pose is that there is only a finite number 
of such ends. (It is not hard to visualize manifolds with even an 
uncountable number of ends.)  With respect to each end we can 
talk of approaching infinity. This means to let $r\rightarrow\infty$ 
if $r$ is the standard radial coordinate on $\reals^3-B$ to which this 
end is diffeomorphic.

A first working definition of asymptotically flat initial data 
\index{asymptotic!flatness according to Regge-Teitelboim}
in Hamiltonian GR was given in 1974 by Regge 
\& Teiltelboim \cite{ReggeTeitelboim:1974}. It was shown by 
Beig \& \'O\,Murchadha in 1986 \cite{Beig.Murchadha:1987}
that this definition is sufficient to allow the implementation 
of the 10-parameter Poincar\'e group 
\index{Poincar\'e!group}
as asymptotic symmetries giving rise to 10 corresponding conserved 
quantities. The definition can be given as follows 
(here we restrict to one end): 

\begin{definition}[Regge-Teitelboim asymptotic flatness]
\index{asymptotic!flatness according to Regge-Teitelboim}
Let $\Sigma$ be a 3-manifold with one end. An initial data 
set $(h,\pi)$ on $\Sigma$ is asymptotically flat in the sense 
of Regge-Teitelboim if there is a coordinate  system 
$\{x^1,x^2,x^3\}$ covering the end, such that 
as $r:=\sqrt{x^ax^b\delta_{ab}}\rightarrow \infty$)
\begin{subequations}
\label{eq:ReggeTeitelboimFalloff}
\begin{alignat}{2}
\label{eq:ReggeTeitelboimFalloff-a}
  &h_{ab}(x)&&=\delta_{ab}+\frac{s_{ab}(\nu)}{r}+O_2(r^{-2})\,,\\
\label{eq:ReggeTeitelboimFalloff-b}
&\pi^{ab}(x)&&=\frac{t^{ab}(\nu)}{r^2}+O_1(r^{-3})\,,
\end{alignat} 
\end{subequations}
where $x=(x^1,x^2,x^3)$ and $\nu=(\nu^1,\nu^2,\nu^3)$ with $\nu^a:=x^a/r$. 
$O_k(r^{-n})$ denotes terms falling off like $1/r^n$ 
and whose $l$-th derivatives fall off like $1/r^{(n+l)}$ for 
$0<l\leq k$. Moreover, $s_{ab}$ and $t^{ab}$ obey the 
\emph{parity conditions}
\index{parity conditions}
\begin{subequations}
\label{eq:ReggeTeitelboimParity}
\begin{alignat}{2}
\label{eq:ReggeTeitelboimParity-a}
&s_{ab}(-\nu)&&=s_{ab}(\nu)\,,\\
\label{eq:ReggeTeitelboimParity-b}
&t^{ab}(-\nu)&&=-t^{ab}(\nu)\,.
\end{alignat} 
\end{subequations} 
\end{definition}

The first thing to observe is that these conditions suffice 
to make the integral \eqref{eq:GR-SymplPotential} for the 
symplectic potential convergent. Note that 
\eqref{eq:ReggeTeitelboimFalloff} merely implies that the 
integrand falls off like $1/r^3$, which could still produce
a logarithmic divergence. But \eqref{eq:ReggeTeitelboimParity}
implies that the $1/r^3$ integrand is of odd parity and hence 
gives no contribution. Next we have a look at the constraint 
functionals \eqref{eq:GR-ConstraintsHamForm}. In 
\eqref{eq:GR-ConstraintsHamForm-a} the first integrand has a 
$1/r^4$ and the second a $1/r^3$ parity-even fall-off. In 
\eqref{eq:GR-ConstraintsHamForm-b} the integrand has also a 
$1/r^3$ parity-even fall-off. Hence the integrals
\eqref{eq:GR-ConstraintsHamForm} certainly converge for those 
lapse and shift fields $\alpha,\beta$ which asymptotically 
either tend to zero or approach direction-dependent constants 
in a parity-odd fashion. As we will see below, the constraints 
for such parameter fields $\alpha$ and $\beta$ are differentiable 
with respect to the canonical variables $h$ and $\pi$ and hence 
generate a Hamiltonian flow that has to be considered as gauge
transformations; compare Section\,\ref{sec:GeometricTheory}. 
Hence we set
\begin{subequations}
\label{eq:LapseShiftFalloff}
\begin{alignat}{2}
\label{eq:LapseShiftFalloff-a}
&\alpha(x)_{\rm gauge}&&=a(\nu)+O_2(r^{-1})\,,\\
\label{eq:LapseShiftFalloff-b}
&\beta^a(x)_{\rm gauge}&&=b^a(\nu)+O_1(r^{-1})\,,
\end{alignat} 
\end{subequations}
where 
\begin{subequations}
\label{eq:LapseShiftParity}
\begin{alignat}{2}
\label{eq:LapseShiftParity-a}
&a(-\nu)&&=-a(\nu)\,,\\
\label{eq:LapseShiftParity-b}
&b^a(-\nu)&&=-b^a(\nu)\,.
\end{alignat} 
\end{subequations}
To see that $C_s(\alpha)[h,\pi]$ and $C_v(\beta)[h,\pi]$ as 
defined in \eqref{eq:GR-ConstraintsHamForm-a} and 
\eqref{eq:GR-ConstraintsHamForm-b} are functionally differentiable 
with respect to $h$ and $\pi$ we make the following observations: 
For \eqref{eq:GR-ConstraintsHamForm-a} the only boundary term 
one might pick up is that from the variation of the scalar 
curvature with respect to $h$, which follows from 
\eqref{eq:ScalarDifferenceFirstOrder} (in which formula we 
have to replace $g_{ab}$ with $h_{ab}$ and $h_{ab}$ with 
$\delta h_{ab}$ in order to match the notation here). It reads  
\begin{equation}
\label{eq:ScalarConstSurfTerm}
\frac{\signature}{2\kappa}\int_{S^2_\infty}d\Omega\,\alpha\sqrt{h}\nu_aG^{abcd}
D_b\delta h_{cd}
\end{equation}
and is thus seen to have an integrand that has $1/r^2$ fall-off and is 
parity-odd. Hence the integral vanishes. Note that here and in what 
follows we used the following shorthand notation
\begin{equation}
\label{eq:IntegralSphereAtInfty}
\int_{S^2_\infty}d\Omega(\,\cdots\,) :=
\lim_{r\rightarrow\infty}\left\{
\int_{S^2(r)}d\Omega_r(\,\cdots\,)\right\}\,,
\end{equation}
where $S^2(r)$ is the sphere of constant ``radius'' $r$ 
(as defined above) and $d\Omega_r$ its induced volume form. 

The vector constraint \eqref{eq:GR-ConstraintsHamForm-b} contains 
$\pi$ as well as $h$ in differentiated form (the latter in $D$),
so that boundary terms may appear in both variation, that with 
respect to $\pi$ as well as that with respect to $h$. For 
both cases it is convenient to rewrite the integral \eqref{eq:GR-ConstraintsHamForm-b} by performing a partial 
integration before variation:
\begin{equation}
 \label{eq:VectorConstSufTerm-1}
\begin{split}
C_v(\beta):
&=\int_\Sigma d^3x\ \beta^a\Bigl[-2ch_{ab}D_c\pi^{bc}\Bigr]\\
&=c\int_\Sigma d^3x (L_\beta h)_{ab}\pi^{ab}\\
&-2c\int_{S^2_\infty}d\Omega\,\beta^ah_{ab}h_{cd}\nu^d\pi^{bc}\,.
\end{split}
\end{equation}
Under the fall-offs and parity conditions mentioned above the 
last (surface) integral is zero since its integrand has 
$1/r^2$ decay and is parity-odd. Hence variation with respect 
to $\pi$ does not lead to surface terms. Variation with respect 
to $h$ is now simply given by varying the $h$ under the Lie 
differentiation $L_\beta$ which itself has no dependency on $h$
(unlike the covariant derivative $D$). Using  
$L_\beta (\delta h_{ab}\pi^{ab})=D_c(\beta^c\delta h_{ab}\pi^{ab})$, partial 
differentiation with respect to the Lie derivative then 
gives the surface term
\begin{equation}
\label{eq:VectorConstSufTerm-2}
2c\int_{S^2_\infty} d\Omega\,h_{ab}\nu^a\beta^b\,\delta h_{cd}\pi^{cd}\,,
\end{equation}
the integrand of which again falls off like $1/r^2$ and is parity 
odd. 

The considerations so far show that the constraints 
$C_s(\alpha)[h,\pi]$ and $C_v(\beta)[h,\pi]$ are differentiable 
with respect to $h$ and $\pi$ whenever $(h,\pi)$ satisfy   
\eqref{eq:ReggeTeitelboimFalloff} and \eqref{eq:ReggeTeitelboimParity}
and the parameter-functions $(\alpha,\beta)$ satisfy \eqref{eq:LapseShiftFalloff} and \eqref{eq:LapseShiftParity}.
From the considerations it also follows that we cannot improve
on the latter two conditions given the fall-offs and parity 
conditions on $(h,\pi)$. This characterizes the lapse and shift 
functions which generate pure \emph{gauge transformations}
in Hamiltonian gravity for asymptotically states.   
\index{gauge!transformations (asympt. flat case)}
We stress that \eqref{eq:LapseShiftFalloff} and 
\eqref{eq:LapseShiftParity} includes  motions that do not 
vanish at infinity. These are called \emph{supertranslations}.
\index{supertranslations, odd parity} Without their careful 
inclusion into the transformations considered as gauge,
we would not obtain the Poincar\'e group as proper physical 
symmetry group but rather an infinite-dimensional extension 
thereof. For more discussion on this conceptually important 
point compare the discussion in~\cite{Giulini:1995d}.      

Motions characterized by functions $(\alpha,\beta)$ outside 
the class \eqref{eq:LapseShiftFalloff} and 
\eqref{eq:LapseShiftParity} do change the physical state. If 
this motion is to be generated by the Hamiltonian 
\eqref{eq:GR-Hamiltonian} we must restrict to those 
$(\alpha,\beta)$ for which suitable boundary terms can be found 
such that $H(\alpha,\beta)[h,\pi]$ is differentiable 
with respect to $h$ and $\pi$ obeying 
\eqref{eq:ReggeTeitelboimFalloff} and 
\eqref{eq:ReggeTeitelboimParity}. To accommodate asymptotic 
Poincar\'e transformations we must worry about asymptotic 
translations in time and space directions, asymptotic 
rotations, and finally asymptotic boosts, all of which we only 
need to specify modulo gauge transformations. Asymptotic time
translations correspond to constant $\alpha$. The surface term 
that results from the variation of $C_s(\alpha)$ is just \eqref{eq:ScalarConstSurfTerm}. It immediately follows that 
the term that has to be added to $C_s(\alpha)$ so as to just 
cancel this surface term upon variation with respect to $h$ 
is just $\alpha\,E_{\rm ADM}$, where 
\begin{equation}
\label{eq:ADM-EnergyMass}
\begin{split}
E_{\rm ADM}&=M_{\rm ADM}c^2\\
&= -\signature(2\kappa)^{-1}
\int_{S^2_\infty}d\Omega\,(\partial_{a}h_{ab}-\partial_bh_{aa})\nu_b\,,
\end{split}
\end{equation}
is called the \emph{ADM energy} and $M_{\rm ADM}$ the \emph{ADM mass}.
\index{ADM!energy}\index{ADM!mass}
\index{energy!ADM}\index{mass!ADM}
Note that we just replaced all non differentiated $h_{ab}$ that 
appear in \eqref{eq:ScalarConstSurfTerm} by $\delta_{ab}$ since 
the difference does not contribute to the surface integral in the 
limit as $r\rightarrow\infty$. Similarly, asymptotic space 
translations corresponds to covariantly constant $\beta$, 
i.e. constant components $\beta^a$ with respect to the preferred 
coordinates that served to define asymptotic flatness. Again we 
immediately read off \eqref{eq:VectorConstSufTerm-1}  the boundary 
term that we need to add in order to cancel that in the last 
line of  \eqref{eq:VectorConstSufTerm-1} upon variation of 
$\pi$. It can be written in the form $cP_{\rm ADM}(\beta)$,
where 
\begin{equation}
\label{eq:ADM-Momentum}
P_{\rm ADM}(\beta)=2\int_{S^2_\infty}d\Omega\,\beta^a\pi_{ab}\nu^b\,.
\end{equation}
This we call the \emph{ADM general momentum}. 
\index{ADM!general momentum}
Note that the integrand has fall-off $1/r^2$ and even parity 
and hence gives a finite contribution. Furthermore, it follows 
from \eqref{eq:VectorConstSufTerm-2} that the variation with 
respect to $h$ does not give rise to a boundary term since the 
integrand in \eqref{eq:VectorConstSufTerm-2} has $1/r^3$ fall
off and hence does not contribute, independently of its (even) 
parity. 

Asymptotic rotations and boosts are a priori more delicate 
since now $\alpha$ and $\beta$ are allowed to grow linearly 
with $r$. We have, up to gauge transformations, 
$\alpha\propto u_ax^a$ for a boost in $\vec u$-direction and 
$\beta_a\propto \varepsilon_{abc}\omega^b x^c$ for a rotation 
around the $\vec\omega$ axis. Here $\varepsilon_{abc}$ are the 
components of the metric volume-form for the asymptotic metric 
$\delta$ with respect to the  coordinates $\{x^a\}$,
so that $\varepsilon_{abc}=\pm 1$, depending on whether $(abc)$
is an even ($+$) or odd $(-)$ permutation of $(123)$. As 
indices are raised and lowered with respect to $\delta$, we 
need not be concerned whether they are upper or lower (as 
long as we work in components with respect to  $\{x^a\}$).
The components $u_a$ and $\omega^a$ are then again constants. 

We start with rotations and read off the last line of \eqref{eq:VectorConstSufTerm-1} 
that the surface integral has an integrand 
$\propto \pi^{ab}\beta_a\nu_b$, which looks dangerous as its naive 
fall-off is $1/r$. However, if we use that $\pi$ actually 
satisfies the constraint, $D_b\pi^{ab}=0$, we can convert this 
surface integral to a bulk integral whose integrand is 
proportional to $D_b(\pi^{ab}\beta_a)=\pi^{ab}D_{(a}\beta_{b)}$.
But $D_{(a}\beta_{b)}=\partial_{(a}\beta_{b)}-\Gamma^c_{ab}\beta_c$
and $\beta$ is Killing with respect to the metric $\delta_{ab}$, 
so that $\partial_{(a}\beta_{b)}=0$. Now, the Christoffel symbols 
$\Gamma^c_{ab}$ for the metric $h$ fall off as $1/r^2$ with odd 
parity, so that $\Gamma^c_{ab}\beta_c$ falls off as $1/r$ with 
even parity. Hence $\pi^{ab}D_{(a}\beta_{b)}$ falls off like 
$1/r^3$ with odd parity, showing that this volume integral 
also converges (a logarithmic divergence being just avoided 
by odd parity). Finally we observe that for asymptotic 
rotations there is still no surface term of the form 
\eqref{eq:VectorConstSufTerm-2}, since its integrand has 
$1/r^2$ fall-off and is of odd parity.  As a result we have 
that even for asymptotic rotations we obtain the same 
formula \eqref{eq:ADM-Momentum} for the (linear or angular) 
momentum as long as the Regge-Teitelboim conditions \eqref{eq:ReggeTeitelboimFalloff} and \eqref{eq:ReggeTeitelboimParity}
are satisfied and $(h,\pi)$ satisfy the constraints. 
In components with respect to the asymptotic 
coordinates the components of the linear momentum are 
\begin{equation}
\label{eq:ADM-LinearMomentum}
P_{\rm ADM}^a=2\int_{S^2_\infty}d\Omega\,\pi^{ab}\nu_b\,,
\end{equation}
\index{ADM!linear momentum}
and for the angular momentum 
\begin{equation}
\label{eq:ADM-AngularMomentum}
J_{\rm ADM}^a=2\int_{S^2_\infty}d\Omega\,\varepsilon_{abc}x^b\pi^{cd}\nu_d\,,
\end{equation}
where $\varepsilon_{abc}$ is as above.
\index{ADM!angular momentum}

Turning now to the boosts, we have 
\begin{equation}
\label{eq:BoostAlpha}
\alpha=u_ax^a+\alpha_{\rm gauge}\,.
\end{equation}
We need to repeat the same steps that previously led us to \eqref{eq:ScalarConstSurfTerm}. Now, according to \eqref{eq:ScalarDifferenceFirstOrder} the space 
integral over the divergence term in the variation of the scalar
curvature is 
\begin{equation}
\label{eq:DivergenceScalCurv-1}
X:=\int_\Sigma d^3x\,\alpha\sqrt{h}G^{abcd}D_aD_b\delta h_{cd}\,.
\end{equation}
A first integration by parts leads to 
\begin{equation}
\label{eq:DivergenceScalCurv-2}
\begin{split}
X
&=\int_{S^2_\infty}d\Omega\,\alpha\sqrt{h}\nu_aG^{abcd}D_b\delta h_{cd}\\
&-\int_\Sigma d^3x\,\sqrt{h}G^{abcd}(D_a\alpha)D_b\delta h_{cd}\,.
\end{split}
\end{equation}
One more integration by parts of the second term gives 
\begin{equation}
\label{eq:DivergenceScalCurv-3}
\begin{split}
X
&=\int_{S^2_\infty}d\Omega\,\alpha\sqrt{h}\nu_aG^{abcd}D_b\delta h_{cd}\\
&-\int_{S^2_\infty}d\Omega\,\sqrt{h}\,(D_a\alpha)\nu_b G^{abcd}\delta h_{cd}\\
&+\int_\Sigma d^3x\,\sqrt{h}G^{abcd}(D_aD_b\alpha)\delta h_{cd}\,.
\end{split}
\end{equation}
Equation \eqref{eq:BoostAlpha} implies that $D^2\alpha$ has 
$1/r^2$ fall-off with odd parity. Hence the last (volume) 
integral in \eqref{eq:DivergenceScalCurv-3} has $1/r^3$ fall-off 
with odd parity and hence converges. It gives rise to a term 
$\propto \sqrt{h}G^{abcd}D_aD_b\alpha$ in the Hamiltonian equation 
for ${\dot\pi}^{cd}$. According to the general strategy the 
surface integrals must be taken care of by adding suitable 
surface integrals to the right-hand side of 
\eqref{eq:GR-Hamiltonian} so as to just cancel $\signature/(2\kappa)$
times the integrals just found as resulting from the variation of 
the scalar curvature in \eqref{eq:GR-ConstraintsHamForm}. Hence the 
right surface terms to be added to \eqref{eq:GR-Hamiltonian} are 
of the form $u_aX^a$, where 
\begin{equation}
\label{eq:ADM-CenterOfMassSurfTerm}
\begin{split}
X^a&=\frac{-\signature}{2\kappa}\biggl\{
 \int_{S^2_\infty}d\Omega\,x^a\bigl(\partial_{b}h_{bc}-\partial_{c}h_{bb}\bigr)\nu_c\\
 &-\int_{S^2_\infty}d\Omega\,\bigl((h_{ab}-\delta_{ab})\nu_b-(h_{bb}-\delta_{bb})\nu_a\bigr)
\biggr\}\,.
\end{split}
\end{equation}
The coordinates $Z^a$ of the \emph{center of mass} are then 
defined by the rescaled forms of 
\eqref{eq:ADM-CenterOfMassSurfTerm}, with rescaling factor 
$E_{\rm ADM}$:
\index{ADM!center of mass} 
\begin{equation}
\label{eq:ADM-CenterOfMassCoordinates}
Z^a:=\frac{X^a}{E_{\rm ADM}}\,.
\end{equation}
In order to arrive at \eqref{eq:ADM-CenterOfMassSurfTerm} we wrote $\delta h_{ab}=\delta(h_{ab}-\delta_{ab})$ and left the difference 
$(h_{ab}-\delta_{ab})$ rather than just $h_{ab}$ under the integral 
in order to not keep the asymptotically constant term of \eqref{eq:ReggeTeitelboimFalloff-a} under 
the integral when pulling the variation $\delta$ outside 
it (cf.~\cite{Beig.Murchadha:1987} Appendix\,C). 

It has been shown in \cite{Corvino.Hu:2008} that the expression 
\eqref{eq:ADM-CenterOfMassSurfTerm} for the (unscaled) center of 
mass coincides with the geometric definition of Huisken and Yau's 
\cite{Huisken.Yau:1996}. The latter is defined by means of 
mean-curvature foliations of $\Sigma$.
\index{foliation!mean curvature}
Its relation to 
alternative definitions, including not only ADM but also a 
definition due to R.\,Schoen, using asymptotically conformal Killing 
fields, is discussed and lucidly summarized in \cite{Huang:2011}.

So far we have been working with the particular asymptotic 
conditions \eqref{eq:ReggeTeitelboimFalloff} and 
\eqref{eq:ReggeTeitelboimParity}. We have been arguing for 
existence of certain quantities to be identified with physical 
quantities of energy, linear and angular momentum, and center 
of mass. But what about uniqueness? All these quantities 
depend a priori on the choice of the asymptotic coordinates 
within the set of all coordinates satisfying the given 
fall-off conditions. Hence one needs to prove that this 
dependence is actually spurious and that, consequently, these 
quantities are geometric invariants. For the ADM mass and 
linear momentum this has been shown in \cite{Bartnik:1986}
and \cite{Chrusciel:1986}. Moreover, ignoring angular momentum 
and center of mass, these proofs were given under much weaker 
asymptotic conditions, in fact the weakest possible ones.
Regarding the latter, we recall that it was shown in 
\cite{Denisov.Solov'ev:1983} by means of explicit coordinate 
transformations that the expression can be made change its 
value if $\vert h_{ab}-\delta_{ab}\vert <r^{-\alpha}$ with 
$\alpha\leq 1/2$; see also the lucid discussion in 
\cite{Dain:SpringerHandbookSpacetime2014}. 
Hence we certainly need $\alpha>1/2$.  That this indeed
suffices to prove existence and uniqueness was established in \cite{Murchadha:1986,Bartnik:1986,Chrusciel:1986}.
This fits nicely with recent generalizations of stability 
results of Minkowski space by Bieri \cite{Bieri.Zipser:StabilityTheorem}, 
which work under the following asymtotic decay conditions, 
where $\alpha>1/2$:
\index{asymptotic!flatness according to Bieri} 
\begin{subequations}
\label{eq:BieriFallOff}
\begin{alignat}{2}
\label{eq:BieriFallOff-a}
&h_{ab}&&=\delta_{ab}+O_2(r^{-\alpha})\,,\\
\label{eq:BieriFallOff-b}
&\pi^{ab}&&=O_1(r^{-1-\alpha})\,.
\end{alignat}
\end{subequations}
These conditions suffice to establish ADM energy and 
linear momentum not only as being well defined, but also 
as being preserved under Hamiltonian evolution. 

At first sight \eqref{eq:BieriFallOff-a} might seem too weak to 
guarantee existence of \eqref{eq:ADM-EnergyMass}. The reason 
why it is not is, in fact, easy to see: If we convert 
\eqref{eq:ADM-EnergyMass} into a bulk integral using Gauss' 
theorem, the integrand contains a combination of 2nd derivatives
of $h$ which just form the 2nd derivative part of the scalar 
curvature. Using the scalar constraint, which schematically
has the form
\begin{equation}
\label{eq:SchematicScalarConstraint}
\pi^2+\partial^2h+(\partial h)^2=0\,,
\end{equation}
this can be written as a bulk integral containing only 
integrands of the form $\propto \pi^2$ and $\propto (\partial h)^2$,
which according to \eqref{eq:BieriFallOff-a} fall off like  
$1/r^{2(1+\alpha)}$, i.e., faster than $1/r^3$. Hence the bulk 
integral converges. But note that the conditions \eqref{eq:BieriFallOff} 
do not suffice to ensure the existence of conserved quantities 
regarding angular momentum or center of mass. Alternative conditions 
to \eqref{eq:ReggeTeitelboimFalloff} and \eqref{eq:ReggeTeitelboimParity} 
for the existence of angular momentum have been discussed in 
\cite{Ashtekar.Magnon:1979} and \cite{Chrusciel:1987}.   

We recall that even in the context of the Regge-Teitelboim conditions  
\eqref{eq:ReggeTeitelboimFalloff} and \eqref{eq:ReggeTeitelboimParity}
we needed to invoke the fact that $(h,\pi)$ satisfy the constraints
in order to conclude sufficiently strong fall-offs. This we have 
already seen explicitly in the discussion on, e.g.,  the existence 
of angular momentum in the paragraph above 
equation \eqref{eq:ADM-LinearMomentum}. From the scalar constraint 
\eqref{eq:SchematicScalarConstraint} we now learn that 
\eqref{eq:ReggeTeitelboimFalloff} and \eqref{eq:ReggeTeitelboimParity}
implies a $1/r^4$ fall-off for $G^{abcd}\partial_a\partial_bh_{cd}$, 
and not just $1/r^3$ as naively  anticipated from 
\eqref{eq:ReggeTeitelboimFalloff}. 

Alternative expressions for mass/energy exist in cases of 
symmetries. For example, for asymptotically flat and 
\emph{stationary} solutions to Einstein's equations, 
the ADM mass $M_{\rm ADM}$ is known to coincide with the 
so-called Komar mass~\cite{Komar:1959}, 
\index{Komar mass}
\index{mass!Komar}
whose simple and coordinate invariant expression is
\begin{equation}
\label{eq:KomarMass}
M_{\rm Komar}=\frac{-\signature}{c^2\kappa}\int_{S^2_\infty}\star dK^\flat\,.
\end{equation}
Here $K$ is the timelike Killing vector field so normalized that 
$\lim_{r\rightarrow\infty}g(K,K)=\signature$. There exist various 
proofs in the literature showing $M_{\rm Komar}=M_{\rm ADM}$; see, e.g.,  
\cite{Beig:1978}\cite{Ashtekar.Magnon:1979}\cite{Chrusciel:1986}
and Theorem~4.13 of \cite{Choquet-Bruhat:GR}.  

Since the Komar mass 
\index{Komar mass}
\index{mass!Komar}
is frequently used in applications, 
let us say a few things about it. From a mathematical point of 
view the main merit of \eqref{eq:KomarMass} is that it allows to 
associate a ``mass'' to any 2-dimensional submanifold $S\in M$, 
independently of any choice of coordinates. We call it the 
Komar mass of $S$:
\index{Komar mass}
\index{mass!Komar} 
\begin{equation}
\label{eq:KomarMassLocal}
M_{\rm Komar}(S)=\frac{-\signature}{c^2\kappa}\int_S\star dK^\flat\,.
\end{equation}
For $S\rightarrow S^2_\infty$ its interpretation is that of the 
ADM mass, whose physical significance as the value of the 
Hamiltonian (divided by $c^2$) endows it with a sound physical 
interpretation. But what might the interpretation be for 
general $S$? Well, suppose $S=\partial B$, where $B$ is a 
3-dimensional spacelike submanifold of $M$. Then, by Stokes' 
theorem: 
\begin{equation}
\label{eq:KomarMassLocalBulk}
\begin{split}
M_{\rm Komar}(S)
&=\frac{-\signature}{c^2\kappa}\int_B d\star dK^\flat\\
&=\frac{1}{c^2\kappa}\int_B \star(\star d\star dK^\flat)\\
&=\frac{\signature}{c^2\kappa}\int_B \star(\nabla\cdot dK^\flat)\\
&=\frac{-2\signature}{c^2\kappa}\int_B \star i_k\Ricci\,.
\end{split}
\end{equation}
Here we used the general identity for the square of the Hodge star
restricted to $p$-forms in $n$ dimensions,
\begin{equation}
\label{eq:HodgeSquared}
\star\circ\star\big\vert_{\Lambda^p(M)}=\signature\,(-1)^{p(n-p)}\
\Id_{\Lambda^p(M)}\,,
\end{equation}
and also the general formula that allows to express
$\star d\star$ in terms of the covariant divergence 
$\nabla\cdot$ on the first index with respect to the 
Levi-Civita connection,
\begin{equation}
\label{eq:Star_d_star}
\star\circ d\circ\star\big\vert_{\Lambda^p(M)}
=\signature\,(-1)^{n(p+1)}\
\nabla\cdot\,.
\end{equation}
In the final step of \eqref{eq:KomarMassLocalBulk} we used that 
any Killing vector-field satisfies the identity (again for the 
Levi-Civita connection): 
\begin{equation}
\label{eq:KillingRiemannIdenity}
\nabla_a\nabla_bK_c=R^d_{\phantom{d}abc}K_d\,.
\end{equation}
If the spacetime satisfies Einstein's equation we can use 
\eqref{eq:EinsteinTensorEq-Alt} to eliminate the Ricci tensor 
in the last line of \eqref{eq:KomarMassLocal} in favor of 
the energy-momentum tensor. This shows that if $S$ has two 
connected components $S_1$ and $S_2$, and if $\EMT\vert_B=0$,
then (choosing the relative orientations of  $S_1$ and $S_2$ 
appropriately) $M_{\rm Komar}(S_1)=M_{\rm Komar}(S_2)M$. For a 
finite-size star or a black hole this means that we may take 
any $S$ to calculate the Komar- and hence the ADM mass, as 
long as $S\cup S^2_\infty$ bounds a 3-dimensional region $B$ 
on which $\EMT$ vanishes. In particular, $S$ may be taken as 
any 2-sphere outside the star's surface. More specifically, 
consider a static star where $K$ is the hypersurface orthogonal 
Killing vector-field. The topology of the hypersurfaces 
orthogonal to $K$ inside the star shall be just that of a ball in 
$\reals^3$; then we express the star's ADM energy by the 
Komar integral over the star's surface $S$ and that, in turn,
by the bulk integral \eqref{eq:KomarMassLocalBulk} over the 
star's interior, where we replace the Ricci tensor by $\EMT$. 
This results in the so-called Tolman mass 
\index{Tolman mass}
\index{mass!Tolman}
(see \cite{Tolman:1930a} and \S\,92 of 
\cite{Tolman:RelThermCosm}), which in our notation 
reads: 
\begin{equation}
\label{eq:TolmanMass}
\begin{split}
& M_{\rm Tolman}=
\frac{-\signature}{c^2}\int_b d^3x\ \sqrt{\det(h)}\ \times\\
& \sqrt{\signature g(K,K)}\  
\Bigl[\EMT(n,n)-\signature\Trace_h(\EMT)\Bigr]\,.
\end{split}
\end{equation}
Here $n:=K/\sqrt{\epsilon g(K,K)}$ is the normal to the hypersurfaces
and $\Trace_h$ the trace with respect to the spatial metric $h$,
where, we recall, $g=\signature\,n^\flat\otimes n^\flat+h$.  In the 
Lorentzian case ($\varepsilon=-1$) we see that the first 
$T(n,n)$-term in \eqref{eq:TolmanMass} is just the integral over the 
spatial energy-density of the matter divided by $c^2$ and weighted 
by the redshift factor $\sqrt{-g(K,K)}$. The additional term is 
absent if the pressures are negligible compared to the energy 
density, but this need not be the case. For example, if $\EMT$ is 
that of an electromagnetic field, we have 
$\Trace_g(\EMT)=\signature T(n,n)+\Trace_h(\EMT)=0$, so that 
the pressure effectively doubles the contribution of the first term 
to the overall mass. This is the origin of the infamous 
``factor-2-anomaly'' of the Komar mass, which, e.g., leads to the 
result that the difference between two Komar masses evaluated on 
two different 2-spheres of spherical symmetry in the 
Reissner-Nordstr{\o}m manifold (electrically charged black-hole)
gives \emph{twice} the electrostatic field energy stored in 
the region between the spheres. On the other hand, for a spherically 
symmetric perfect-fluid star, Tolman has shown in \cite{Tolman:1930a}
that in a weak-field approximation the leading-order difference 
between \eqref{eq:TolmanMass} and the integrated mass-density 
of the fluid is just the Newtonian binding energy, which makes 
perfect sense.  

At this point we should mention the positive-mass theorem (for Lorentzian 
signature $\signature =-1$), which states that for any pair $(h,\pi)$ of 
initial data satisfying the constraints $M_{\rm ADM}\geq 0$, with 
equality only if the data are that of Minkowski space. 
Note that the expression \eqref{eq:ADM-EnergyMass} for $M_{\rm ADM}$ is a 
functional of $h$ alone, but that in the formulation of the 
positive-mass theorem given here it is crucial that for $h$ 
there exists a $\pi$ so that the pair $(h,\pi)$ solves the 
constraint.
\index{constraints!solve (methods, meaning)}
 Otherwise it is easy to write down 3-metrics 
with negative ADM mass; take e.g. \eqref{eq:OneHole-Metric} 
(see below) with $r_0$ replaced by $-r_0$, where $r_0>0$, 
suitably smoothed out for smaller radii so as to avoid the 
singularity at $r=r_0$. Since 
the ADM mass only depends on the asymptotic behavior it is 
completely independent of any alterations to the metric in 
the interior. If one wishes to make the positive mass theorem 
a statement about metrics alone without any reference to the 
constraints one has to impose positivity conditions on the 
scalar curvature. But that also imposes topological restrictions 
due to the result of Gromov and Lawson \cite{Gromov.Lawson:1983}
mentioned at the end of Section\,\ref{sec:DecEinstEq}. 
For a recent up-to-date survey on the positive-mass theorem 
we refer to \cite{Dain:SpringerHandbookSpacetime2014}. 

Note that $M_{\rm ADM}=M_{\rm Komar}$
implies the positivity of $M_{\rm Komar}\equiv M_{\rm Komar}(S^2_\infty)$.
But this does not imply that $M_{\rm Komar}(S^2)$ is also positive 
for general $S^2$. In fact, explicit examples of regular, 
asymptotically flat spacetimes with matter satisfying the 
hypotheses of the positive-mass theorem are known in which 
 $M_{\rm Komar}(S^2)<0$ for suitably chosen 
2-spheres~\cite{Ansorg.Petroff:2006}. The recipe here is 
to regard two concentric counter-rotating objects in an 
axially-symmetric and stationary spacetime, e.g., 
an outside perfect-fluid ring and an inner rigidly rotating 
disk of dust. The Komar mass of the disk (i.e. $S^2$ encloses 
the disk but not the ring) may then turn out negative if 
the frame-dragging effect of the ring is large enough so as 
to let the angular velocity and the (Komar) angular 
momentum of the central object have \emph{opposite} signs.  

In passing we remark that the positive mass theorem in combination 
with the equality $M_{\rm ADM}=M_{\rm Komar}$ gives a simple proof 
of the absence of ``gravitational solitons'', i.e. stationary 
asymptotically flat solutions to Einstein's equations on 
$\Sigma=\reals^3$. This follows from \eqref{eq:KomarMass} and 
$d\star dK^\flat\propto \star i_K\Ricci$. The vaccum equation $\Ricci=0$
then implies $M_{\rm ADM}=M_{\rm Komar}=0$ which implies that 
spacetime is flat Minkowski. This theorem was originally shown 
for static spacetimes (i.e. hypersurface orthogonal $K$) by 
Pauli and Einstein \cite{Einstein.Pauli:1943} and later generalized
to the stationary case by Lichnerowicz 
\cite{Lichnerowicz:TheoriesRelativistes1955}. The result of this 
theorem cannot be circumvented by trying more complicated topologies 
for $\Sigma$. As soon as $\Sigma$ becomes non-simply connected 
(which in view of the validity of the Poincar\'e conjecture will 
be the case for any one-ended manifold other than $\reals^3$)
we know from Gannon's theorem \cite{Gannon:1975} that the evolving 
spacetime will inevitably develop singularities.

Finally we mention that under suitable fall-off conditions we can 
find the Poincar\'e group as 
\emph{asymptotic symmetry group}~\cite{Beig.Murchadha:1987}.
\index{asymptotic!symmetry group}
\index{symmetry group, asymptotic} 
It will emerge from \eqref{eq:H-Rep} as equivalence classes of all
hypersurface deformations, including those in which $\alpha$ and 
$\beta$ asymptotically approach rigid translations, rotations, or 
boosts. The quotient is taken with respect to those deformations
which are generated by the constraints, in which $\alpha$ and 
$\beta$ tend to zero at spatial infinity. There are various 
subtleties and fine tunings involved for the precise fall-off
conditions that are necessary in order to exactly obtain a 
10-dimensional symmetry as a quotient of two infinite-dimensional
objects. This is particularly true for asymptotic boosts, for which 
one needs to tilt the hypersurface, corresponding to asymptotic 
lapse functions $\alpha\propto r$. (Boosted hypersurfaces are known 
to exist in the development of asymptotically flat initial data 
\cite{Christodoulou.Murchadha:1981}.) 
But leaving the analytic details aside, the qualitative picture is 
quite generic for gauge field theories with long-ranging field 
configurations~\cite{Giulini:1995d}: A proper physical symmetry group 
arises as quotient of a general covariance group with respect to a 
proper normal subgroup, the latter being defined to be that object 
that is generated by the constraints.

\section{Black-Hole data}
\label{sec:BH-Data}
\begin{figure}[htb]
\centering\includegraphics[width=0.98\linewidth]{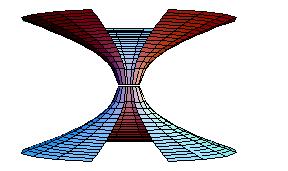}
\put(-170,61){$r=r_0\,\rightarrow$}
\caption{\footnotesize\label{fig:FigOneHole}
Cauchy surface with time symmetric initial data and two 
isometric asymptotically flat ends separated by a totally geodesic 
2-sphere.}
\end{figure}

In this section we discuss some simple solutions to the vacuum 
Einstein equations without cosmological constant. We first specify 
to the simplest case of time symmetric conformally flat data. 
\index{time symmetric data}
\index{conformally flat data}
Time symmetry means that the initial extrinsic curvature vanishes, 
$K=0$. The corresponding Cauchy surface will then be totally 
geodesic in the spacetime that emerges from it. The vector 
constraint \eqref{eq:EinsteinEquation-0a} is identically satisfied
and the scalar constraint \eqref{eq:EinsteinEquation-00}
reduces to scalar flatness
\begin{equation}
\label{eq:ScalarFlatness}
R(h)=\Scalar^D=0\,.
\end{equation}
Conformal flatness means that 
\begin{equation}
\label{eq:ConformalFlatness}
h=\Omega^4\,\delta\,,
\end{equation}
where $\delta$ is the flat metric. From \eqref{eq:ConfTransScal-a}
we infer that \eqref{eq:ScalarFlatness} is equivalent to 
$\Omega$ being harmonic
\begin{equation}
\label{eq:OmegaHarmonic} 
\Delta_\delta\Omega=0
\end{equation}
where $\Delta_\delta$ is the Laplacian with respect to the 
flat metric $\delta$. We seek solutions $\Omega$ which are 
asymptotically flat for $r\rightarrow\infty$ and give rise 
to complete manifolds in the metric structure defined by $g$.
The only spherically symmetric such solution is 
\begin{equation}
\label{eq:OneHole-Omega}
\Omega(r)=1+\frac{r_0}{r}\,,
\end{equation}
where the integration constant $r_0$ can be related to the 
ADM mass \eqref{eq:ADM-EnergyMass} by 
\begin{equation}
\label{eq:ADM-Radius} 
M_{\rm ADM}=2c^2 r_0/G\,.
\end{equation}
This solution is defined on $\Sigma=\reals^3-\{0\}$. The metric on 
$\Sigma$ so obtained is 
\begin{equation}
\label{eq:OneHole-Metric}
h=\left(1+\frac{r_0}{r}\right)^4\bigl(dr^2+r^2(d\theta^2+\sin^2\theta\,d\varphi^2)\bigr)\,.
\end{equation}
It admits the following isometries
\begin{subequations}
\label{eq:OneHole-Isometries}
\begin{alignat}{2}
\label{eq:OneHole-Isometries-a}
&I_1(r,\theta,\varphi)&&:=(r_0^2/r,\theta,\varphi)\,,\\
\label{eq:OneHole-Isometries-b}
&I_2(r,\theta,\varphi)&&:=(r_0^2/r,\pi-\theta,\varphi+\pi)\,.
\end{alignat}
\end{subequations}
Note that the second is just a composition of the first with 
the antipodal map 
$(r,\theta,\varphi)\mapsto (r,\pi-\theta,\varphi+\pi)$
which is well defined on $\reals^3-\{0\}$. This makes $I_2$
a fixed-point free action. The fixed-point set of $I_1$ 
is the 2-sphere $r=r_0$. Note that generally a submanifold 
that is the fixed-point set of an isometry is necessarily 
totally geodesic (has vanishing extrinsic curvature). To see 
this, consider  a geodesic that starts on and tangentially to 
this submanifold. Such a geodesic cannot leave the submanifold, 
for if it did we could use the isometry to map it to a different 
geodesic with identical initial conditions, in contradiction 
to the uniqueness of solutions for the geodesic equation. 
Hence the 2-sphere $r=r_0$ has vanishing extrinsic curvature 
and is therefore, in particular, a minimal surface (has 
vanishing trace of the extrinsic curvature).  The geometry 
inside the sphere $r=r_0$ is isometric to that outside it. 
This is depicted in Fig.\,\ref{fig:FigOneHole}.

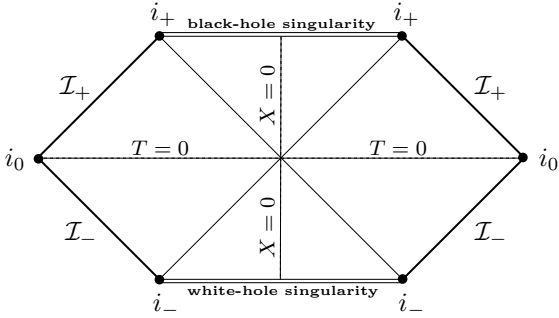
\begin{figure}[htb]
\centering\includegraphics[width=.85\linewidth]{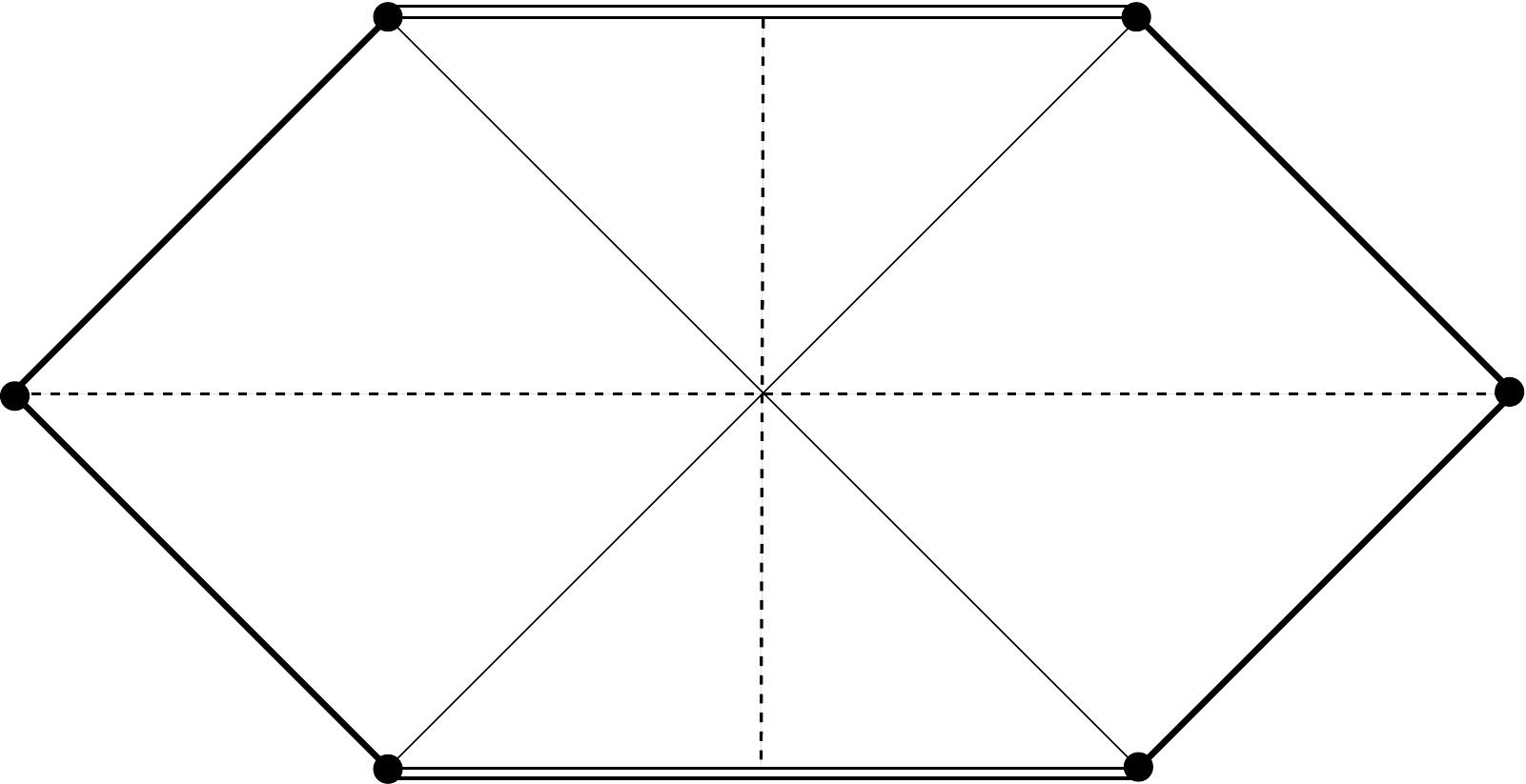}
\put(4,45){$i_0$}
\put(-198,45){$i_0$}
\put(-51,100){$i_+$}
\put(-143,100){$i_+$}
\put(-49,-10){$i_-$}
\put(-142,-10){$i_-$}
\put(-20,72){$\mathcal{I}_+$}
\put(-177,72){$\mathcal{I}_+$}
\put(-20,17){$\mathcal{I}_-$}
\put(-175,17){$\mathcal{I}_-$}
\put(-129,97){\tiny\textbf{black-hole singularity}}
\put(-129,-4){\tiny\textbf{white-hole singularity}}
\put(-60,49){\footnotesize $T=0$}
\put(-150,49){\footnotesize $T=0$}
\put(-96,10){\begin{rotate}{90}\footnotesize $X=0$\end{rotate}}
\put(-96,60){\begin{rotate}{90}\footnotesize $X=0$\end{rotate}}
\caption{\footnotesize\label{fig:FigKruskalConf}
Conformally compactified Kruskal spacetime. The $T$ axis 
points up vertically, the $X$ axis horizontally to the right.  
The Cauchy surface of Fig.\,\ref{fig:FigOneHole} corresponds 
trio the hypersurface $T=0$. The various infinities are: 
$i_0$ spacelike, $i_\pm$ future/past timelike, and $\mathcal{I}_\pm$ 
future/past lightlike infinity. The right diamond-shaped region 
corresponds to the usual exterior Schwarzschild solution 
containing one asymptotically flat end.}
\end{figure}

For the data $(h=\eqref{eq:OneHole-Metric}\,,\,K=0)$ on 
$\Sigma=\reals^3-\{0\}$ we actually know its maximal 
time evolution: It is the Kruskal 
spacetime~\cite{Kruskal:1960}\cite{Hawking.Ellis:TLSSOS}
\index{Kruskal spacetime}
which maximally extends the exterior Schwarzschild spacetime.
\index{Schwarzschild spacetime}
Figure\,\ref{fig:FigKruskalConf} shows a conformal diagram of 
Kruskal spacetime.

In Kruskal coordinates (Kruskal~\cite{Kruskal:1960} uses $(v,u)$, 
Hawking Ellis \cite{Hawking.Ellis:TLSSOS} $(t',x')$ for what we 
call $(T,X)$) $(T,X,\theta,\varphi)$, where $T$ and $X$ 
each range in $(-\infty,\infty)$ obeying $T^2-X^2<1$, the Kruskal 
metric reads (as usual, we write 
$d\Omega^2$ for $d\theta^2+\sin^2\theta\,d\varphi^2$): 
\begin{equation}
\label{eq:KruskalMetric1}
g=\frac{8 r_0^2}{r}\,\exp(-r/r_0)\,\bigl(-dT^2+dX^2\bigr)+r^2d\Omega^2\,,
\end{equation}
where $r$ is a function of $T$ and $X$, implicitly 
defined by 
\begin{equation}
\label{eq:KruskalMetric2}
\bigl((r/r_0)-1\bigr)\,\exp(r/r_0)=X^2-T^2\,.
\end{equation}
The metric is spherically symmetric and allows for the additional 
Killing field
\begin{equation}
\label{eq:KruskalKilling}
K=\bigl(X\partial_T+T\partial_X\bigr)\,,
\end{equation}
which is timelike for $\vert X\vert>\vert T\vert$ and spacelike for 
$\vert X\vert<\vert T\vert$. 

Both maps \eqref{eq:OneHole-Isometries} extend to the Kruskal manifold. 
The fixed-point free action  \eqref{eq:OneHole-Isometries-b} has the 
extension  
\begin{equation}
\label{eq:KrskalZ2Isometry}
J:(T,X,\theta,\varphi)\mapsto (T,-X,\pi-\theta,\varphi+\pi)\,.
\end{equation}
It generates a freely acting group $\mathbb{Z}_2$ of smooth 
isometries which preserve space- as well as time-orientation. 
Hence the quotient is a smooth space- and time-orientable manifold, 
that is sometimes called the $\RP^3$-\emph{geon}.
\index{geon}%
It represents the maximal time evolution 
of the data $(h=\eqref{eq:OneHole-Metric},K=0)$ as above, 
but now defined on the initial quotient manifold 
$\Sigma=(\reals^3-\{0\})/I_2$. It has only one asymptotically 
flat end and the topology of a once punctured real projective 
space $\RP^3$. Note that the map $J$ preserves the Killing 
field \eqref{eq:KruskalKilling} only up to sign. Had one chosen 
$J':(T,X,\theta,\varphi)\mapsto(-T,-X,\pi-\theta,\varphi+\pi)$
as in \cite{Misner.Wheeler:1957} and \cite{Gibbons:1986}, one would 
have preserved $K$ but lost time orientability. 

Within the set of conformally-flat and time symmetric initial 
data we can easily generalize the solution \eqref{eq:OneHole-Omega} 
to \eqref{eq:OmegaHarmonic} to include more than one monopole term 
on a multi-punctured $\reals^3$. For two terms we get  
\begin{equation}
\label{eq:TwoHole-Omega}
\Omega(r)=1+\frac{a_1}{r_1}+\frac{a_2}{r_2}\,,
\end{equation}
where $r_i=\Vert\vec x-\vec c_i\Vert$. This represents two black 
holes without spin and orbital angular momentum momentarily at 
rest, with $\vec c_i\in\reals^3$ representing the hole's ``positions''.
The manifold has three ends, one for $r\rightarrow\infty$ 
and one each for $r_i\rightarrow 0$. For each end we can calculate 
the ADM mass and get 
\begin{subequations}
\label{eq:TwoHoleMasses}
\begin{alignat}{2}
\label{eq:TwoHoleMasses-a}
&M  &&=2(a_1+a_2)c^2/G\,,\\
\label{eq:TwoHoleMasses-b}
&M_1&&=2\left(a_1+\frac{a_1a_2}{r_{12}}\right)\frac{c^2}{G}\,,\\
\label{eq:TwoHoleMasses-c}
&M_2&&=2\left(a_2+\frac{a_1a_2}{r_{12}}\right)\frac{c^2}{G}\,,
\end{alignat}
\end{subequations}
where $r_{12}:=\Vert\vec c_1-\vec c_2\Vert$.
Here $M$ is the total mass associated with the end 
$r\rightarrow\infty$ and $M_i$ is the individual hole 
mass associated with the end $r_i\rightarrow 0$.
The binding energy is the overall energy minus the individual 
ones. One obtains 
\begin{equation} 
\label{eq:BindingEnergy}
\Delta E:=(M-M_1-M_2)c^2=-G\frac{M_1M_2}{r_{12}}+\cdots
\end{equation}
where the dots stand for corrections of quadratic and higher 
powers in $GM_i/c^2r_{12}$. This can be easily generalized to 
any finite number of poles. Note that the initial manifolds
are all complete, i.e. all punctures lie at infinite metric 
distance from any interior point.  
 
Other generalizations consist in adding linear and angular momentum. 
This can be done using the \emph{conformal method}, 
\index{conformal method}
which we now briefly describe. 
Recall that we wish to solve the constraints \eqref{eq:Constraints-1} 
for $\EMT=0$ but now with $K\ne 0$. 
\index{constraints!solve (methods, meaning)}
Encouraged by previous experience 
with the simplifying effect of conformal transformations, we now 
study the general conformal transformation properties of the 
left-hand sides of \eqref{eq:Constraints-1}. 
Generalizing \eqref{eq:ConformalFlatness}, we write
\begin{subequations}
\label{eq:ConfTransGen}
\begin{alignat}{2}
\label{eq:ConfTransGen-a}
&h_{ab}&&=\Omega^4{\bar h}_{ab}\,,\\
\label{eq:ConfTransGen-b}
&K^{ab}&&=\Omega^{-s}{\bar K}^{ab}\,.
\end{alignat}
Note that in view of \eqref{eq:ConfTransGen-a} the second equation 
is equivalent to 
\begin{equation}
\label{eq:ConfTransGen-c}
K_{ab}=\Omega^{8-s}{\bar K}_{ab}\,.
\end{equation}
\end{subequations}
We first wish to determine the power $s$ that is most suitable
for simplifying \eqref{eq:EinsteinEquation-0a} if written in 
terms of $\bar h$ and $\bar K$.  A slightly lengthy but 
straightforward computation gives
\begin{equation}
\label{eq:ConfTransVectConstr}
\begin{split}
D_a\bigl(K^{ab}-h^{ab}K^c_c\bigr)
&=\Omega^{-s}{\bar D}_a\bigl({\bar K}^{ab}-{\bar h}^{ab}{\bar K}^c_c\bigr)\\
&+(10-s)\Omega^{-(s+1)}({\bar D}_a\Omega){\bar K}^{ab}\\
&+(s-6)\Omega^{-(s+1)}{\bar D}^b{\bar K}^c_c\,.
\end{split}
\end{equation}
Here $\bar D$ is the Levi-Civita covariant derivative with respect 
to $\bar h$ and indices on barred quantities are moved with the 
barred metric. A suitable simplification in the sense of conformal 
covariance would occur if only the first line in 
\eqref{eq:ConfTransVectConstr} survived. The other two lines cannot 
be made to vanish simultaneously on account of a suitable choice of 
$s$. The best one can do is to choose $s=10$ and restrict to traceless 
$\bar K$, i.e. ${\bar K}^c_c=0$. From \eqref{eq:ConfTransGen} this 
might seem as if we had to restrict to traceless $K$. But note that 
as the left-hand side of \eqref{eq:EinsteinEquation-0a} is linear 
in $K$ we can always add to any traceless solution $K^{(1)}$ a pure
trace part 
\begin{equation}
\label{eq:PureTraceK}
K^{(2)}_{ab}=\tfrac{1}{3}\tau h_{ab}\,,
\end{equation}
which satisfies the vector constraint as long as $\tau$ is 
constant. 
Putting all this together we see that we get a solution to the 
vector constraint if we maintain \eqref{eq:ConfTransGen-a} 
but replace \eqref{eq:ConfTransGen-b} with 
\begin{subequations}
\label{eq:ConfMethodExtCurv}
\begin{alignat}{3}
\label{eq:ConfMethodExtCurv-a}
&K^{ab}&&\,=
\,\Omega^{-10}\bar K^{ab}&&+\,\tfrac{1}{3}\Omega^{-4}\,{\bar h}^{ab}\,\tau\,,\\
\label{eq:ConfMethodExtCurv-b}
&K_{ab}&&\,=
\,\Omega^{-2} \bar K_{ab}&&+\,\tfrac{1}{3}\Omega^4\, {\bar h}_{ab}\,\tau\,,
\end{alignat}
\end{subequations}
where $\tau$ is a constant and ${\bar K}$ is transverse traceless 
in the metric $\bar h$:  
\begin{subequations}
\label{eq:TransverseTraceless}
\begin{alignat}{1}
\label{eq:TransverseTraceless-a}
{\bar h}_{ab}{\bar K}^{ab}&=0\,,\\
\label{eq:TransverseTraceless-b}
{\bar D}_a{\bar K}^{ab}&=0\,.
\end{alignat}
\end{subequations}
Note that $K^a_a=\tau$ so that the method as presented here only 
produces initial data of constant mean curvature.
\index{mean curvature, constant} 
It can be generalized to non-constant $\tau$, see e.g. 
\cite{Isenberg:SpringerHandbookSpacetime2014}. 

As before, the idea is now to let the remaining scalar 
constraint determine the conformal factor. 
Inserting  \eqref{eq:ConfTransGen-a} and \eqref{eq:ConfMethodExtCurv} 
into the scalar constraint \eqref{eq:EinsteinEquation-00} 
and using \eqref{eq:ConfTransScal}, we obtain the following 
elliptic \emph{York equation} for $\Omega$
\index{York!equation}
\begin{equation}
\label{eq:YorkEquation}
-\signature\Bigl(\Delta_{\bar h}-\tfrac{1}{8}\Scalar^{\bar D}\Bigr)\Omega
+\tfrac{1}{8}\Omega^{-7}\bar K_{ab}\bar K^{ab}
-\tfrac{1}{12}\Omega^5\tau^2=0\,.
\end{equation}
Existence and  uniqueness of this equation for the Lorentzian 
case $\signature=-1$ is discussed in the survey 
\cite{Isenberg:SpringerHandbookSpacetime2014}. 
It may be further simplified if, as before,  we assume conformally 
flat intitial data, i.e. 
\begin{equation}
\label{eq:SpatialFlatMetric}
\bar h=\delta=\text{flat metric}\,.
\end{equation}
Then  
\begin{equation}
\label{eq:YorkEquationFlat}
-\signature\Delta_\delta\Omega
+\tfrac{1}{8}\Omega^{-7}\bar K_{ab}\bar K^{ab}
-\tfrac{1}{12}\Omega^5\tau^2=0\,.
\end{equation}
where $\bar K$ is now transverse-traceless with respect to the 
flat connection (partial derivatives in suitable coordinates).   

It is remarkable that the ADM momenta \eqref{eq:ADM-Momentum} can 
be calculated without knowing $\Omega$. Hence we can parametrize 
solutions to \eqref{eq:TransverseTraceless} directly by the 
momenta without solving \eqref{eq:YorkEquation} first.
Two solutions of particular interest for $\bar h=\delta$ are 
the 
\emph{Bowen-York data}~\cite{Bowen.York:1980}\cite{York:1979}.
\index{Bowen York initial data}
\index{initial data, Bowen York} 
In Cartesian coordinates and corresponding components they read 
\begin{subequations}
\label{eq:BowenYorkData}
\begin{alignat}{2}
\label{eq:BowenYorkData-a}
&{\bar K}^{(1)}_{ab}
&&=r^{-2}\bigl(\nu_a A_b+\nu_bA_a-(\delta_{ab}-\nu_a\nu_b)\nu^cA_c\bigr)\,,\\
\label{eq:BowenYorkData-b}
&{\bar K}^{(2)}_{ab}
&&=r^{-3}\bigl(\nu_a\varepsilon_{bcd}+\nu_b\varepsilon_{acd}\bigr)B^c\nu^d\,,
\end{alignat}
\end{subequations}
where $\nu^a:=x^a/r$ and where all indices are raised and 
lowered with the flat metric $\delta$. $A$ and $B$ are covariantly 
constant vector fields with respect to the Levi Civita connection 
for the flat metric $\delta$, which are here represented by constant components $A^b$ and $B^c$. One verifies by direct computation that they satisfy \eqref{eq:TransverseTraceless} with ${\bar h}_{ab}=\delta_{ab}$ and 
${\bar D}_a=\partial_a$. Furthermore, using \eqref{eq:ADM-Momentum} 
one shows that \eqref{eq:BowenYorkData-a} has vanishing angular 
momentum and a linear momentum with components 
\begin{equation}
\label{eq:BowenYorkLinMom-1}
P^a=\frac{2c^3}{3G}A^a\,,
\end{equation}
whereas \eqref{eq:BowenYorkData-b} has vanishing linear momentum and an 
angular momentum with components 
\begin{equation}
\label{eq:BowenYorkLinMom-2}
J^a=\frac{c^3}{3G}B^a\,.
\end{equation}
They can be combined to give data for single holes with non-zero 
linear and angular momenta and also be superposed in order to give 
data for multi black-hole configurations. Such data, and certain 
modifications of them,  form the essential ingredient for present-day 
numerical simulations of black hole scattering and the subsequent 
emission of gravitational radiation. 

Let us now return to equation \eqref{eq:YorkEquation}, which 
we have to solve once suitable expressions for the components 
${\bar K}_{ab}$ have been found. As the metric is flat at the 
end representing spatial infinity, i.e. $\Omega(r\rightarrow\infty)=1$, 
it is  clear that $\Omega$ cannot be bounded in the interior 
region. The idea of the \emph{puncture method}\index{puncture method}, 
first proposed in \cite{Brandt.Bruegmann:1997}, is to restrict 
the type of singularities of $\Omega$ to be, in some sense, as 
simple as possible, which here means to be of the pure monopole 
type that we already encountered in \eqref{eq:OneHole-Omega} for 
a single hole and generalized to two holes in \eqref{eq:TwoHole-Omega}. 
This amounts to the following: Take 
$\Sigma=\reals^3-\{\vec c_1,\cdots,\vec c_n\}$ and assume we 
are given suitable $\bar K_{ab}$ which are regular in $\Sigma$, e.g., 
a sum of York data of the form \eqref{eq:BowenYorkData}, each 
centered at one of the punctures $\vec c_i$. Accordingly, we write  
\begin{subequations}
\label{eq:PunctureAnsatz}
\begin{equation}
\label{eq:PunctureAnsatz-a}
\Omega=u+\frac{1}{\omega}\,,
\end{equation}
where
\begin{equation}
\label{eq:PunctureAnsatz-b}
\frac{1}{\omega}:=\sum_{i=1}^n\frac{a_i}{r_i}\,,
\end{equation}
\end{subequations}
again with $r_i:=\Vert\vec x-\vec c_i\Vert$. The crucial 
assumption now is that the function $u$ is smooth, at least $C^2$, 
on \emph{all} of $\reals^3$, \emph{including} the points $\vec c_i$.
Since $1/\omega$ is annihilated by $\Delta_\delta$, 
\eqref{eq:YorkEquation} applied to \eqref{eq:PunctureAnsatz}
then leads to a second order elliptic differential equation 
for $u$. In the simple case of conformally flat maximal 
data, i.e. $\bar h_{ab}=\delta_{ab}$ and $\tau=0$, we get 
in the Lorentzian case ($\signature=-1$)
\begin{equation}
\label{eq:PunctureEquation}
\Delta_\delta u
=-\,\tfrac{1}{8}\bigl(1+\omega u\bigr)^{-7}\,
 \omega^7\bar K_{ab}\bar K^{ab}\,,
\end{equation} 
where $u\rightarrow 1$ at spatial infinity. 
Now, at the $i$-th puncture, $\bar K_{ab}$ diverges as $(1/r_i)^2$ 
for the data \eqref{eq:BowenYorkData-a} and as $(1/r_i)^3$ 
for the data \eqref{eq:BowenYorkData-b}. This means that  
$\bar K_{ab}\bar K^{ab}$ diverges at most as $(1/r_i)^6$. 
But from~\eqref{eq:PunctureAnsatz-b} we see that $\omega$ 
vanishes as $r_i$ at $\vec c_i$ so that 
$\omega^7\bar K_{ab}\bar K^{ab}$ also vanishes at least as fast
as $r_i$ at $\vec c_i$ and is hence continuous on all of 
$\reals^3$. Standard elliptic theory now allows to conclude 
existence and uniqueness of $C^2$ solutions to 
\eqref{eq:PunctureEquation}; see \cite{Brandt.Bruegmann:1997}
for more details. It is then not difficult to see that the 
Riemannian manifold $(\Sigma,h)$ so obtained has $n+1$ 
asymptotically flat ends whose ADM masses are readily calculated. 
Similar to \eqref{eq:TwoHoleMasses},
one obtains 
\begin{subequations}
\label{eq:PunctureMasses}
\begin{alignat}{1}
\label{eq:PunctureMasses-a}
M&\,=\,\frac{2c^2}{G}\,\sum_{i=1}^n a_i\,,\\
\label{eq:PunctureMasses-b}
M_i&\,=\,\frac{2c^2}{G}\,
\biggl(a_iu_i+\sum_{\genfrac{}{}{0pt}{}{j=1}{j\ne i}}^n\frac{a_ia_j}{r_{ij}}\biggr)\,,
\end{alignat}
\end{subequations}
where $u_i:=u(\vec c_i)$ and $r_{ij}:=\Vert\vec c_i-\vec c_j\Vert$.
Comparing this to \eqref{eq:TwoHoleMasses} (and its obvious 
generalization from 2 to $n$ punctures) shows that the only 
difference in the analytic expression for the masses is the 
appearance of $u_i$ (instead of $1$) in the first term on the 
right-hand side of \eqref{eq:PunctureMasses-b}. Thus, formally,
for fixed monopole parameters $a_i$, the switching-on of  
linear and angular momentum (here represented locally by 
trace-free extrinsic curvatures) adds to each individual 
mass a term $a_i(u_i-1)$. This contribution is non-negative
as a consequence of $u\geq 1$. The latter equation follows 
immediately from standard elliptic theory. Indeed, 
\eqref{eq:PunctureEquation} implies  $\Delta_\delta u\leq 0$, 
i.e. that $u$ is \emph{superharmonic}, and hence that $u\geq f$ 
for any continuous harmonic $f$ with the same boundary values, 
i.e. $f\equiv 1$. Note that this argument relies on 
$\signature=-1$. 

Finally we wish to point out an interesting type on non-uniqueness
in writing down initial data of the form \eqref{eq:BowenYorkData}. 
It has to do with the question of whether we wish to enforce the 
inversion symmetries of the type \eqref{eq:OneHole-Isometries} 
to become isometries of the initial geometry. Let us focus on 
$I_1$ as defined in \eqref{eq:OneHole-Isometries}, where we now
denote the radius of inversion $a$ (rather than $r_0$). Dropping 
the subscript 1, we have 
\begin{equation}
\label{eq:OneHoleInversion}
[I(x)]^a=(a^2/r^2)\,x^a\,.
\end{equation}
Its Jacobian is 
\begin{equation}
\label{eq:OneHoleInversion-Jacobian}
[I_*(x)]_b^a=(a^2/r^2)\bigl(\delta^a_b-2\nu^a\nu_b\bigr)\,.
\end{equation}
We note in passing that the matrix in round brackets is 
orthogonal with determinant $-1$. 

It follows that the conformally flat metric $h=\Omega^4\,\delta$ 
satisfies $I^*h=h$ iff
\begin{equation}
\label{eq:OneHoleInversion-Isometry}
(a/r) (\Omega\circ I)=\Omega\,.
\end{equation}
In such a metric the sphere $r=a$ is the fixed-point
set of the isometry $I$ and hence totally geodesic. 
In particular, this implies that it is a stationary 
point of the area function which is equivalent to 
$\partial_r(r^2\Omega^4)=0$ and hence to 
\begin{equation}
\label{eq:OneHoleInversion-InnerBC}
\left[\frac{\partial\Omega}{\partial r}+\frac{\Omega}{2a}\right]_{r=a}=0\,,
\end{equation}
which may also be directly verified from differentiating \eqref{eq:OneHoleInversion-Isometry} with respect to $r$ at $r=a$. 
This would be the condition for \eqref{eq:YorkEquationFlat} at the 
``inner boundary'' in order to produce a solution that gives rise 
to a metric $h$ that has $I$ as an isometry. But that, clearly, 
also puts conditions on the extrinsic curvature $\bar K$, for 
$h$ and $K$ have to satisfy the coupled system of constraints
\eqref{eq:Constraints-1} in the vacuum case $\EMT=0$. 
A sufficient condition is
\begin{subequations}
\label{eq:OneHoleInversion-ExtCurv}
\begin{equation}
\label{eq:OneHoleInversion-ExtCurv-a}
I^*K=\pm K\,.
\end{equation}
Using \eqref{eq:ConfMethodExtCurv-b} and restricting to 
maximal data, $K^c_c=\tau=0$, this is equivalent to 
\begin{equation}
\label{eq:OneHoleInversion-ExtCurv-b}
(a^2/r^2)\,I^*{\bar K}=\pm{\bar K}\,,
\end{equation}
\end{subequations}
where we also made use of \eqref{eq:OneHoleInversion-Isometry}.

Using the global chart $\{x^1,x^2,x^3\}$ on $\reals^3-\{0\}$ 
together wit the flat metric $\bar h=\delta$, we have the 
function $r$ whose value at $x$ is the $\delta$-geodesic 
distance ($(x^1)^2+(x^2)^2+(x^3)^2$) and the following vector 
fields and volume form ($A^a,B^a,\epsilon_{abc}$ being constant 
component functions)
\begin{subequations}
\label{eq:YorkDataStructures}
\begin{alignat}{2}
\label{eq:YorkDataStructures-a}
&\nu&&\,=\,\frac{x^a}{r}\frac{\partial}{\partial x^a}\,,\\
\label{eq:YorkDataStructures-b}
&A&&\,=\,A^a\frac{\partial}{\partial x^a}\,,\\
\label{eq:YorkDataStructures-c}
&B&&\,=\,B^a\frac{\partial}{\partial x^a}\,,\\
\label{eq:YorkDataStructures-d}
&\varepsilon &&\,=\,\tfrac{1}{3!}
\varepsilon_{abc}dx^a\wedge dx^b\wedge dx^c\,.
\end{alignat}
\end{subequations} 
The co-vector fields that arise from these vector fields 
via the isomorphism induced by the flat metric $\delta$ 
are called $\nu^\flat, A^\flat,B^\flat$. Under the 
inversion-map \eqref{eq:OneHoleInversion}, making also 
use of \eqref{eq:OneHoleInversion-Jacobian}, these 
structures behave as follows:
\begin{equation}
\label{eq:YorkDataStructuresTrans-r}
I^*r=r\circ I=\frac{a^2}{r}\,,
\end{equation}
\begin{subequations}
\label{eq:YorkDataStructuresTrans-nu}
\begin{alignat}{2}
\label{eq:YorkDataStructuresTrans-nu-cont}
&I_*\nu&&\,=\,-(a/r)^{-2}\nu\,,\\
&I^*\nu^\flat&&\,=\,-(a/r)^2\nu^\flat\,,
\end{alignat}
\end{subequations}
\begin{subequations}
\label{eq:YorkDataStructuresTrans-X}
\begin{alignat}{2}
\label{eq:YorkDataStructuresTrans-A-cont}
&I_*A&&\,=\,(a/r)^{-2}\bigl(A-2\,\nu(\nu\cdot A)\bigr)\,,\\
&I^*A^\flat&&\,=\,(a/r)^2\bigl(A^\flat-2\,\nu^\flat(\nu\cdot A)\bigr)\,,
\end{alignat}
\end{subequations}
and identically for $B$, where a dot denotes the scalar 
product with respect to the flat metric, 
i.e. $\nu\cdot A=\delta(\nu,A)$, 
\begin{equation}
\label{eq:YorkDataStructuresTrans-delta}
I^*\delta=(a/r)^4\,\delta\,,
\end{equation}
and%
\footnote{A straightforward calculation using \eqref{eq:OneHoleInversion-Jacobian} first yields 
$
I^*\varepsilon
=(a/r)^6 \bigl(
\varepsilon-2\,\nu^\flat\wedge \star\nu^\flat\bigr)
$, where $\star$ denotes the Hodge dual with respect 
to $\delta$. But for any vector field $\nu$ one trivially has 
$\nu^\flat\wedge\varepsilon =0$ and hence, 
now assuming $\nu$ to be also normalized,  
$
0=i_\nu(\nu^\flat\wedge\varepsilon)
=\varepsilon-\nu^\flat\wedge i_\nu\varepsilon
$. 
Using $i_\nu\varepsilon=\star\nu^\flat$, this gives 
$\nu^\flat\wedge\star\nu^\flat=\varepsilon$ and hence \eqref{eq:YorkDataStructuresTrans-epsilon}.}
\begin{equation}
\label{eq:YorkDataStructuresTrans-epsilon}
I^*\varepsilon=-(a/r)^6\varepsilon\,.
\end{equation}

These formulae allow to immediatly write down the 
$I$-transforms of the data \eqref{eq:BowenYorkData}.   
We have, in components, 
\begin{subequations}
\label{eq:YorkDataTrans}
\begin{equation}
\label{eq:YorkDataTrans-1}
\begin{split}
&{I^*\bar K}^{(1)}_{ab}\\
&
=-r^{-2}\bigl(\nu_a A_b+\nu_bA_a+(\delta_{ab}-5\,\nu_a\nu_b)\nu^cA_c\bigr).
\end{split}
\end{equation}
and 
\begin{equation}
\label{eq:YorkDataTrans-2}
{I^*\bar K}^{(2)}_{ab}=-(a/r)^{-2}{\bar K}^{(2)}\,.
\end{equation}
\end{subequations}
This means that ${\bar K}^{(2)}$ as given by \eqref{eq:BowenYorkData-b}
is already antisymmetric in the sense of 
\eqref{eq:OneHoleInversion-ExtCurv-b}, but \eqref{eq:BowenYorkData-a}
is neither symmetric nor antisymmetric. Symmetric or 
antisymmetric data can be obtained by forming the symmetric 
or antisymmetric combination 
\begin{equation}
\label{eq:YorkDataSymmAnti}
{\bar K}^{(1)}_\pm:={\bar K}^{(1)}\pm (a/r)^2 I^*{\bar K}^{(1)}\,.
\end{equation}
which satisfies 
\begin{equation}
\label{eq:YorkDataSymmAnti-expl}
(a/r)^2I^*{\bar K}^{(1)}_\pm=\pm{\bar K}^{(1)}_\pm\,.
\end{equation}
In components they read
\begin{equation}
\label{eq:YorkDataSymmAnti-Comp}
\begin{split}
&\bigl[{\bar K}^{(1)}_\pm\bigr]_{ab}\\
&=1/r^2\bigl(\nu_a A_b+\nu_bA_a-(\delta_{ab}-\nu_a\nu_b)\nu^cA_c\bigr)\\
&\mp a^2/r^4\bigl(\nu_a A_b+\nu_bA_a+(\delta_{ab}-5\,\nu_a\nu_b)\nu^cA_c\bigr).
\end{split}
\end{equation}

Having enforced symmetry with respect to the inversion \eqref{eq:OneHoleInversion} we will obtain an initial-data 
3-manifold with two isometric asymptotically flat ends
whose Poincar\'e charges 
\index{Poincar\'e!charges} 
coincide (possibly up to sign) and are given by those of 
the original data. This is trivially true for angular
momentum and follows for linear momentum from the $1/r^4$ 
fall-off of the second term in \eqref{eq:YorkDataSymmAnti-Comp}. 
It is interesting to note that a term proportional to the 
second term in \eqref{eq:YorkDataSymmAnti-Comp} follows in
case of spherically symmetric extended matter sources~\cite{Bowen:1979}.


\section{Further developments, problems, and outlook}
In this contribution we have explained in some detail the dynamical 
and Hamiltonian formulation of GR. We followed the traditional ADM 
approach in which the basic variables are the Riemannian metric $h$
of space and its conjugate momentum $\pi$, which is essentially 
the extrinsic curvature that $\Sigma$ will assume once the spacetime 
is developed and $\Sigma$ is isometrically embedded in it. 
Attempts to establish a theory of \emph{Quantum Gravity}
\index{Quantum Gravity}
based on the Hamiltonian formulation of GR  suggest that other canonical 
variables are better suited for the mathematical implementation of 
the constraints and the ensuing construction of spaces of states and 
observables~%
\cite{Thiemann:MCQGR}%
\cite{Rovelli:QuantumGravity}%
\cite{Bojowald:CanonicalGravity}.
These variables are a (suitably densitized) orthonormal 3-bein field 
$E$ on $\Sigma$ and the \emph{Ashtekar-Barbero connection}.
\index{Ashtekar!-Barbero connection}%
We have already seen that orientable $\Sigma$  are parallelizable 
so that global fields $E$ do indeed exist. Any field $E$ determines 
a Riemannian metric $h$, which in turn determines its Levi-Civita 
connection. The Ashtekar-Barbero covariant derivative, $\D$, 
differs from the Levi-Civita connection $D$ of $h$ by the 
endomorphism-valued 1-form which associates to each tangent 
vector $X$ the tangent-space endomorphism 
$Y\mapsto\gamma\Wein(X)\times Y$, where $\gamma$ is a dimensionless 
constant, the so-called Barbero-Immirzi-parameter, which was first 
introduced by Immirzi in \cite{Immirzi:1997} on the basis of 
Barbero's generalization \cite{Barbero:1995} of Ashtekar's 
variables. 
\index{Barbero-Immirzi parameter}
Hence we have 
\begin{equation}
\label{eq:AshtekarConnection}
\D_XY=D_XY+\gamma\Wein(X)\times Y\,.
\end{equation}
The multiplication $\times$ is the standard 3-dimensional vector 
product with respect to the metric $h$. It is defined as follows
\begin{equation}
\label{eq:VectorProduct-h}
X\times Y:=\bigl[\star(X^\flat\wedge Y^\flat)\bigr]^\sharp\,,
\end{equation}
where the isomorphisms $\flat$ and $\sharp$ are with respect to $h$
(cf. \eqref{eq:MusicalIsomorphisms}).
The product $\times$ obeys the standard rules: It is bilinear, 
antisymmetric, and $X\times (Y\times Z)=h(X,Z)Y-h(X,Y)Z$. 
Moreover, for any $X$, the endomorphism $Y\mapsto X\times Y$ is
antisymmetric with respect to $h$, i.e. $h(X\times Y,Z)=-h(Y,X\times Z)$, 
and hence it is in the Lie algebra of the orthogonal group of $h$. 
\index{Lie!algebra}\index{algebra!Lie}
In particular this is true for $Y\mapsto \Wein(X)\times Y$,
showing that $\D$ is again metric, i.e. obeys $\D h=0$ 
once its unique extension to all tensor fields is understood. 
Clearly, unlike $D$, the torsion of $\D$ cannot be zero: 
\begin{equation}
\label{eq:AshtekarTorsion}
\begin{split}
T^\mathcal{D}(X,Y)
&=\mathcal{D}_XY-\mathcal{D}_YX-[X,Y]\\
&=\gamma\bigl(\Wein(X)\times Y-\Wein(Y)\times X\bigr)\,.
\end{split}
\end{equation}
Using \eqref{eq:RelWeingartenMapExtCurv} and index notation, 
the curvature tensor for $\D$ is 
\begin{equation} 
\label{eq:AshtekarCurvature-Index}
\begin{split}
R^\D_{abcd}
&=R^D_{abcd}\\
&+\signature\gamma\bigl(D_cK_{dn}-D_dK_{cn}\bigr)\varepsilon^n_{\phantom{n}ab}\\
&-\gamma^2\bigl(K_{ac}K_{bd}-K_{ad}K_{bc}\bigr)\,.
\end{split}
\end{equation}
From this the scalar curvature follows
\begin{equation} 
\label{eq:AshtekarScalarCurvature}
\Scalar^\D=\Scalar^D+\gamma^2\,G^{abcd}K_{ab}K_{cd}\,.
\end{equation}
Camparison with \eqref{eq:EinsteinEquation-00} shows that for 
$\gamma^2=\signature$ the gravitational part of the scalar 
constraint is just ($\signature$ times) the scalar curvature of 
$\D$. This striking simplification of the scalar constraint 
formed the original motivation for the introduction of $\D$ by 
Ashtekar~\cite{Ashtekar:1987}. However, for  $\signature=-1$ 
one needs to complexify the tensor bundle over $\Sigma$ for 
$\gamma=\pm i$ to make sense, and subsequently impose reality 
conditions which re-introduce a certain degree of complication; 
see, e.g.,~\cite{Giulini:1994b} for a compact account not using 
spinors. The usage of $\D$ in the real case was then proposed 
by Barbero in \cite{Barbero:1995} and forms the basic tool in 
\emph{Loop Quantum Gravity} \cite{Thiemann:MCQGR},
\index{Loop Quantum Gravity} 
which has definite technical advantages over the 
metric-based traditional approach.

At this point we wish to inject one word of caution concerning
the possible geometric interpretation of the of Barbero 
connection $\D$, depending on the value of $\gamma$. 
From the defining 
equation \eqref{eq:AshtekarConnection} it is clear that any 
$\D$ explicitly contains extrinsic information, i.e. information 
that refers to the way $\Sigma$ is embedded into spacetime $M$. 
Moreover, unless $\gamma^2=\signature$, this dependence on the 
embedding is such that $\D$ cannot be considered as pull-back 
of a connection defined on (the bundle of linear frames over) 
$M$, as has been pointed out in \cite{Samuel:2000a}. The argument 
is simple: If it were the pull-back of such a connection on $M$, 
its holonomy along a loop in spacetime would be the same for all 
$\Sigma$ containing that loop. That this is not the case for 
$\gamma^2\ne\signature$ can, e.g., be checked for the simple 
example where $M$ is Minkowski space and the loop is a planar 
unit circle that is contained in a flat spacelike hyperplane as well as  
in the constant-curvature spacelike hyperboloid of unit 
future-pointing timelike vectors. The holonomy in the 
latter case turns out to be non-trivial (see \cite{Samuel:2000a}
for the explicit calculation). For the Ashtekar connection, 
i.e. for $\gamma^2=\signature$, we know from its original 
construction that it is the pull-back of a spacetime connection 
(see, e.g., the derivation in  \cite{Giulini:1994b}). But for 
Barbero's generalizations \eqref{eq:AshtekarConnection} with 
$\gamma^2\ne\signature$, and in particular for all real
values of $\gamma$ in the Lorentzian case ($\signature=-1$),
this means that it is impossible to attach a gauge-theoretic 
\emph{spacetime} interpretation to $\D$.

Despite this conceptual shortcoming, the technical advantages 
over the metric-based approach remain. On the other hand, the 
latter is well suited to address certain conceptual 
problems~\cite{Kiefer:QuantumGravity}, 
like e.g. the problem of time that emerges in those cases 
where the Hamiltonian \eqref{eq:GR-Hamiltonian} has no boundary 
terms and is therefore just a sum of constraints. This happens 
in cosmology based on closed $\Sigma$. The motions generated by 
the Hamiltonian are then just pure gauge transformations and the 
question arises whether and how `motion' and `change' are to be 
recovered; see, e.g., \cite{Rovelli:QuantumGravity,Rovelli_ForgetTime:2009}.

Dynamical models in cosmology often start from symmetry 
assumptions that initially reduce the infinitely many 
degrees of freedom to finitely many ones (so-called 
mini-superspace models). Other modes are then treated 
perturbatively in an expansion around the symmetric 
configurations. In these cases quantization in the metric 
representation can be performed, with potentially 
interesting consequences for observational cosmology, 
like the modification of the anisotropy spectrum of the 
cosmic microwave background 
\cite{Kiefer.Kraemer:2012}\cite{Bini.etal:2013}. 
All these attempts make essential use of the Hamiltonian 
theory as described in this contribution. 

\section{Appendix: Group actions on manifolds}
\index{group action}
Let $G$ be a group and $M$ a set. An \emph{action} of $G$ on $M$ 
is a map
\begin{equation}
\label{eq:GroupAction_Def-1}
\Phi: G\times M\rightarrow M
\end{equation}
such that, for all $m\in M$ and $e\in G$ the neutral element, 
\begin{equation}
\label{eq:GroupAction_Def-2}
\Phi(e,m)=m\,,
\end{equation}
and where, in addition, one of the following two conditions 
holds:
\begin{subequations}
\label{eq:GroupAction_Def-3}
\begin{alignat}{2}
\label{eq:GroupAction_Def-3a}
&\Phi\bigl(g,\Phi(h,m)\bigr)
&&\,=\,\Phi(gh,m)\,,\\
\label{eq:GroupAction_Def-3b}
&\Phi\bigl(g,\Phi(h,m)\bigr)
&&\,=\,\Phi(hg,m)\,.
\end{alignat}
\end{subequations}
If \eqref{eq:GroupAction_Def-1}, \eqref{eq:GroupAction_Def-2},
and \eqref{eq:GroupAction_Def-3a} hold we speak of a 
\emph{left action}. A \emph{right action} satisfies   
\index{group action!left and right}%
\index{action!left, of group}\index{action!right, of group}%
\eqref{eq:GroupAction_Def-1}, \eqref{eq:GroupAction_Def-2},
and \eqref{eq:GroupAction_Def-3b}. For a left action we also 
write 
\begin{subequations}
\label{eq:GroupActions_DotNotation}
\begin{equation}
\label{eq:GroupActions_DotNotation-a}
\Phi(g,m)=:g\cdot m
\end{equation}
and for a right action 
\begin{equation}
\label{eq:GroupActions_DotNotation-b}
\Phi(g,m)=:m\cdot g\,.
\end{equation}
\end{subequations}
Equations \eqref{eq:GroupAction_Def-3} then simply become
(group multiplication is denoted by juxtaposition without 
a dot) 
\begin{subequations}
\label{eq:GroupAction_AltDef-3}
\begin{alignat}{2}
\label{eq:GroupAction_AltDef-3a}
&g\cdot(h\cdot m)
&&\,=\,(gh)\cdot m\,,\\
\label{eq:GroupAction_AltDef-3b}
&(m\cdot h)\cdot g
&&\,=\,m\cdot(hg)\,.
\end{alignat}
\end{subequations}
Holding either of the two arguments of $\Phi$ fixed we obtain 
the families of maps 
\begin{equation}
\label{eq:GroupAction_ArgFixed-1}
\begin{split}
\Phi_g:\ 
M&\rightarrow M\\
m&\mapsto\Phi(g,m)
\end{split}
\end{equation}
for each $g\in G$, or
\begin{equation}
\label{eq:GroupAction_ArgFixed-2}
\begin{split}
\Phi_m:\
G&\rightarrow M\\
g&\mapsto\Phi(g,m)
\end{split}
\end{equation}
for each $m\in M$. Note that \eqref{eq:GroupAction_Def-2} and \eqref{eq:GroupAction_Def-3} imply that 
$\Phi_{g^{-1}}=\bigl(\Phi_g)^{-1}$. Hence each $\Phi_g$ is a 
bijection of $M$.  The set of bijections of $M$ will be denoted 
by $\Bij(M)$. It is naturally a group with group multiplication 
being given by composition of maps and the neutral element being 
given by the identity map.  Conditions \eqref{eq:GroupAction_Def-2} and \eqref{eq:GroupAction_Def-3a} are then equivalent to the statement 
that the map $G\rightarrow\Bij(M)$, given by $g\mapsto\Phi_g$, is 
a group homomorphism. Likewise, \eqref{eq:GroupAction_Def-2} and \eqref{eq:GroupAction_Def-3b} is equivalent to the statement that 
this map is a group anti-homomorphism.  

The following terminology is standard: The set $\Stab(m):=\{g\in G :\Phi(g,m)=m\}\subset G$ is called the \emph{stabilizer} of $m$. 
It is easily proven to be a normal subgroup of $G$ satisfying 
$\Stab(g\cdot m)=g\bigl(\Stab(g\cdot m)\bigr)g^{-1}$ for left and 
$\Stab(m\cdot g)=g^{-1}\bigl(\Stab(g\cdot m)\bigr)g$ for right 
actions. The \emph{orbit} of $G$ through $m\in M$ is the set 
$\Orb(m):=\{\Phi(g,m):g\in G\}=:\Phi(G,m)$ (also written 
$G\cdot m$ for left and $m\cdot G$ for right action). 
It is easy to see that two 
orbits are either disjoint or identical. Hence the orbits 
partition $M$.  A point $m\in M$ is  called a \emph{fixed point} 
of the action $\Phi$ iff $\Stab(m)=G$.  An action $\Phi$ is called 
\emph{effective} iff $\Phi(g,m)=m$ for all $m\in M$ implies $g=e$;
i.e., ``only the group identity moves nothing''. Alternatively, 
we may say that effectiveness is equivalent to the map 
$G\mapsto\Bij(M)$, $g\mapsto\Phi_g$, being injective; i.e., 
$\Phi_g=\Id_M$ implies $g=e$. The action $\Phi$ is called \emph{free}
iff $\Phi(g,m)=m$ for some $m\in M$ implies $g=e$; i.e., 
``no $g\ne e$ fixes a point''. This is equivalent to the 
injectivity  of all maps $\Phi_m:G\rightarrow M$, 
$g\mapsto\Phi(g,m)$, which can be expressed by saying that 
all orbits of $G$ in $M$ are faithful images of $G$. 

Here we are interested in \emph{smooth} actions. For this we need to 
assume that $G$ is a Lie group, 
\index{Lie!group}
that $M$ is a  differentiable manifold, and that the map \eqref{eq:GroupAction_Def-1} is smooth. We denote 
by $\exp: T_eG\rightarrow G$ the exponential map. For each 
$X\in T_eG$ there is a vector field $V^X$ on $M$, given by 
\begin{equation}
\label{eq:GroupAction_FundVecField}
\begin{split}
V^X(m)
&=\frac{d}{dt}\Big\vert_{t=0}\Phi\bigl(\exp(tX),m\bigr)\\
&=\Phi_{m*e}(X)\,.
\end{split}
\end{equation}
Recall that $\Phi_{m*e}$ denotes the differential of the map $\Phi_{m}$
evaluated at $e\in G$. $V^X$ is also called the \emph{fundamental vector 
field} on $M$ associated to the action $\Phi$ of $G$ and to $X\in T_eG$.
(We will later write $\Lie(G)$ for $T_eG$, after we have discussed 
\emph{which} Lie structure on $T_eG$ we choose.)    

In passing we note that from \eqref{eq:GroupAction_FundVecField}
it already follows that the flow map of $V^X$ is given by 
\begin{equation}
\label{eq:GroupAction_FundVF-FlowMap}
\Fl^{V^X}_t(m)=\Phi(\exp(tX),m)\,.
\end{equation}
This follows from $\exp(sX)\exp(tX)=\exp\bigl((s+t)X\bigr)$
and \eqref{eq:GroupAction_Def-3} (any of them), which imply 
\begin{equation}
\label{eq:GroupAction_FundVF-FlowMapComp}
\Fl_s^{V^X}\circ\Fl_t^{V^X}=\Fl_{s+t}^{V^X}
\end{equation}
on the domain of $M$ where all three maps appearing in \eqref{eq:GroupAction_FundVF-FlowMapComp} are defined. 
Uniqueness of flow maps for vector fields then suffices to 
show that \eqref{eq:GroupAction_FundVF-FlowMap} is indeed 
the flow of $V^X$. 
 
Before we continue with the general case, we have a closer 
look at the special cases where $M=G$ and $\Phi$ is 
either the left translation of $G$ on $G$, $\Phi(g,h)=L_g(h):=gh$, 
or the right translation, $\Phi(g,h)=R_g(h):=hg$. The corresponding 
fundamental vector fields \eqref{eq:GroupAction_FundVecField} 
are denoted by $V^X_R$ and $V^X_L$ respectively:
\begin{subequations}
\label{eq:GroupActions_DefFundVF-OnGroups}
\begin{alignat}{2}
\label{eq:GroupActions_DefFundVF-OnGroups-a}
&V_R^X(h)
&&\,=\,\frac{d}{dt}\Big\vert_{t=0}\Bigl(\exp(tX)\,h\Bigr)\,,\\
\label{eq:GroupActions_DefFundVF-OnGroups-b}
&V_L^X(h)
&&\,=\,\frac{d}{dt}\Big\vert_{t=0}\Bigl(h\,\exp(tX)\Bigr)\,.
\end{alignat}
\end{subequations}
The seemingly paradoxical labeling of $R$ for left and $L$ 
for right translation finds its explanation in the fact 
that $V^X_R$ is right and  $V^X_L$ is left 
invariant, i.e., $R_{g*}V^X_R=V^X_R$ and $L_{g*}V^X_L=V^X_L$.
Recall that the latter two equations are shorthands for 
\begin{subequations}
\label{eq:GroupAction_LRinvariantVF}
\begin{alignat}{2}
\label{eq:GroupAction_LRinvariantVF-a}
& R_{g*h}V^X_R(h)&&\,=\,V^X_R(hg)\,,\\
\label{eq:GroupAction_LRinvariantVF-b}
& L_{g*h}V^X_L(h)&&\,=\,V^X_L(gh)\,.
\end{alignat}
\end{subequations}
The proofs of \eqref{eq:GroupAction_LRinvariantVF-a} 
only uses \eqref{eq:GroupActions_DefFundVF-OnGroups-a} 
and the chain rule: 
\begin{subequations}
\label{eq:GroupAction_LeftRightInvarianceGroupFundVF}
\begin{equation}
\label{eq:GroupAction_LeftRightInvarianceGroupFundVF-a}
\begin{split}
R_{g*h}V^X_R(h)
&=R_{g*h}\frac{d}{dt}\Big\vert_{t=0}\Bigl(\exp(tX)\,h\Bigr)\\
&=\frac{d}{dt}\Big\vert_{t=0}R_g\Bigl(\exp(tX)\,h\Bigr)\\
&=\frac{d}{dt}\Big\vert_{t=0}\Bigl(\exp(tX)\,hg\Bigr)\\
&=V^X_R(hg)\,.
\end{split}
\end{equation}
Similarly, the proof of 
\eqref{eq:GroupAction_LRinvariantVF-b} starts from 
\eqref{eq:GroupActions_DefFundVF-OnGroups-b}: 
\begin{equation}
\label{eq:GroupAction_LeftRightInvarianceGroupFundVF-b}
\begin{split}
L_{g*h}V^X_L(h)
&=L_{g*h}\frac{d}{dt}\Big\vert_{t=0}\Bigl(h\,\exp(tX)\Bigr)\\
&=\frac{d}{dt}\Big\vert_{t=0}L_g\Bigl(h\,\exp(tX)\Bigr)\\
&=\frac{d}{dt}\Big\vert_{t=0}\Bigl(gh\,\exp(tX)\Bigr)\\
&=V^X_L(gh)\,.
\end{split}
\end{equation}
\end{subequations}
In particular, we have
\begin{subequations}
\begin{alignat}{2}
&V^X_R(g)=R_{g*e}V^X_L(e)&&\,=\,R_{g*e}X\,,\\
&V^X_L(g)=L_{g*e}V^X_R(e)&&\,=\,L_{g*e}X\,,
\end{alignat}
\end{subequations}
showing that the vector spaces of right/left invariant vector
fields on $G$ are isomorphic to $T_eG$. Moreover, the vector spaces 
of right/left invariant vector fields on $G$ are Lie algebras, 
\index{Lie!algebra}\index{algebra!Lie}
the Lie product being their ordinary commutator 
(as vector fields). This is true because the operation of commuting
vector fields commutes with push-forward maps of diffeomorphisms: 
$\phi_*[V,W]=[\phi_*V,\phi_*W]$. This implies that the commutator 
of right/left invariant vector fields is again right/left invariant. 
Hence the isomorphisms can be used to turn $T_eG$ into a Lie algebra,
\index{Lie!algebra}\index{algebra!Lie}
identifying it either with the Lie algebra of right- or left-invariant    
vector fields. The standard convention is to choose the latter.  
Hence, for any $X,Y\in\Lie(G)$, one defines 
\begin{equation}
\label{eq:GroupAction-DefLieAlg-a}
[X,Y]:=[V_L^X,V_L^Y](e)\,.
\end{equation}
$T_eG$ endowed with \emph{that} structure is called $\Lie(G)$.
Clearly, this turns $V_L:\Lie(G)\rightarrow\ST G$, $X\mapsto V_L^X$,  
into a Lie homomorphism:
\index{Lie!homomorphism}
\begin{equation}
\label{eq:GroupAction-LeftCommutator}
V_L^{[X,Y]}=[V_L^X,V_L^Y]\,.
\end{equation}
As a consequence, $V_R:\Lie(G)\rightarrow\ST G$, $X\mapsto V_R^X$, 
now turns out to be an \emph{anti} Lie homomorphism, i.e., 
\index{Lie!anti-isomorphism}
to contain an extra minus sign: 
\begin{equation}
\label{eq:GroupAction-RightCommutator}
V_R^{[X,Y]}:=-\,[V_R^X,V_R^Y]\,.
\end{equation}
This can be proven directly but will also follow from the 
more general considerations below.

On $G$ consider the map 
\begin{equation}
\label{eq:GoupAction_DefConjugation}
\begin{split}
C: G\times G & \rightarrow G\\
(h,g) & \mapsto hgh^{-1}\,.
\end{split}
\end{equation} 
For fixed $h$ this map, $C_h:G\rightarrow G$, $g\mapsto C_h(g)=hgh^{-1}$, 
is an automorphism (i.e., self-isomorphism) of $G$. Automorphisms of 
$G$ form a group (multiplication being composition of maps) which we
denote by $\mathrm{Aut}(G)$. It is immediate that the map $C\rightarrow\mathrm{Aut}(G)$, $h\mapsto C_h$, is a 
homomorphism of groups; i.e., 
\begin{subequations}
\label{eq:GroupAction_InnerAut}
\begin{alignat}{1}
\label{eq:GroupAction_InnerAut-a}
C_e&\,=\,\Id_G\,,\\
\label{eq:GroupAction_InnerAut-b}
C_h\circ C_k&\,=\,C_{hk}\,.
\end{alignat}
\end{subequations}

Taking the differential at $e\in G$ of $C_h$ we obtain a linear 
self-map of $T_eG$, which we call $\Ad_h$:
\begin{subequations}
\label{eq:GroupAction_AdjointMap}
\begin{equation}
\label{eq:GroupAction_AdjointMap-a}
\Ad_h:=C_{h*e}: T_eG\rightarrow T_eG\,.
\end{equation}
Differentiating both sides of both equations 
\eqref{eq:GroupAction_InnerAut} at $e\in G$, using the 
chain rule together with $C_k(e)=e$ for the second, 
we infer that 
\begin{alignat}{2}
\label{eq:GroupAction_AdjointMap-b}
\Ad_e&\,=\,\Id_{T_eG}\,,\\
\label{eq:GroupAction_AdjointMap-c}
\Ad_h\circ\Ad_k&\,=\,\Ad_{hk}\,.
\end{alignat}
\end{subequations}
This implies, firstly, that each linear map 
\eqref{eq:GroupAction_AdjointMap-a} is invertible, i.e. 
an element of the \emph{general linear group} $\GL(T_eG)$ 
of the vector space $T_eG$, and, secondly, that the map
\begin{equation}
\label{eq:GroupAction_AdjointRep}
\begin{split}
\Ad: G&\rightarrow\GL(T_eG)\\
h&\mapsto\Ad_h
\end{split}
\end{equation}
is a group homomorphism. In other words, $\Ad$ is a 
linear representation of $G$ on $T_eG$, called the 
\emph{adjoint representation}.
\index{group representation!adjoint} 

In \eqref{eq:GroupAction_LRinvariantVF} we saw that $V^X_R$ 
and $V^X_L$ are invariant under the action of right and left 
translations respectively (hence their names). But what happens 
if we act on $V^X_R$ with left and on  $V^X_L$ with right 
translations? The answer is obtained from straightforward 
computation. In the first case we get: 
\begin{subequations}
\label{eq:GroupActions_AdEquivRightLeftVF}
\begin{equation}
\label{eq:GroupActions_AdEquivRightLeftVF-a}
\begin{split}
L_{g*h}\bigl(V^X_R(h)\bigr)
&=L_{g*h}\frac{d}{dt}\Big\vert_{t=0}\Bigl(\exp(tX)\,h\Bigr)\\
&=\frac{d}{dt}\Big\vert_{t=0}\Bigl(g\,\exp(tX)\,h\Bigr)\\
&=\frac{d}{dt}\Big\vert_{t=0}\Bigl(C_g\bigl(\exp(tX)\bigr)\,gh\Bigr)\quad\\
&=V_R^{\Ad_g(X)}(gh)\,,
\end{split}
\end{equation}
where we used \eqref{eq:GroupAction_AdjointMap} in the last and 
the definition of $V_R^X$ in the first and last step. Similarly, 
in the second case we have 
\begin{equation}
\label{eq:GroupActions_AdEquivRightLeftVF-b}
\begin{split}
R_{g*h}\bigl(V^X_L(h)\bigr)
&=R_{g*h}\frac{d}{dt}\Big\vert_{t=0}\Bigl(h\,\exp(tX)\,h\Bigr)\\
&=\frac{d}{dt}\Big\vert_{t=0}\Bigl(h\,\exp(tX)\,g\Bigr)\\
&=\frac{d}{dt}\Big\vert_{t=0}\Bigl(hg\,C_{g^{-1}}\bigl(\exp(tX)\bigr)\Bigr)\\
&=V_L^{\Ad_{g^{-1}}(X)}(gh)\,.
\end{split}
\end{equation}
\end{subequations}
Taking the differential of $\Ad$ at $e\in G$ we obtain a linear 
map from $T_eG$ into $\End(T_eG)$, the linear space of 
endomorphisms of $T_eG$ (linear self-maps of $T_eG$).
\begin{equation}
\label{eq:GroupAction_adointRep}
\begin{split}
\ad:=\Ad_{*e}: T_eG&\rightarrow\End(T_eG)\\
X&\mapsto \ad_X\,.
\end{split}
\end{equation}
Now, we have 
\begin{equation}
\label{eq:GroupAction_adjointRepIsComm}
\ad_X(Y)=[X,Y]
\end{equation}
where the right-hand side is defined in 
\eqref{eq:GroupAction-DefLieAlg-a}. The proof of \eqref{eq:GroupAction_adjointRepIsComm} starts from 
the fact that the commutator of two vector fields can be 
expressed in terms of the Lie derivative of the second 
with respect to the first vector field in the commutator, 
and the definition of the Lie derivative. We recall from \eqref{eq:GroupAction_FundVF-FlowMap} that the flow of 
the left invariant vector fields is given by right 
translation: $\Fl^{V^X_L}_t(g)=g\,\exp(tX)$. Then we have 
\begin{subequations}
\label{eq:GroupAction_ProofOfLeftRightComm}
\begin{equation}
\label{eq:GroupAction_ProofOfLeftRightComm-a}
\begin{split}
[X,Y]
&=[V^X_L,V^Y_L](e)\\
&=(L_{V^X_L}V^Y_L)(e)\\
&=\frac{d}{dt}\Big\vert_{t=0}\Fl^{V^X_L}_{(-t)*}
\Bigl(V^Y_L(\Fl_t^{V^X_L}(e))\Bigr)\\
&=\frac{d}{dt}\Big\vert_{t=0}\Fl^{V^X_L}_{(-t)*}
\frac{d}{ds}\Big\vert_{s=0}\Fl^{V^Y_L}_{s}\Bigl(\Fl_t^{V^X_L}(e)\Bigr)\\
&=\frac{d}{dt}\Big\vert_{t=0}\frac{d}{ds}\Big\vert_{s=0}
\exp(tX)\exp(sY)\exp(-tX)\\
&=\frac{d}{dt}\Big\vert_{t=0}\Ad_{\exp(tX)}(Y)\\
&=\ad_X(Y)\,.
\end{split}
\end{equation}
A completely analogous consideration, now using 
$\Fl^{V^X_R}_{t}(g)=\exp(tX)\,g$, allows to compute the 
commutator of the right-invariant vector fields 
evaluated at $e\in G$: 
\begin{equation}
\label{eq:GroupAction_ProofOfLeftRightComm-b}
\begin{split}
[V^X_R,V^Y_R](e)
&=(L_{V^X_R}V^Y_R)(e)\\
&=\frac{d}{dt}\Big\vert_{t=0}\Fl^{V^X_R}_{(-t)*}
\Bigl(V^Y_R(\Fl_t^{V^X_R}(e))\Bigr)\\
&=\frac{d}{dt}\Big\vert_{t=0}\Fl^{V^X_R}_{(-t)*}
\frac{d}{ds}\Big\vert_{s=0}\Fl^{V^Y_R}_{s}\Bigl(\Fl_t^{V^X_R}(e)\Bigr)\\
&=\frac{d}{dt}\Big\vert_{t=0}\frac{d}{ds}\Big\vert_{s=0}
\exp(-tX)\exp(sY)\exp(tX)\\
&=\frac{d}{dt}\Big\vert_{t=0}\Ad_{\exp(-tX)}(Y)\\
&=-\ad_X(Y)\\
&=-[X,Y]\,.
\end{split}
\end{equation}
\end{subequations}
Equation \eqref{eq:GroupAction-RightCommutator} now follows 
if we act on both sides of $[V^X_R,V^Y_R](e)=-[X,Y]$
with $R_{g*e}$ and use \eqref{eq:GroupAction_LRinvariantVF-a}. 

We now return to the general case where $M$ is any manifold and 
the vector field $V^X$ is defined by an action $\Phi$ as in 
\eqref{eq:GroupAction_FundVecField} and whose flow map is given by \eqref{eq:GroupAction_FundVF-FlowMap}. Now, given that 
$\Phi$ is a \emph{right action}, we obtain
\begin{subequations}
\label{eq:GroupAction_GenCommRightLeftAction}
\begin{equation}
\label{eq:GroupAction_GenCommRightLeftAction-a}
\begin{split}
&\bigl[V^X,V^Y\bigr](m)\\
&=(L_{V^X}V^Y)(m)\\
&=\frac{d}{dt}\Big\vert_{t=0}\Fl^{V^X}_{(-t)*}
\Bigl(V^Y(\Fl_t^{V^X}(m))\Bigr)\\
&=\frac{d}{dt}\Big\vert_{t=0}\Fl^{V^X}_{(-t)*}
\frac{d}{ds}\Big\vert_{s=0}\Fl^{V^Y}_{s}\Bigl(\Fl_t^{V^X}(m)\Bigr)\\
&=\frac{d}{dt}\Big\vert_{t=0}\frac{d}{ds}\Big\vert_{s=0}
\Phi\bigl(\exp(tX)\exp(sY)\exp(-tX),m\bigr)\\
&=\frac{d}{dt}\Big\vert_{t=0}\Phi_{m*e}\bigl(\Ad_{\exp(tX)}(Y)\bigr)\\
&=V^{\ad_X(Y)}(m)\\
&=V^{[X,Y]}(m)
\end{split}
\end{equation}
where we used \eqref{eq:GroupAction_FundVF-FlowMap} and \eqref{eq:GroupAction_Def-3b} at the fourth and \eqref{eq:GroupAction_adjointRepIsComm} at the 
last equality. 
Similarly, if $\Phi$ is a \emph{left action}, we have 
\begin{equation}
\label{eq:GroupAction_GenCommRightLeftAction-b}
\begin{split}
&\bigl[V^X,V^Y\bigr](m)\\
&=(L_{V^X}V^Y)(m)\\
&=\frac{d}{dt}\Big\vert_{t=0}\Fl^{V^X}_{(-t)*}
\Bigl(V^Y(\Fl_t^{V^X}(m))\Bigr)\\
&=\frac{d}{dt}\Big\vert_{t=0}\Fl^{V^X}_{(-t)*}
\frac{d}{ds}\Big\vert_{s=0}\Fl^{V^Y}_{s}\Bigl(\Fl_t^{V^X}(m)\Bigr)\\
&=\frac{d}{dt}\Big\vert_{t=0}\frac{d}{ds}\Big\vert_{s=0}
\Phi\bigl(\exp(-tX)\exp(sY)\exp(tX),m\bigr)\\
&=\frac{d}{dt}\Big\vert_{t=0}\Phi_{m*e}\bigl(\Ad_{\exp(-tX)}(Y)\bigr)\\
&=-V^{\ad_X(Y)}(m)\\
&=-V^{[X,Y]}(m)
\end{split}
\end{equation}
\end{subequations}
where we used~\eqref{eq:GroupAction_FundVF-FlowMap} and \eqref{eq:GroupAction_Def-3a} at the fourth and again  \eqref{eq:GroupAction_adjointRepIsComm} at the 
last equality.

Finally we derive the analog of 
\eqref{eq:GroupActions_AdEquivRightLeftVF} in the general case. 
This corresponds to computing the push-forward of $V^X$ under 
$\Phi_g$. If $\Phi$ is a left action we will obtain the analog 
of \eqref{eq:GroupActions_AdEquivRightLeftVF-a}, and the analog 
of \eqref{eq:GroupActions_AdEquivRightLeftVF-b}  if $\Phi$ is
a right action. For easier readability we shall also make use of 
the notation \eqref{eq:GroupActions_DotNotation}. For a left 
action we then get 
\begin{subequations}
\label{eq:GroupActions_AdEquivLeftRightaction}
\begin{equation}
\label{eq:GroupActions_AdEquivLeftRightaction-a}
\begin{split}
\Phi_{g*m}\bigl(V^X(m)\bigr)
&=\Phi_{g*m}\frac{d}{dt}\Big\vert_{t=0}\Phi\bigl(\exp(tX),m\bigr)\\
&=\frac{d}{dt}\Big\vert_{t=0}\Phi\bigl(g\,\exp(tX),m\bigr)\\
&=\frac{d}{dt}\Big\vert_{t=0}\Phi\bigr(C_g(\exp(tX)),g\cdot m\bigr)\\
&=\Phi_{(g\cdot m)*e}\frac{d}{dt}\Big\vert_{t=0}C_g\bigl(\exp(tX)\bigr)\\
&=\Phi_{(g\cdot m)*e}\bigl(\Ad_g(X)\bigr)\\
&=V^{\Ad_g(X)}(g\cdot m)\\
&=V^{\Ad_g(X)}\bigl(\Phi(g,m)\bigr)\,.
\end{split}
\end{equation}
Similarly, if $\Phi$ is a right action, 
\begin{equation}
\label{eq:GroupActions_AdEquivLeftRightaction-b}
\begin{split}
\Phi_{g*m}\bigl(V^X(m)\bigr)
&=\Phi_{g*m}\frac{d}{dt}\Big\vert_{t=0}\Phi\bigl(\exp(tX),m\bigr)\\
&=\frac{d}{dt}\Big\vert_{t=0}\Phi\bigl(\exp(tX)\,g,m\bigr)\\
&=\frac{d}{dt}\Big\vert_{t=0}\Phi\bigr(C_{g^{-1}}(\exp(tX)), m\cdot g\bigr)\\
&=\Phi_{(m\cdot g)*e}\frac{d}{dt}\Big\vert_{t=0}C_{g^{-1}}\bigl(\exp(tX)\bigr)\\
&=\Phi_{(m\cdot g)*e}\bigl(\Ad_{g^{-1}}(X)\bigr)\\
&=V^{\Ad_{g^{-1}}(X)}(m\cdot g)\\
&=V^{\Ad_{g^{-1}}(X)}\bigl(\Phi(g,m)\bigr)\,.
\end{split}
\end{equation}
\end{subequations}

\vspace{2.65cm}
\noindent
\textbf{Acknowledgements:} I thank Lukas Brunkhorst, 
Christian Pfeifer and Timo Ziegler for carefully 
reading the manuscript and pointing out errors.

\newpage
\addcontentsline{toc}{section}{References}
\setlength{\itemsep}{2ex}
\bibliographystyle{plain}
\bibliography{RELATIVITY,HIST-PHIL-SCI,MATH,QM} 

\newpage
\addcontentsline{toc}{section}{Index}
\printindex

\end{document}